\PassOptionsToPackage{pdfpagelabels=false}{hyperref}
\documentclass[fleqn,usenatbib]{mnras}
\usepackage{amsmath}
\usepackage{xcolor, graphicx}
\usepackage{amssymb}
\usepackage{soul}
\usepackage{txfonts}
\usepackage{journal_names}
\usepackage{caption}

\newcommand{\Msun}{\rm M_\odot}
\newcommand{\vlos}{V_{\rm los}}
\newcommand{\kms}{\mbox{km s$^{-1}$}}
\newcommand{\re}{R_{\rm e}}
\newcommand{\rin}{r_{\rm in}}
\newcommand{\rout}{r_{\rm out}}
\newcommand{\ml}{\rm M_\odot/ L_{\odot, K}}
\def\Sersic/{{S\'ersic}}

\title[The mass distribution in ETGs within 5$R_e$ and beyond]
{The SLUGGS Survey: The mass distribution in early--type galaxies within five effective radii and beyond}
\author[Alabi~et~al.~ ]
{Adebusola B. Alabi$^{1}$\thanks{Email: aalabi@swin.edu.au} , Duncan A. Forbes$^{1}$, Aaron J. Romanowsky$^{2,3}$, Jean P. Brodie$^{3}$,\\
\\
\normalfont{\LARGE Jay Strader$^{4}$, Joachim Janz$^{1}$, Vincenzo Pota$^{5}$, Nicola Pastorello$^{1}$, Christopher Usher$^{1,6}$,} \\
\\
\normalfont{\LARGE Lee R. Spitler$^{7,8}$, Caroline Foster$^{7}$, Zachary G. Jennings$^{3}$, Alexa Villaume$^{3}$, Sreeja Kartha$^{1}$}\\
\\
$^1$ Centre for Astrophysics \& Supercomputing, Swinburne University, Hawthorn VIC 3122, Australia\\
$^2$ Department of Physics and Astronomy, San Jos\'e State University, San Jose, CA 95192, USA\\
$^3$ University of California Observatories, 1156 High Street, Santa Cruz, CA 95064, USA\\
$^4$ Department of Physics and Astronomy, Michigan State University, East Lansing, Michigan 48824, USA\\
$^5$ INAF -- Observatorio Astronomico di Capodimonte, Salita Moiariello, 16, 80131 Napoli, Italy\\
$^6$ Astrophysics Research Institute, Liverpool John Moores University, Liverpool L3 5RF, United Kingdom\\
$^7$ Australian Astronomical Observatory, PO Box 915, North Ryde, NSW 1670, Australia\\
$^8$ Department of Physics and Astronomy, Macquarie University, North Ryde NSW 2109, Australia\\
}

\begin{document}

\date{Accepted today}

\pagerange{\pageref{firstpage}--\pageref{lastpage}} \pubyear{2015}

\maketitle

\label{firstpage}
\begin{abstract}
We study mass distributions within and beyond 5~effective radii ($\re$) in 23 early--type galaxies from the SLUGGS survey, using their globular cluster (GC) kinematic data. The data are obtained with Keck/DEIMOS spectrograph, and consist of line--of--sight velocities for ${\sim}3500$ GCs, measured with a high precision of ${\sim}15\ \kms$ per GC and extending out to ${\sim}13\ \re$. We obtain the mass distribution in each galaxy using the tracer mass estimator of Watkins et al. and account for kinematic substructures, rotation of the GC systems and galaxy flattening in our mass estimates.

The observed scatter between our mass estimates and results from the literature is less than 0.2 dex. The dark matter fraction within $5~\re$ ($f_{\rm DM}$) increases from ${\sim}0.6$ to ${\sim}0.8$ for low-- and high--mass galaxies, respectively, with some intermediate--mass galaxies ($M_*{\sim}10^{11}\Msun$) having low $f_{\rm DM}\sim0.3$, which appears at odds with predictions from simple galaxy models. We show that these results are independent of the adopted orbital anisotropy, stellar mass--to--light ratio, and the assumed slope of the gravitational potential. However, the low $f_{\rm DM}$ in the ${\sim}10^{11}\Msun$ galaxies agrees with the cosmological simulations of Wu et al. where the pristine dark matter distribution has been modified by baryons during the galaxy assembly process. We find hints that these $M_*{\sim}10^{11}\Msun$ galaxies with low $f_{\rm DM}$ have very diffuse dark matter haloes, implying that they assembled late. Beyond $5~\re$, the $M/L$ gradients are steeper in the more massive galaxies and shallower in both low and intermediate mass galaxies.

\end{abstract}


\section{Introduction}
One of the fundamental properties of galaxies is their total mass (baryonic + dark matter). The total mass profiles of giant  galaxies 
are dominated by baryons in the central parts, with the dark matter (DM) component becoming more dominant at large radii, eventually dominating the total mass budget. Studying the 
distribution of these mass components provides a viable way of testing galaxy formation and evolution models. For example, at the same stellar mass, early--type galaxies (ETGs) are
thought to have a higher DM concentration compared to spiral galaxies. This is because the central portions of the haloes in ETGs are already in place at a higher redshift compared to spiral galaxies for the same galaxy mass \citep[e.g.][]{Thomas_2009}. 

For late--type galaxies, it is relatively easy to determine the total mass distribution out to large radii using the motions of the readily available HI gas as a tracer of the galaxy potential. However, this exercise is more difficult for (individual) ETGs. This is because ETGs are generally poor in cold gas, their stellar motions are predominantly random by nature and at large galactocentric radii, they are optically faint. These properties combine to make studies of the mass distribution in ETGs challenging. Yet, to properly understand the DM content in ETGs, one needs to probe out to at least five effective radii ($\re$), where DM is expected to begin dominating the enclosed mass \citep{Romanowsky_2003, Napolitano_2005, Cappellari_2015}.

Various mass tracers such as planetary nebulae \citep[PNe; e.g.,][]{Morganti_2013}, globular clusters  \citep[GCs; e.g.,][]{Pota_2015} and diffuse X--ray gas \citep[e.g.,][]{Su_2014} have been used to explore the mass distribution in ETGs out to large radii. For PNe and GC based studies, their orbital distributions 
are usually not known, and are notoriously difficult to determine due to the mass--anisotropy degeneracy \citep{Binney_1982}. The discrete kinematic data are often binned and smoothed in order to determine the mass profile, leading to loss of vital information. Since binning is impracticable for sparse samples, only galaxies with relatively rich systems of bright tracers, i.e., massive ETGs, are usually studied. This limitation also extends to X--ray based studies, where X--ray haloes are observed mostly around massive galaxies that usually reside in dense environments. Hence most ETGs with radially extended mass modelling results in the literature are the more massive ones, with the low and intermediate mass ETGs usually overlooked. Furthermore, ETGs tend to be studied one at a time, with different methods and assumptions. This makes it problematic to compare the results in a systematic way. 

Apart from the observational difficulties, results at large galactocentric radii in some intermediate mass ETGs ($M_*{\sim}10^{11} \Msun$) have suggested inconsistencies with the predictions from 
$\rm \Lambda CDM$ cosmology (e.g., \citealt{Romanowsky_2003}; \citealt{Napolitano_2009}; \hypertarget{D+12}{\citealt{Deason_2012}}, \hyperlink{D+12}{D+12} hereafter). While results from the well studied massive ETGs agree with the prediction that in the outer halo, DM dominates the galaxy mass budget, the same is less clear in intermediate mass ETGs, as different mass modelling techniques using the same tracers seem to produce contradictory results (see \citealt{Romanowsky_2003, Napolitano_2009}, \hyperlink{D+12}{D+12}, \citealt{Morganti_2013} for the peculiar case of NGC~4494). The situation is even worse for low stellar mass ETGs, since they have hardly been studied out to large radii. It is therefore imperative to probe the DM halo in these galaxies systematically. 

The traditional methods of mass modelling are difficult to apply to GC kinematic data for sub--$L_{*}$ ETGs. It is therefore desirable to have mass estimators that use the projected kinematic information directly without the need for binning -- an approach that lends itself to relatively sparse tracer populations. Examples include the Virial Mass Estimator (VME) from \citet{Limber_1960} and the Projected Mass Estimator (PME) from \citet{Bahcall_1981}, later modified by \citet{Heisler_1985}. These assume that the tracers (e.g. GCs, PNe, satellite galaxies) have a number density distribution -- $n(r)$, that directly follows the total mass density of the galaxy -- $\rho(r)$, i.e. $n(r) \propto \rho(r)$. This is not usually true since the total mass density is dominated by the dark matter component, especially at large radii. The VME and PME are in principle similar to earlier attempts at estimating mass in a spherically symmetric, self--gravitating system where the tracers orbit a central point mass \citep[e.g.,][]{Zwicky_1937, Schwarzschild_1954}.

A more recent class of mass estimators, the Tracer Mass Estimators (TMEs), however, allows for the more general case where the tracers and total mass densities, while both assumed to be scale--free, have different distributions. They were first introduced by \citet{Evans_2003} and later modified by \hypertarget{W+10}{\citet{Watkins_2010}}, hereafter \hyperlink{W+10}{W+10}, and \cite{An_2011} (see also \citealt{Watkins_2013} for an axisymmetric Jeans modelling of discrete kinematic tracers). A tracer population with number density $n(r) \propto r^{-\gamma}$ resides in a 
power--law gravitational potential of the form $\Phi(r) \propto r^{-\alpha}$. The total mass density, $\rho$, is directly related to the gravitational potential via Poisson's equation and hence it has the power--law form 
$\rho(r) \propto r^{-\alpha-2}$. Also, the TMEs assume that the tracer population is spherically symmetric and that galaxies are in steady state equilibrium, i.e., virialized. 

This paper uses the GC kinematic data from the SLUGGS \footnote{http://sluggs.swin.edu.au} (SAGES Legacy Unifying Globulars and Galaxies Survey, \citealt{Brodie_2014}) and TMEs to study in a homogeneous way the mass distribution within and beyond 5~$\re$ in ETGs. The galaxies we study cover a stellar mass range of $1.9\times10^{10}-4.0\times10^{11} \Msun$ and include galaxies from cluster, group and field environments. We therefore extend the range of galaxies with mass profiles beyond 5~$\re$ into the low stellar mass galaxy regime. 
The science questions we seek to answer are straightforward -- Are TMEs appropriate mass estimators using GCs as the tracers? How is mass distributed between baryons and DM in the outer haloes of ETGs, especially in intermediate and low stellar mass ETGs? Are ETGs always DM dominated in their outer parts? If they are not always DM dominated, as some results from the literature seem to suggest, then why? Are the measured mass and DM content estimates consistent with predictions from $\Lambda$CDM models?

In Section \ref{observation} we describe the observations, data reduction and data preparation. 
Section \ref{analy} starts by introducing in detail the TMEs, defines the mass estimator parameters and quantifies the sensitivity of the mass estimators to these parameters. In this section, we also quantify the effects of galaxy flattening, rotation and kinematic substructures on our mass estimates. We study the deviation of ETGs from isotropy. We obtain the DM fractions within 5~$\re$ and beyond, and compare with expectations from a simple galaxy model, composed of DM and stars only. In 
Section \ref{discu} we discuss how predictions and observations compare. We complete this section by studying correlations between the DM fraction and various galaxy properties. In Section \ref{concl} we summarise our results. 

\begin{table}
\centering
{\small \caption{Summary of the spectroscopic observations for our galaxy sample.} \label{tab:observ}}
\begin{tabular}{@{}r c c c c c c}
\hline
\hline
Galaxy & Masks & Exp. Time & $N_{\rm GC}$ & $N_{\rm sub}$ & $R_{\rm max}$\\
 $\rm [NGC]$ & 		& [hrs] &  & & $[\re]$ &\\
\hline
720    &   5    &    10.6   &  69 & -- & 19.05 &\\
821    &   7    &    11.2   &  69 & -- &	 8.70 &\\
1023   &   4    &     8.8   & 115 & 21 & 16.15 &\\
1400   &   4    &     9.0   &  69 & -- & 20.62 &\\
1407   &  10    &    22.0   & 372 & -- & 14.14 &\\
2768   &   5    &    13.9   & 107 & -- & 11.36 &\\
3115   &   5    &    14.0   & 150 & 12 & 18.35 &\\
3377   &   4    &     8.3   & 122 & -- & 14.34 &\\
3608   &   5    &     9.9   &  36 & -- &  9.75 &\\
4278   &   4    &     8.8   & 270 & -- & 14.87 &\\
4365   &   6    &     9.0   & 251 & -- & 12.90 &\\
4374   &   3    &     5.5   &  41 & -- &  9.22 &\\
4473   &   4    &     2.8   & 106 & -- & 17.35 &\\
4486   &   5    &     5.0   & 702 & 60 & 30.52 &\\
4494   &   5    &     4.6   & 107 & 10 &  8.52 &\\
4526   &   4    &     8.0   & 107 & 25 & 12.06 &\\
4564   &   3    &     4.5   &  27 & -- &  8.33 &\\
4649   &   4    &     8.0   & 431 & 21 & 24.25 &\\
4697   &   1    &     2.0   &  20 & -- &  4.66 &\\
5846   &   6    &     9.1   & 191 & -- & 13.68 &\\
7457   &   4    &     7.5   &  40 & 6  &  6.26 &\\
\hline
3607   &   5    &     9.9   &  36 & -- & 20.72 &\\
5866   &   1    &     2.0   &  20 & -- &  5.75 &\\
\hline
\end{tabular}
\begin{flushleft}
{\small Notes: The last two galaxies, NGC~3607 and NGC~5866, are \textit{bonus} galaxies, in the sense that they were not originally included in the SLUGGS survey but we have obtained and analysed their data using the standard SLUGGS procedure. 
$N_{\rm GC}$ is the number of spectroscopically confirmed globular clusters per galaxy and $N_{\rm sub}$ is the number of globular clusters identified as belonging to kinematic substructures in Section \ref{subs:iso_subx}. $R_{\rm max}$ shows the radial extent probed per galaxy in units of the effective radius, $\re$.}
\end{flushleft}
\end{table}

\section{Observations, data reduction and data pruning}
\label{observation}
\subsection{Observations and data reduction}
\label{Obs}
The GC kinematic data used in this work were obtained through spectroscopic observations, mostly as part of the SLUGGS survey, with the DEIMOS (DEep Imaging Multi-Object Spectrograph, \citealt{Faber_2003}) instrument on the 10~m Keck--II telescope. For NGC~3115, NGC~4486 and NGC~4649, we have supplemented our catalogue with data from some external sources (see \citealt{Arnold_2011, Strader_2011, Pota_2015}, respectively, for details of these externally sourced kinematic data and the re--calibration of their uncertainties to match with those of DEIMOS). Spectroscopic data collection with DEIMOS began in 2006 and we have now obtained ${\sim}3500$ GC radial velocities in 25 carefully chosen ETGs \citep{Brodie_2014}. Here, we only consider 23 galaxies from the SLUGGS survey with 20 or more spectroscopically confirmed GCs. Readers interested in a detailed explanation of our DEIMOS data reduction method are encouraged to check \citet{Pota_2013} though we give a brief description here.
 
We design masks with 1 arcsec--wide slits targeting GC candidates and integrate per mask for an average of 2~hrs. We set up DEIMOS with the 1200~lines~$\rm mm^{-1}$ centred on 7800~$\rm {\AA}$. This ensures we have a wavelength resolution of ${\sim}1.5 \rm {\AA}$ and cover the CaT absorption lines in the near--infrared (8498, 8542, 8662~$\rm {\AA}$) and often the H$\alpha$ line at 6563~$\rm {\AA}$. We reduce our raw spectra using the \textsc{IDL SPEC2D} data reduction pipeline \citep{Cooper_2012} and obtain radial velocities by measuring the doppler shifts of the CaT absorption lines using \textsc{FXCOR} task in \textsc{IRAF}. We cross--correlate our science spectra with spectral templates of 13 carefully chosen Galactic stars, obtained with the same instrument and set--up. The final radial velocity for each object is the average from the cross--correlation. The uncertainties on our radial velocities are obtained by adding in quadrature the uncertainty outputs from \textsc{FXCOR} to the standard deviation among the templates, typically ${\sim}3 \ \kms$. Finally, our science spectra are redshift--corrected.

\begin{figure*}
    		\includegraphics[width=0.32\textwidth]{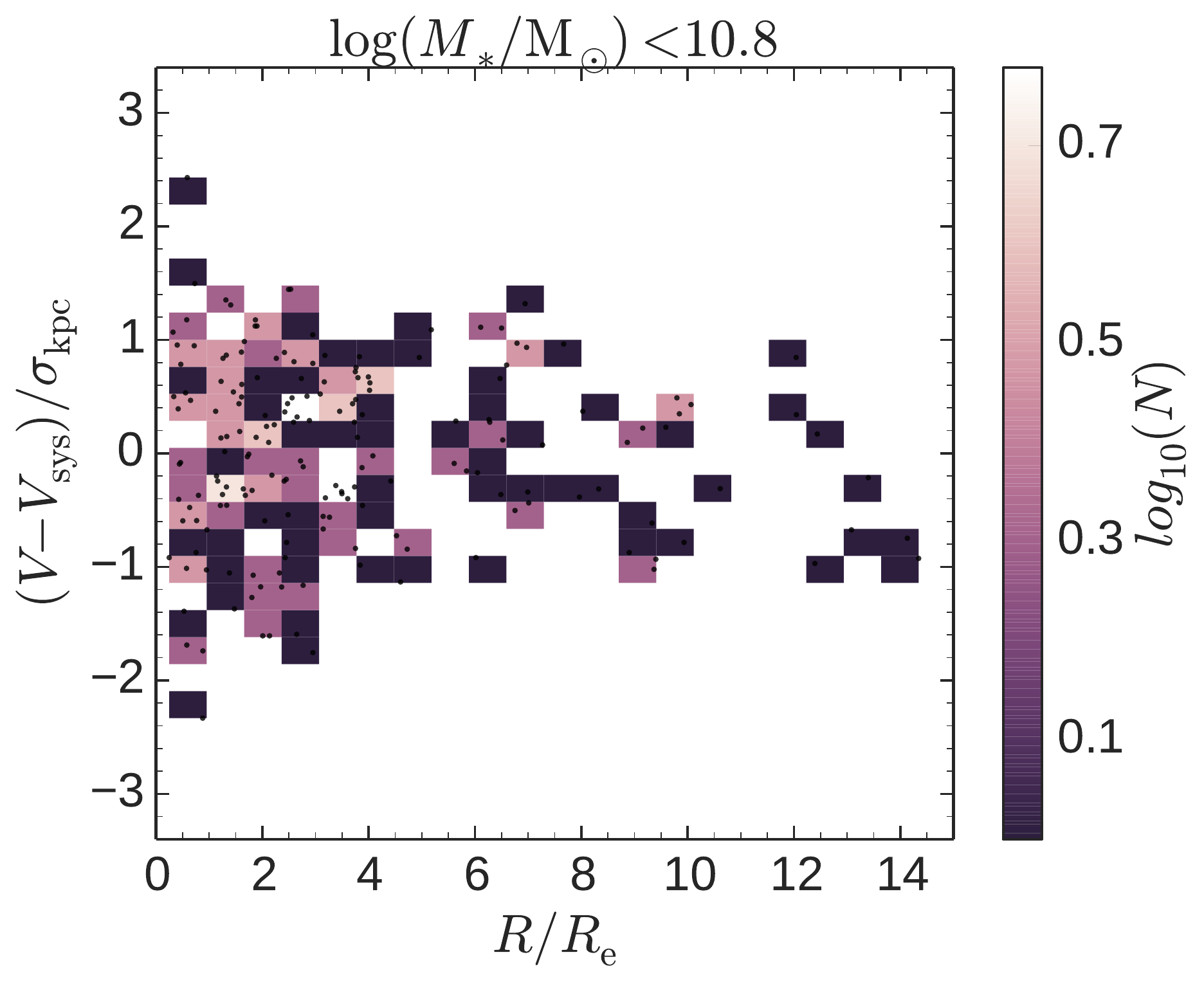}\hspace{0.01\textwidth}%
    		\includegraphics[width=0.32\textwidth]{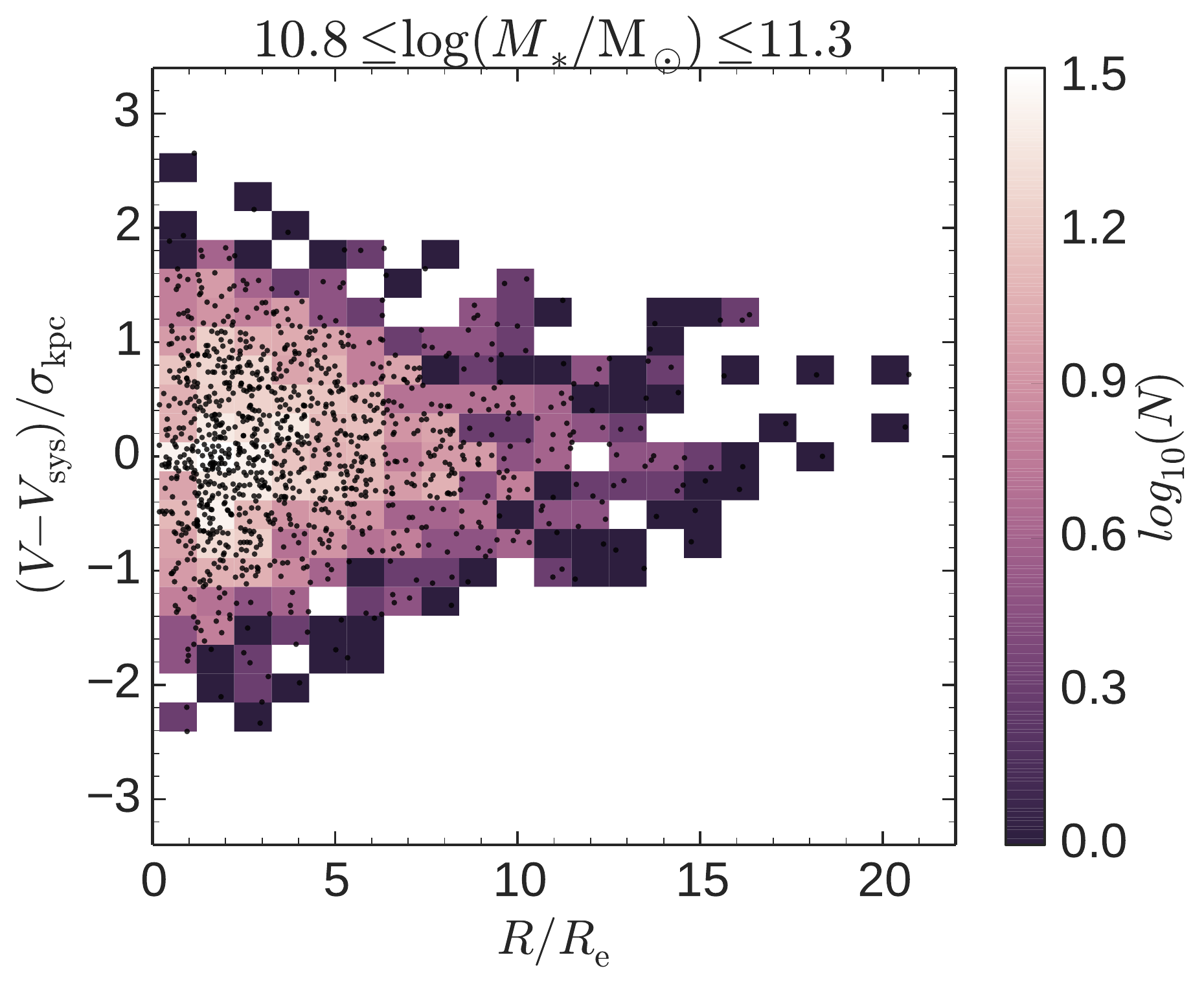}\hspace{0.01\textwidth}
    		\includegraphics[width=0.32\textwidth]{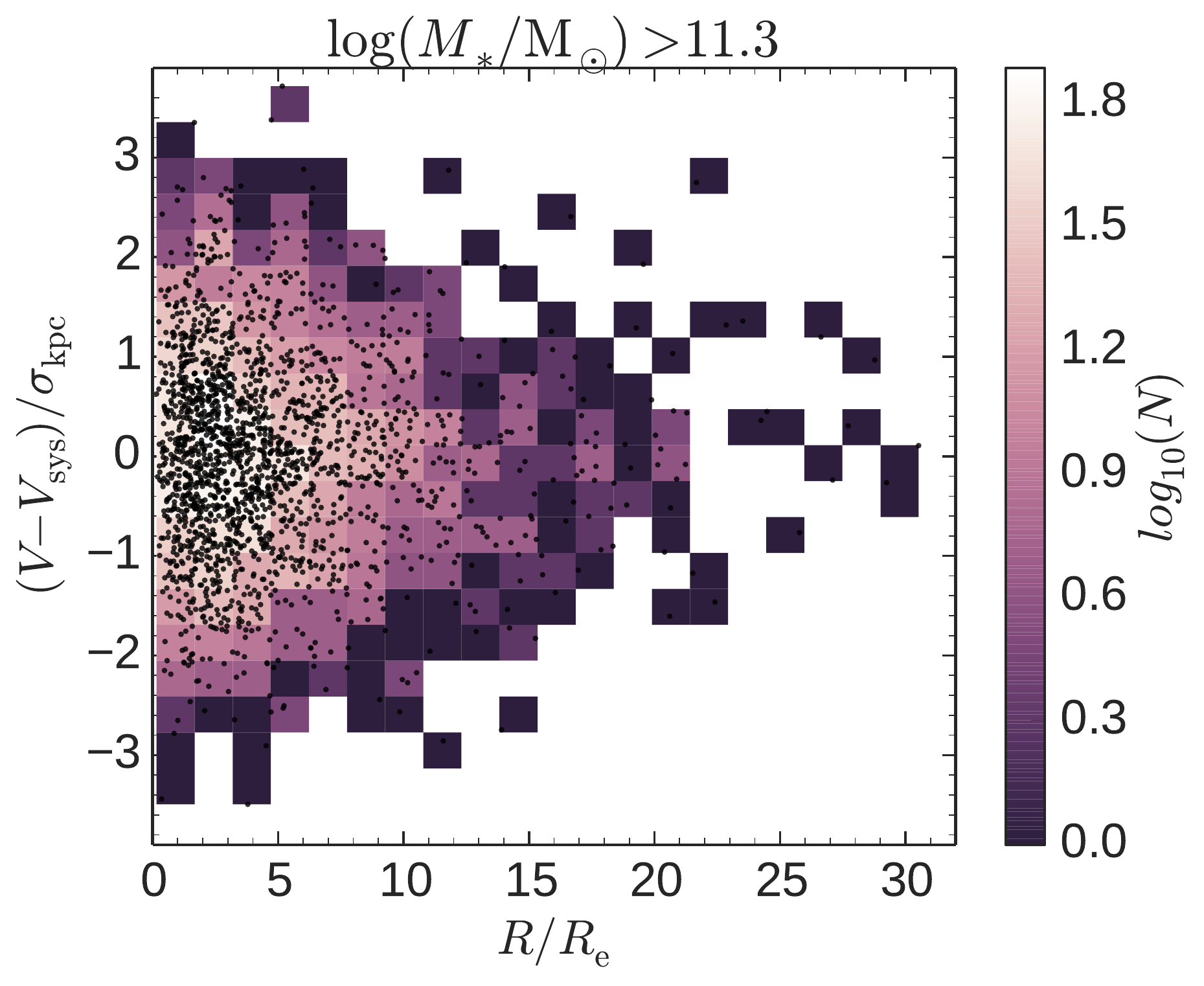}\\
    		\caption{\label{fig:RV_rad} Line--of--sight velocities of the ${\sim}3500$ GCs in our sample of 23 galaxies normalised by their respective galaxy central velocity dispersion ($\sigma_{\rm kpc}$ from Table \ref{tab:summary}) versus galactocentric radius (in effective radius). The \textit{left} panel shows the low--mass galaxies (NGC~7457, NGC~3377 and NGC~4564), the \textit{middle} panel shows intermediate mass galaxies (NGC~3608, NGC~4473, NGC~4278, NGC~821, NGC~3115, NGC~5866, NGC~1023, NGC~4494, NGC~4697, NGC~4697, NGC~1400, NGC~4526, NGC~2768 and NGC~3607) while the \textit{right} panel shows the high mass galaxies (NGC~720, NGC~5846, NGC~4374, NGC~4365, NGC~4486, NGC~4649 and NGC~1407). GCs belonging to kinematic substructures have been excluded from this 2D histogram. The black dots are the individual GCs while the colour bar shows the density of the points. On average, the GC line--of--sight velocities extend out to 13~$\re$ per galaxy.}  
\end{figure*}

To classify an object as a GC, we ensure that the CaT features in the rest--frame spectra are seen at the expected rest wavelength and the radial velocity is consistent with the host galaxy's systemic velocity (through a 3$\sigma$ clipping implemented via the friendless--algorithm of \citet{Merrett_2003}). For secure classification as a GC,
we require that at least the 8542 and 8662 $\rm {\AA}$ CaT lines are observed, as well as the H$\alpha$ line (when the H$\alpha$ wavelength region is probed). In addition, we obtain a consensus from at least two members of the SLUGGS team on the status of our GC candidates. Objects with contentious status, but radial velocities consistent with the host galaxy's systemic velocity, are classified as \textit{marginal} GCs. We do not use such objects in this work. Figure \ref{fig:RV_rad} shows the composite galactocentric distribution of our homogeneous sample of ${\sim}3500$ GC line--of--sight velocities ($\vlos$) with well understood errors used in this work. On average, our GC data extends to 10, 13 and 15 $\re$ in the low (${\rm log}(M_*/\Msun) < 10.8 $)--, intermediate ($10.8 \leq {\rm log}(M_*/\Msun) \leq 11.3$)-- and high (${\rm log}(M_*/\Msun) > 11.3 $)-- stellar mass galaxies in our sample, respectively.

\begin{table*}
{\small \caption{General properties of our galxies.} \label{tab:summary}} 
\begin{tabular}{@{}l l l l l l l l l l l l l l l} 
\hline
\hline
Galaxy & $M_K$ & Dist. & $V_{\rm sys}$ & $\re$ & $\sigma_{\rm kpc}$ & $\epsilon$ & $\rho_{\rm env}$ & $\log(M_{*}/\Msun)$ & $p-val$ & $\alpha$ & $\gamma$ & corr & $V_{\rm rot}/\sigma$\\
$\rm [NGC]$ &  [mag] & [Mpc]  & [$\kms$] & [$\arcsec$] & [$\kms$]	&	& [$\rm Mpc^{-3}$]	&  & 	&	&	&  &\\
(1) & (2) &	(3) & (4) & (5) & (6) & (7) & (8) & (9) & (10) &	(11) & (12) & (13) & (14)  \\
\hline
720   & $-25.09$ & 26.9 & 1745 & 35 & 227 & 0.49 & 0.25 & 11.35 & 0.051     & 0.058  & 2.66 & 0.92 & $0.42_{-0.17} ^{+0.24}$\\[1.3ex]
821   & $-24.14$ & 23.4 & 1718 & 40 & 193 & 0.35 & 0.08 & 10.97 & 0.411     & 0.234  & 2.90 & 0.98 & $0.40_{-0.18} ^{+0.20}$\\[1.3ex]
1023  & $-24.16$ & 11.1 &  602 & 48 & 183 & 0.63 & 0.57 & 10.98 & $<$~0.009 & 0.230  & 2.89 & 0.85  & $0.65_{-0.18} ^{+0.21}$\\[1.3ex]
1400  & $-24.53$ & 26.8 &  558 & 28 & 236 & 0.13 & 0.07 & 11.12 & 0.288     & 0.163  & 2.80 & 1.01 & $0.22_{-0.15} ^{+0.20}$\\[1.3ex]
1407 & $-25.72$ & 26.8 & 1779 & 63 & 252 & 0.07 & 0.42 & 11.60  & 0.106     & -0.056 & 2.60 & 1.01 & $0.04_{-0.07} ^{+0.08}$\\[1.3ex]
2768  & $-24.91$ & 21.8 & 1353 & 63 & 206 & 0.57 & 0.31 & 11.28 & 0.364     & 0.092  & 2.70 & 0.88 & $0.50_{-0.15} ^{+0.15}$ \\[1.3ex]
3115  & $-24.15$ &  9.4 &  663 & 35 & 248 & 0.66 & 0.08 & 10.97 & 0.043     & 0.232  & 2.89 & 0.83 & $0.94_{-0.16} ^{+0.15}$\\[1.3ex]
3377 & $-22.83$ & 10.9 &  690 & 36 & 135 & 0.33 & 0.49 & 10.44  & 0.419     & 0.477  & 3.23 & 0.98 & $0.23_{-0.10} ^{+0.14}$\\[1.3ex]
3608  & $-23.78$ & 22.3 & 1226 & 30 & 179 & 0.20 & 0.56 & 10.82 & 0.953     & 0.301  & 2.99 & 1.01 & $0.21_{-0.18} ^{+0.26}$\\[1.3ex] 
4278  & $-23.93$ & 15.6 &  620 & 32 & 228 & 0.09 & 1.25 & 10.88 & 0.73      & 0.273  & 2.95 & 1.01 & $0.13_{-0.07} ^{+0.08}$\\[1.3ex]
4365  & $-25.43$ & 23.1 & 1243 & 53 & 253 & 0.24 & 2.93 & 11.48 & 0.195     & -0.003 & 2.57 & 1.00 & $0.15_{-0.08} ^{+0.10}$\\[1.3ex]
4374  & $-25.36$ & 18.5 & 1017 & 53 & 284 & 0.05 & 3.99 & 11.46 & 0.472     & 0.009  & 2.59 & 1.01 & $0.45_{-0.24} ^{+0.25} $\\[1.3ex]
4473  & $-23.90$ & 15.2 & 2260 & 27 & 189 & 0.43 & 2.17 & 10.87 & 0.537     & 0.279  & 2.96 & 0.95 & $0.23_{-0.11} ^{+0.15}$\\[1.3ex] 
4486  & $-25.55$ & 16.7 & 1284 & 81 & 307 & 0.16 & 4.17 & 11.53 & $<$~0.001 & -0.027 & 2.54 & 1.01 & $0.14_{-0.05} ^{+0.06}$\\[1.3ex]
4494 & $-24.27$ & 16.6 & 1342 & 49 & 157 & 0.14 & 1.04 & 11.02  & 0.018     & 0.210  & 2.86 & 1.01 & $0.51_{-0.14} ^{+0.15}$\\[1.3ex]
4526  & $-24.81$ & 16.4 &  617 & 45 & 233 & 0.76 & 2.45 & 11.23 & $<$~0.001 & 0.111  & 2.73 & 0.77 & $0.61_{-0.24} ^{+0.23}$\\[1.3ex]
4564 & $-23.17$ & 15.9 & 1155 & 20 & 153 & 0.53 & 4.09 & 10.58  & 0.054     & 0.414  & 3.14 & 0.90 & $1.80_{-0.33} ^{+0.51}$\\[1.3ex]
4649  & $-25.61$ & 16.5 & 1110 & 66 & 308 & 0.16 & 3.49 & 11.56 & $<$~0.001 & -0.037 & 2.53 & 1.01 & $0.34_{-0.08} ^{+0.07}$\\[1.3ex]
4697  & $-24.29$ & 12.5 & 1252 & 62 & 180 & 0.32 & 0.60 & 11.03 & 0.394     & 0.206  & 2.86 & 0.98 & $2.37_{-0.86} ^{+0.83}$\\[1.3ex]
5846  & $-25.22$ & 24.2 & 1712 & 59 & 231 & 0.08 & 0.84 & 11.40 & 0.553     & 0.034  & 2.62 & 1.01 & $0.08_{-0.07} ^{+0.09}$\\[1.3ex]
7457  & $-22.42$ & 12.9 &  844 & 36 &  74 & 0.47 & 0.13 & 10.28 & 0.014     & 0.552  & 3.33 & 0.93 & $1.90_{-0.42} ^{+0.53}$\\[1.3ex]
\hline
3607  & $-24.96$ & 22.2 &  942 & 39 & 229 & 0.13 & 0.34 & 11.29 & 0.227     & 0.084  & 2.69 & 1.01 & $0.18_{-0.15} ^{+0.22}$\\[1.3ex]
5866 & $-24.15$ & 14.9 &  755 & 36 & 163 & 0.58 & 0.24 & 10.97  & 0.978     & 0.232  & 2.89 & 0.88 & $0.16_{-0.36} ^{+1.06}$\\[1.3ex]

 \hline
\end{tabular}
\begin{flushleft}
{\small Column Description: (1) galaxy name; (2) total extinction--corrected $K$--band magnitude, obtained using the absolute $K$--band magnitude from 2MASS \citep{Jarrett_2000}, dust extinction correction from \citet{Schlegel_1998} and the correction to the 2MASS photometry due to sky over--subtraction from \citet{Scott_2013}; (3)--(8) are from \citet{Brodie_2014} and include  (3) distance; (4) systemic velocity; (5) effective (half--light) radius; (6) central stellar velocity dispersion within 1~kpc; (7) ellipticity and (8) environmental density of neighbouring galaxies; (9) total logarithmic stellar mass, obtained from the absolute $K$--band magnitude, assuming ${M/L}_K=1$ (here and elsewhere in the paper, stellar mass--to--light ratio is quoted in units of $\ml$); typical uncertainties on our stellar masses are 
${\sim}0.15$ dex.; (10) statistical significance of having kinematic substructures in globular cluster system [see Section. \ref{subs:iso_subx} for derivation of column (10)]; (11) the power--law slope of the gravitational potential; (12) the power--law slope of the de--projected globular cluster density profile [see Section. \ref{subs:parameters} for derivation of columns (11) and (12)]; (13) normalising factor to correct for effect of galaxy flattening on dynamical mass estimate and (14) rotation dominance parameter for the globular cluster system, after removing kinematic substructures where relevant [see Section. \ref{subs:flat_rot} for columns (13) and (14)]} \end{flushleft}
\end{table*}

\subsection{Kinematic substructures in GC systems}
\label{subs:iso_subx}
A fundamental assumption of mass modelling methods is that the system of tracers is in dynamical equilibrium. However, if galaxies 
assembled their mass hierarchically via mergers and accretion events, a lumpy ``outer" halo is expected, especially in position--velocity phase space \citep{Bullock_2005, Helmi_2008, Cooper_2013}. The fossils of the accreted galaxies or satellite galaxies undergoing disruption that have not been totally phase--mixed can sometimes be isolated in position--velocity phase space, even when the coherent structures are no longer evident in photometric studies.
For the immediate task of mass modelling, it is important to isolate tracers that show correlations in position--velocity phase space, i.e. kinematic substructures, in order to avoid spurious mass estimates. 

For each galaxy, we use the Dressler--Schectman (DS) test \citep{Dressler_1988, Ashman_1993, Pinkney_1996, Mendel_2008, Einasto_2012} to detect substructures 
in position--velocity phase space and to quantify the significance of the substructures. For each GC, we compute the local 
average velocity ($\bar{V}_{\rm local}$) and velocity dispersion ($\sigma _{\rm local}$) using the $N_{\rm {nn}} = \sqrt{N_{\rm GC}}$ 
nearest
\begin{figure*}
\includegraphics[width=0.24\textwidth, height=0.19\textwidth]{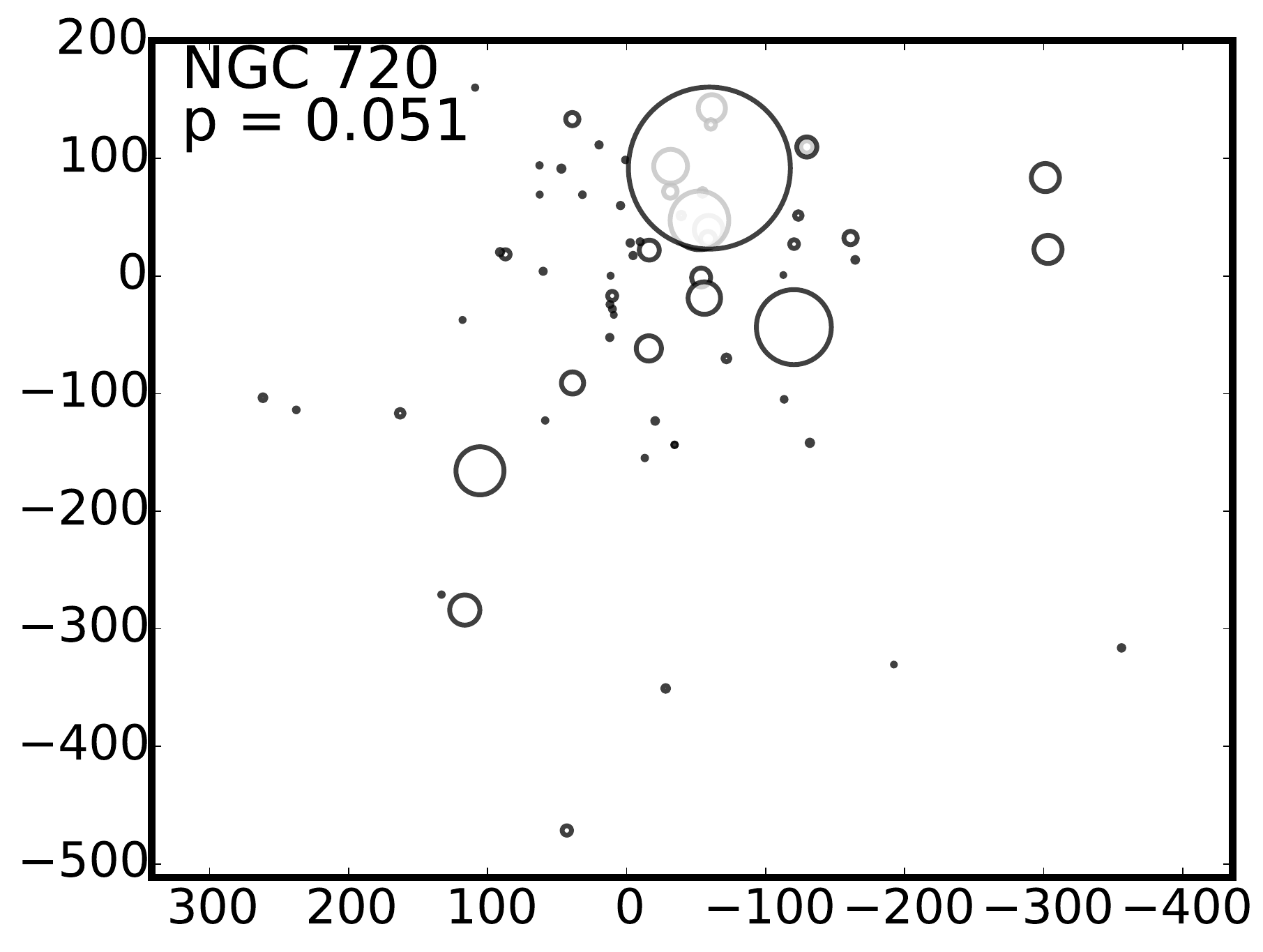}\hspace{0.001\textwidth}%
\includegraphics[width=0.24\textwidth, height=0.19\textwidth]{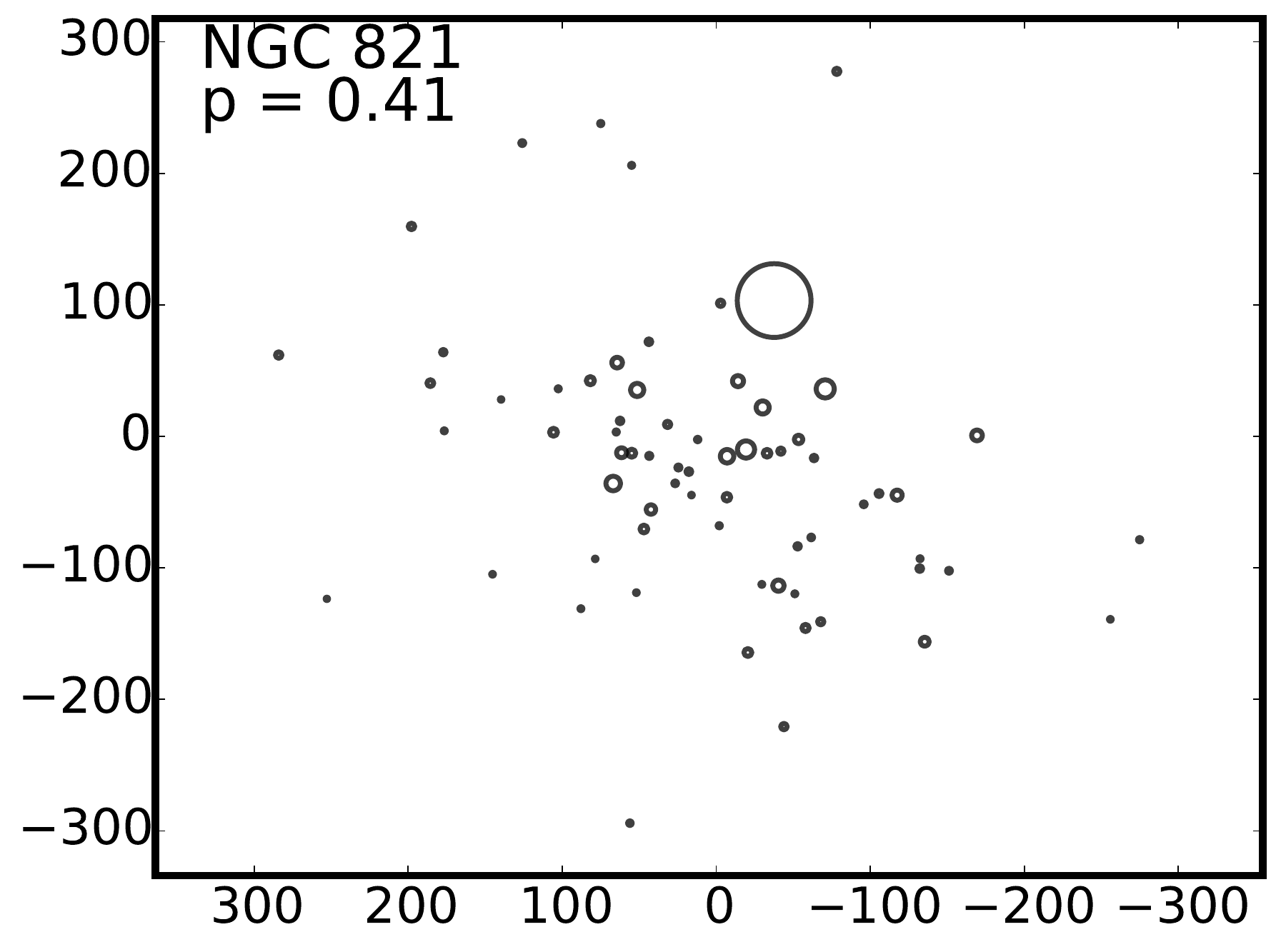}\hspace{0.001\textwidth}%
\includegraphics[width=0.24\textwidth, height=0.19\textwidth]{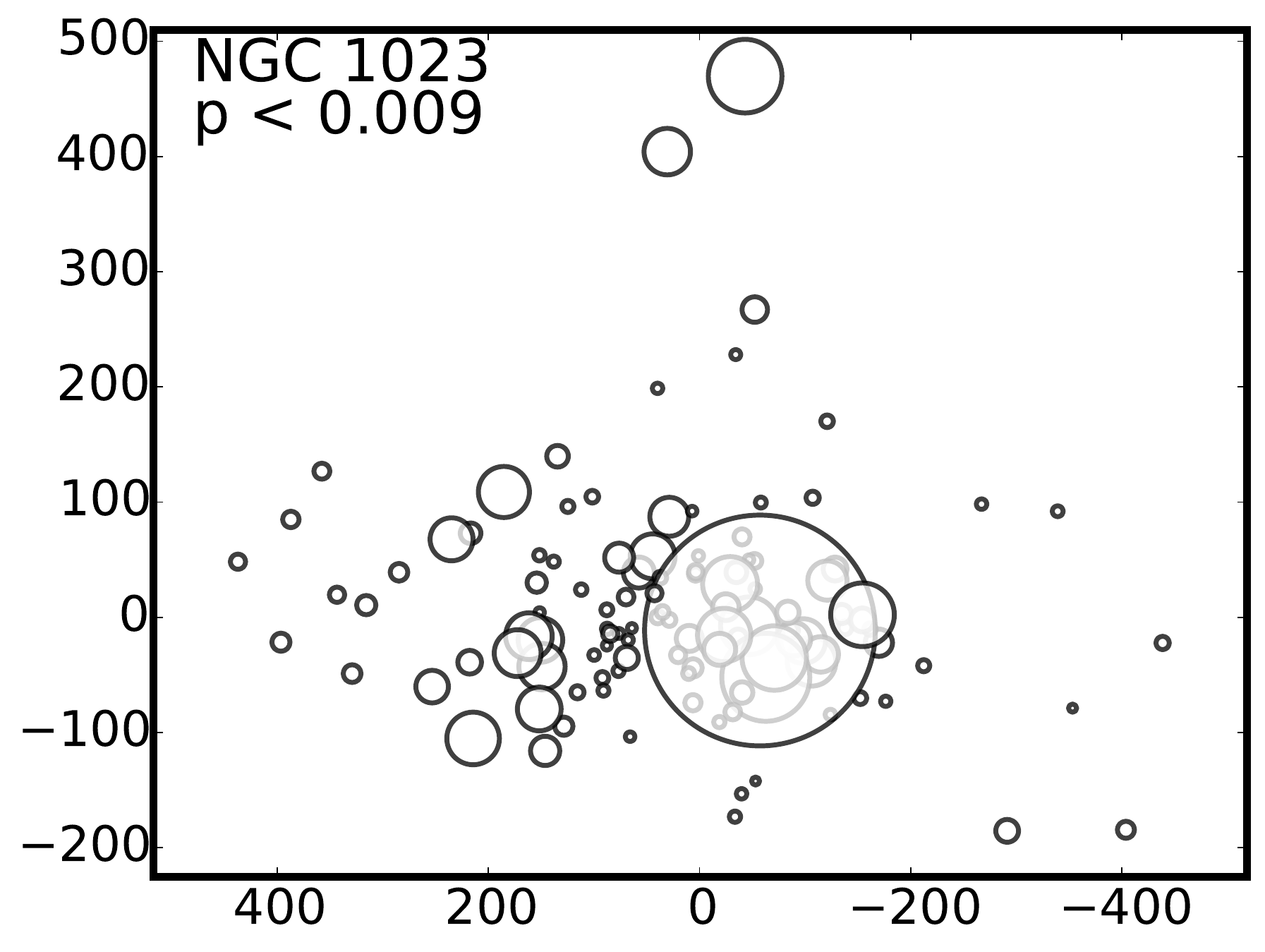}\hspace{0.001\textwidth}%
\includegraphics[width=0.24\textwidth, height=0.19\textwidth]{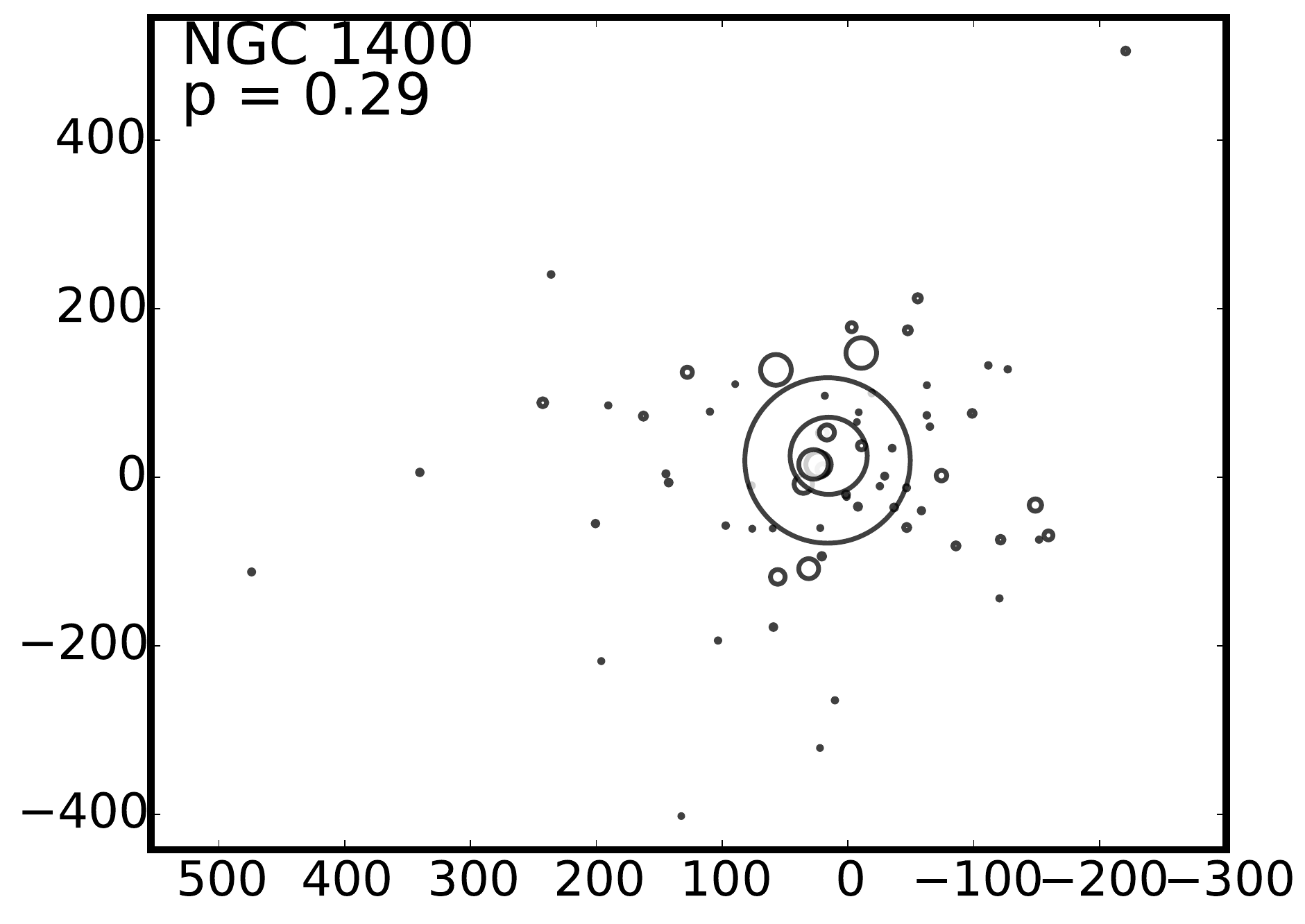}\\[0.05em]
\includegraphics[width=0.24\textwidth, height=0.19\textwidth]{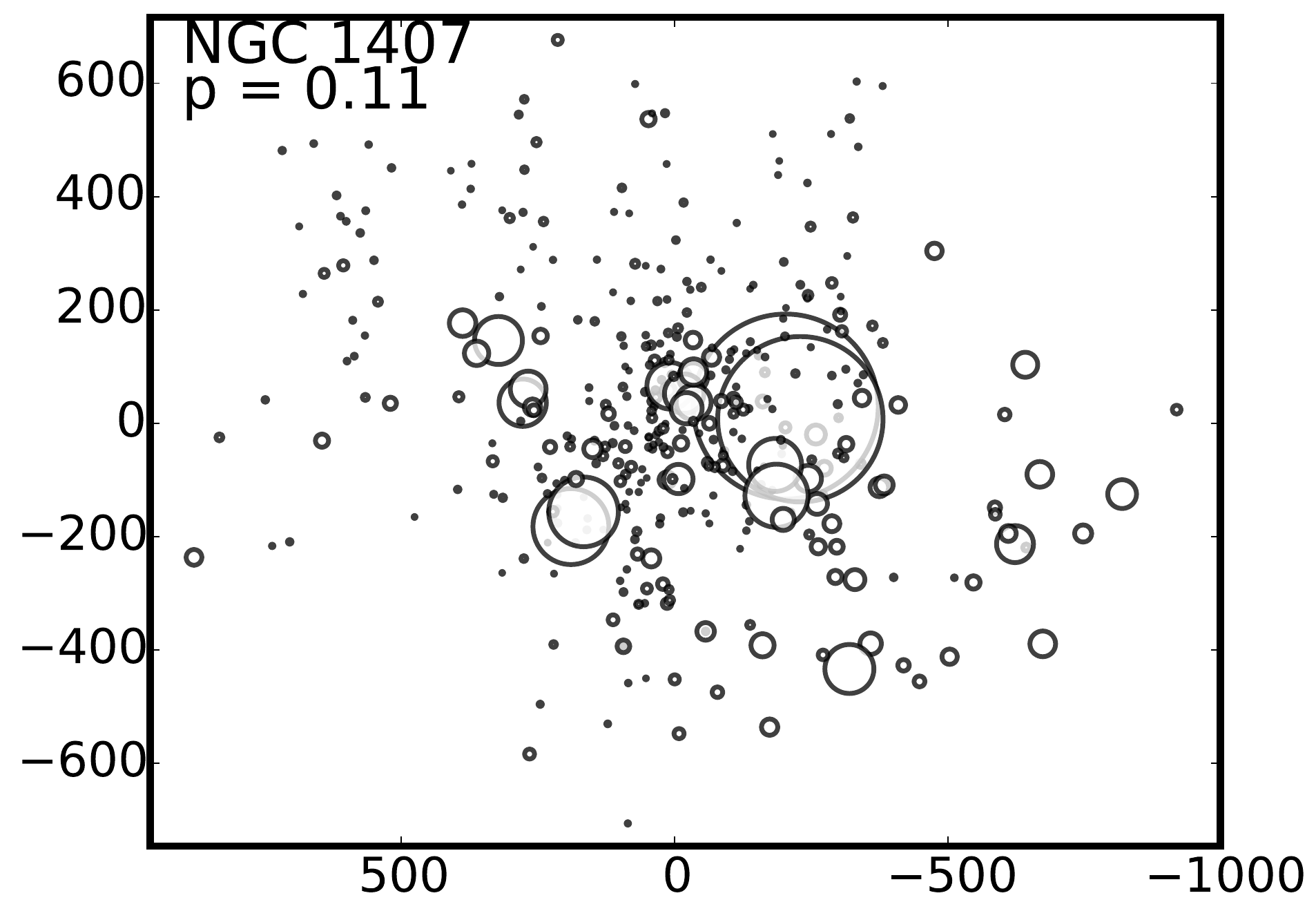}\hspace{0.001\textwidth}%
\includegraphics[width=0.24\textwidth, height=0.19\textwidth]{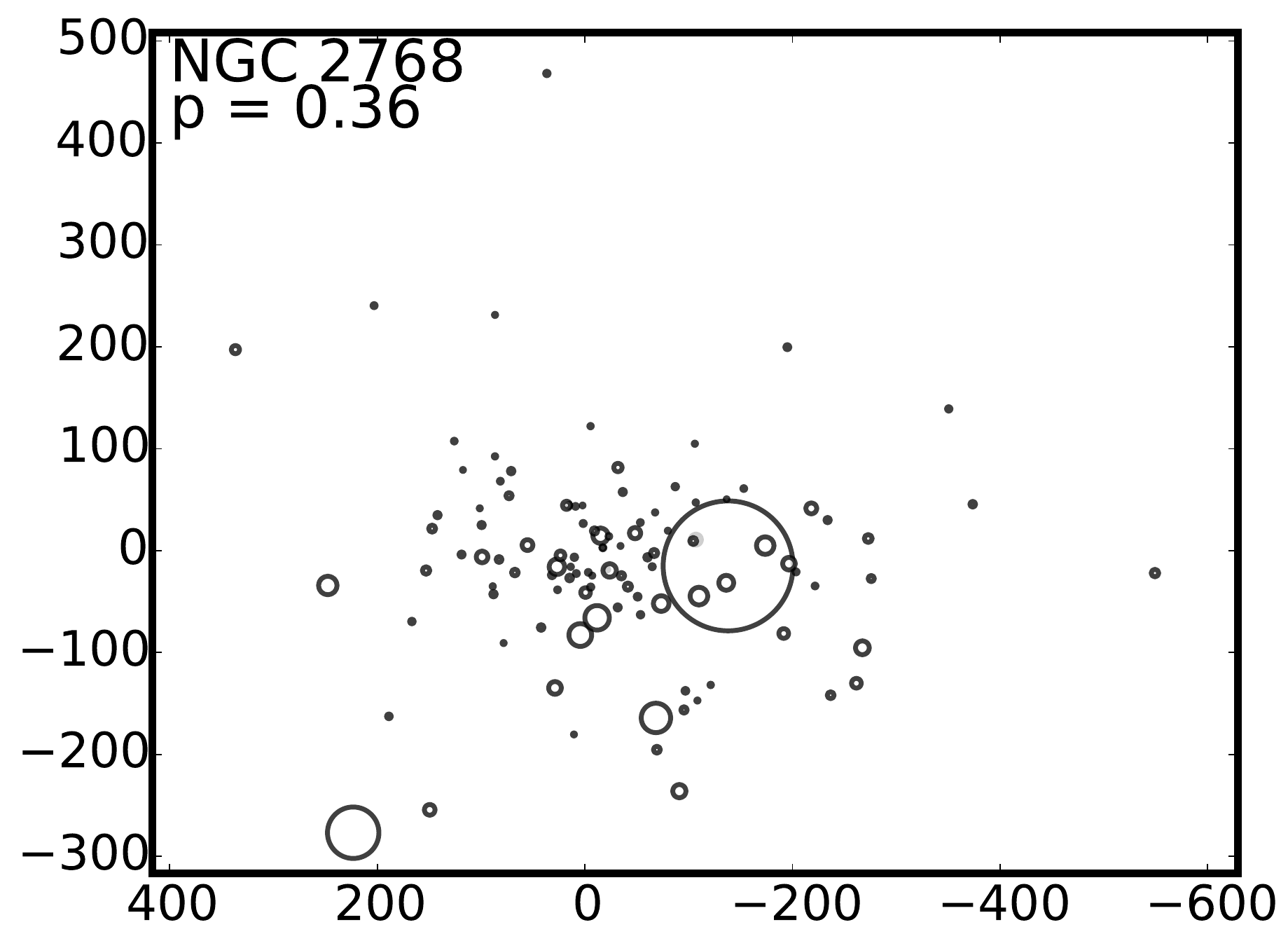}\hspace{0.001\textwidth}%
\includegraphics[width=0.24\textwidth, height=0.19\textwidth]{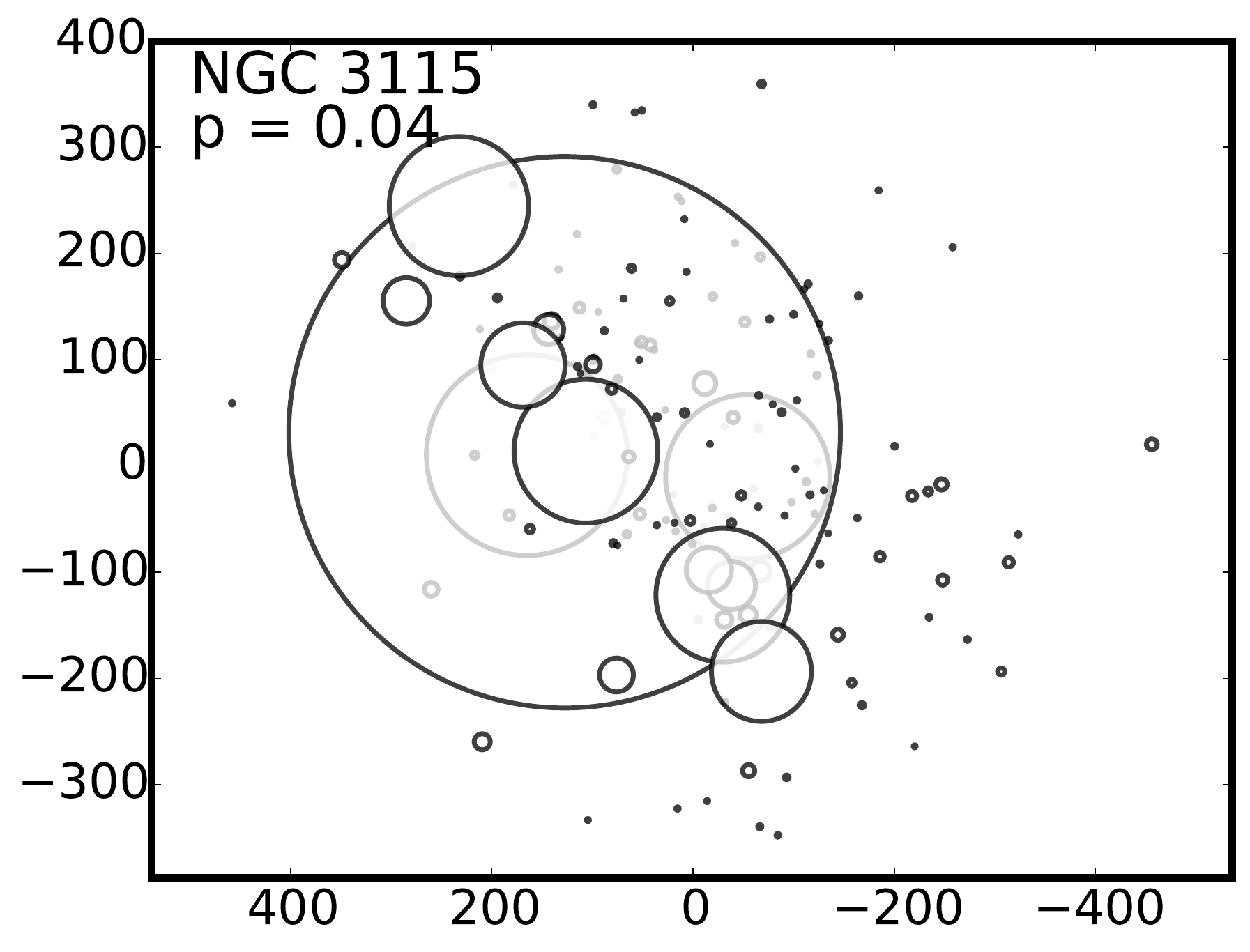}\hspace{0.001\textwidth}%
\includegraphics[width=0.24\textwidth, height=0.19\textwidth]{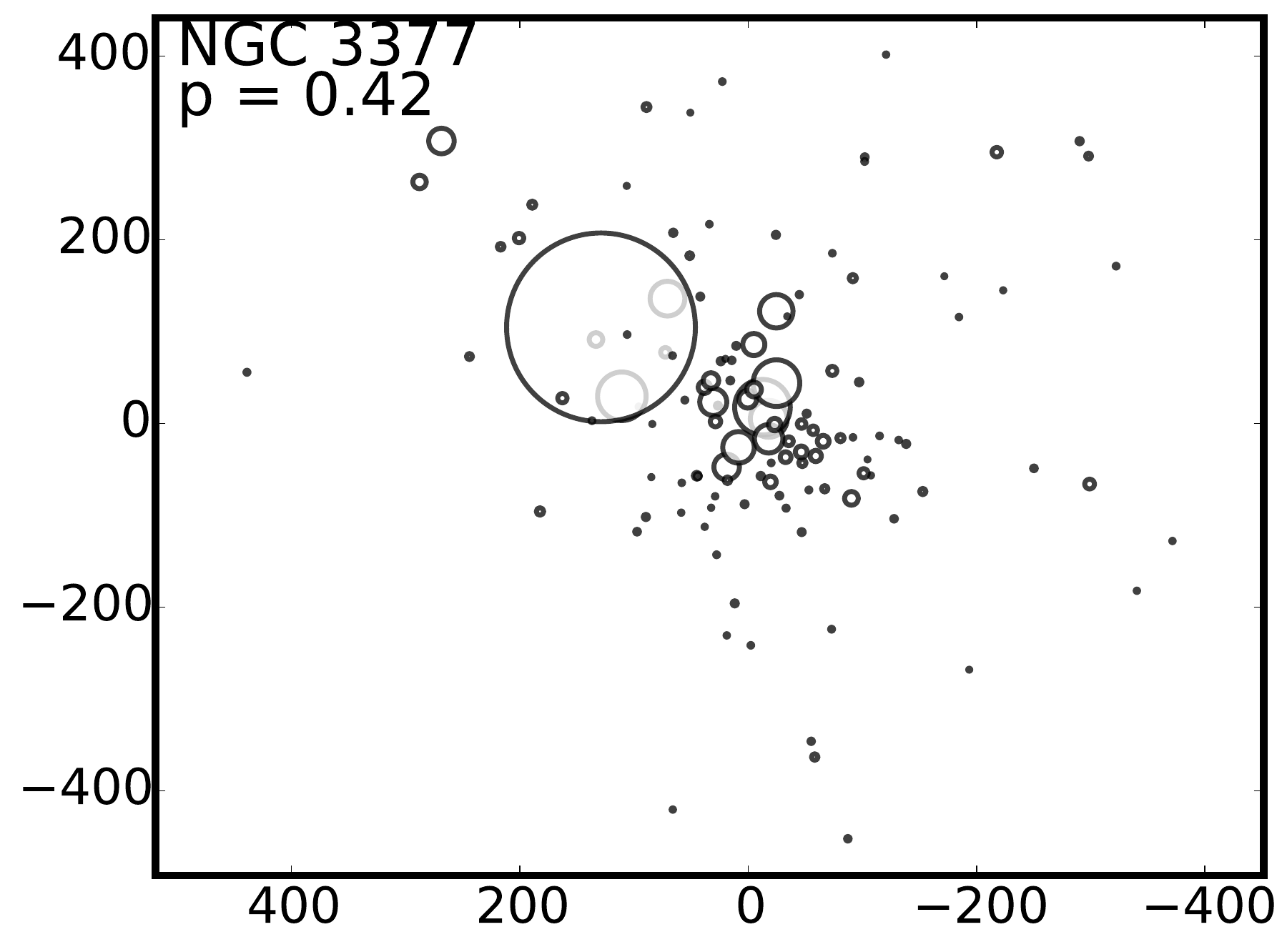}\\[0.05em]
\includegraphics[width=0.24\textwidth, height=0.19\textwidth]{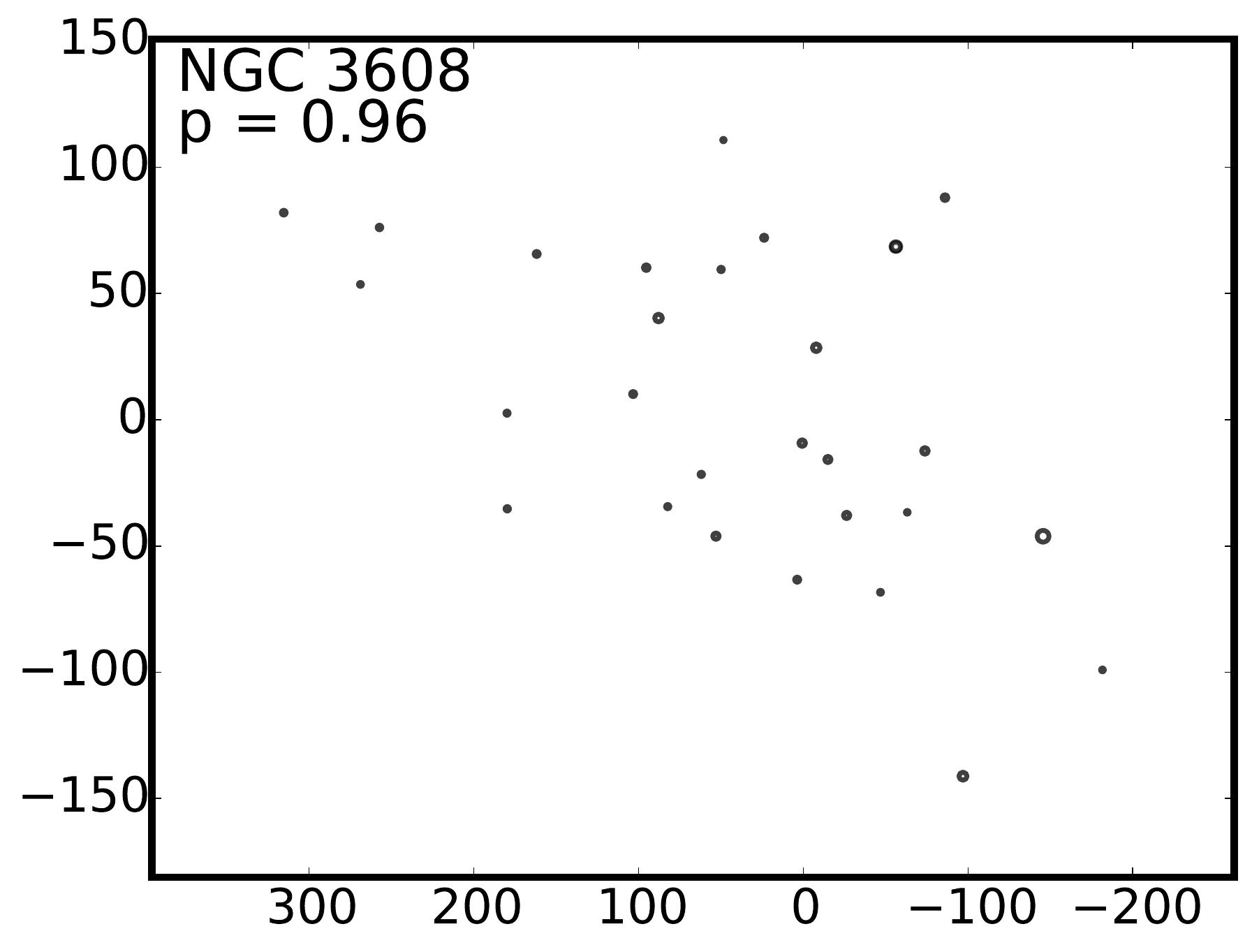}\hspace{0.001\textwidth}%
\includegraphics[width=0.24\textwidth, height=0.19\textwidth]{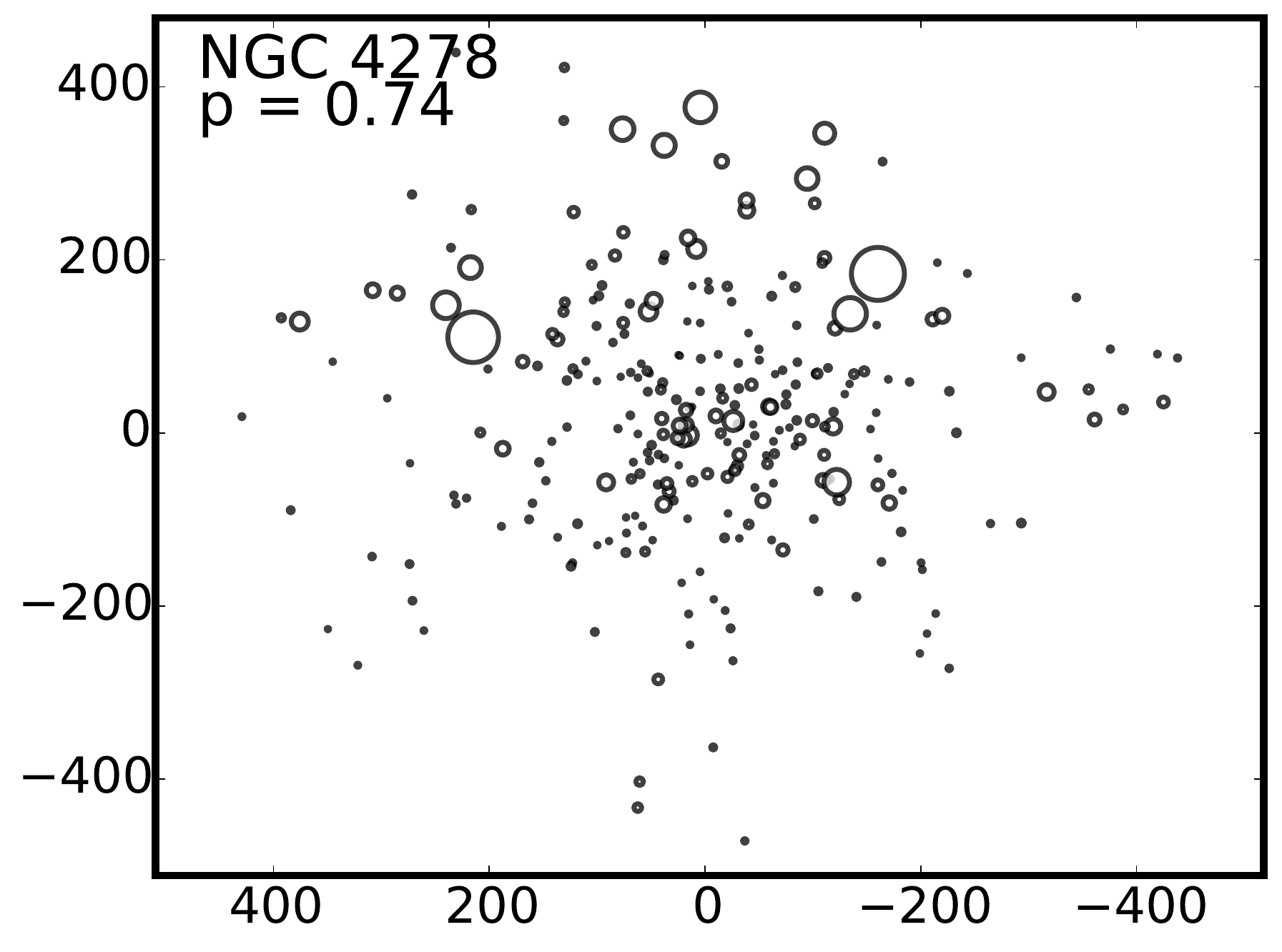}\hspace{0.001\textwidth}%
\includegraphics[width=0.24\textwidth, height=0.19\textwidth]{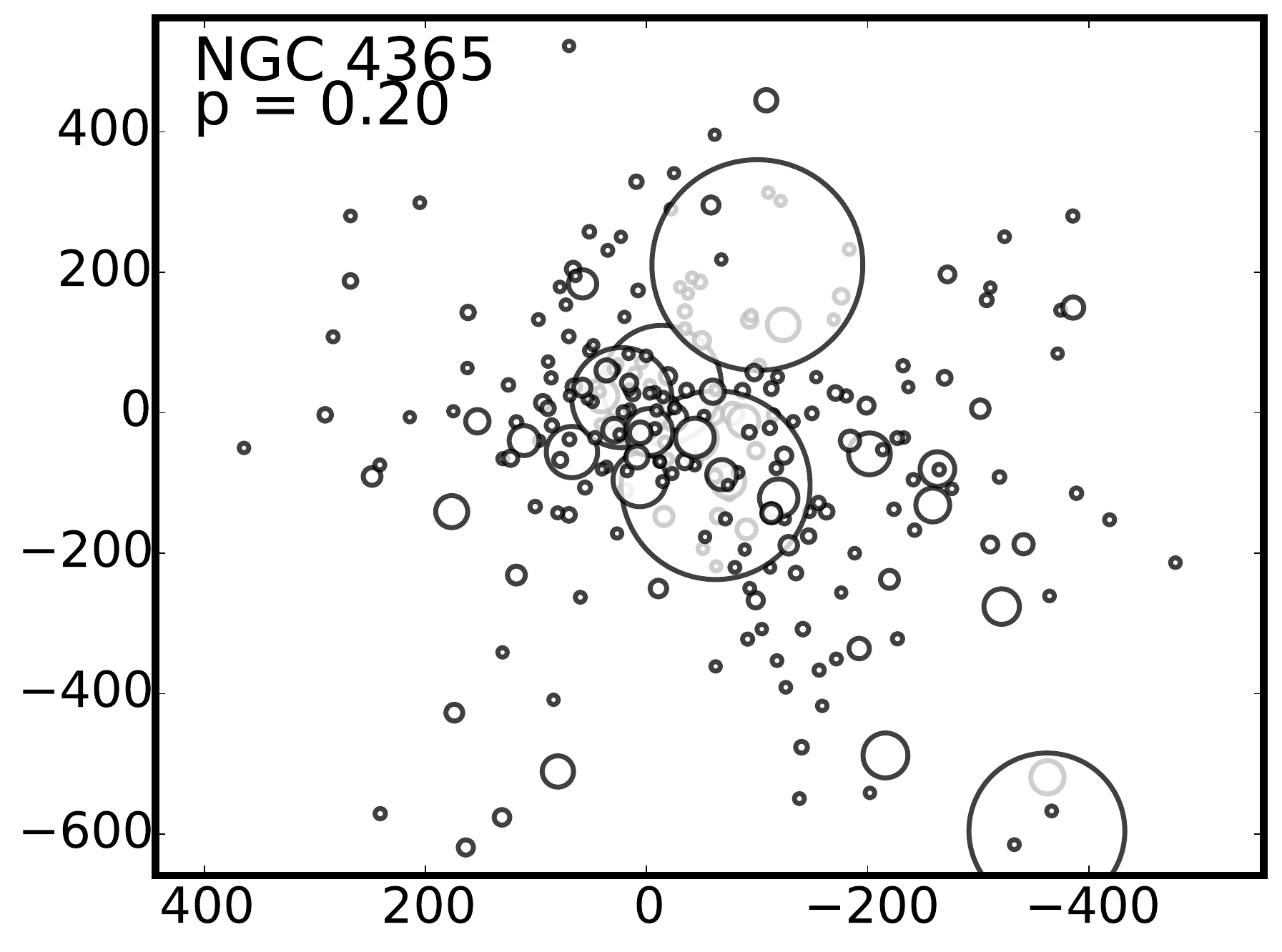}\hspace{0.001\textwidth}%
\includegraphics[width=0.24\textwidth, height=0.19\textwidth]{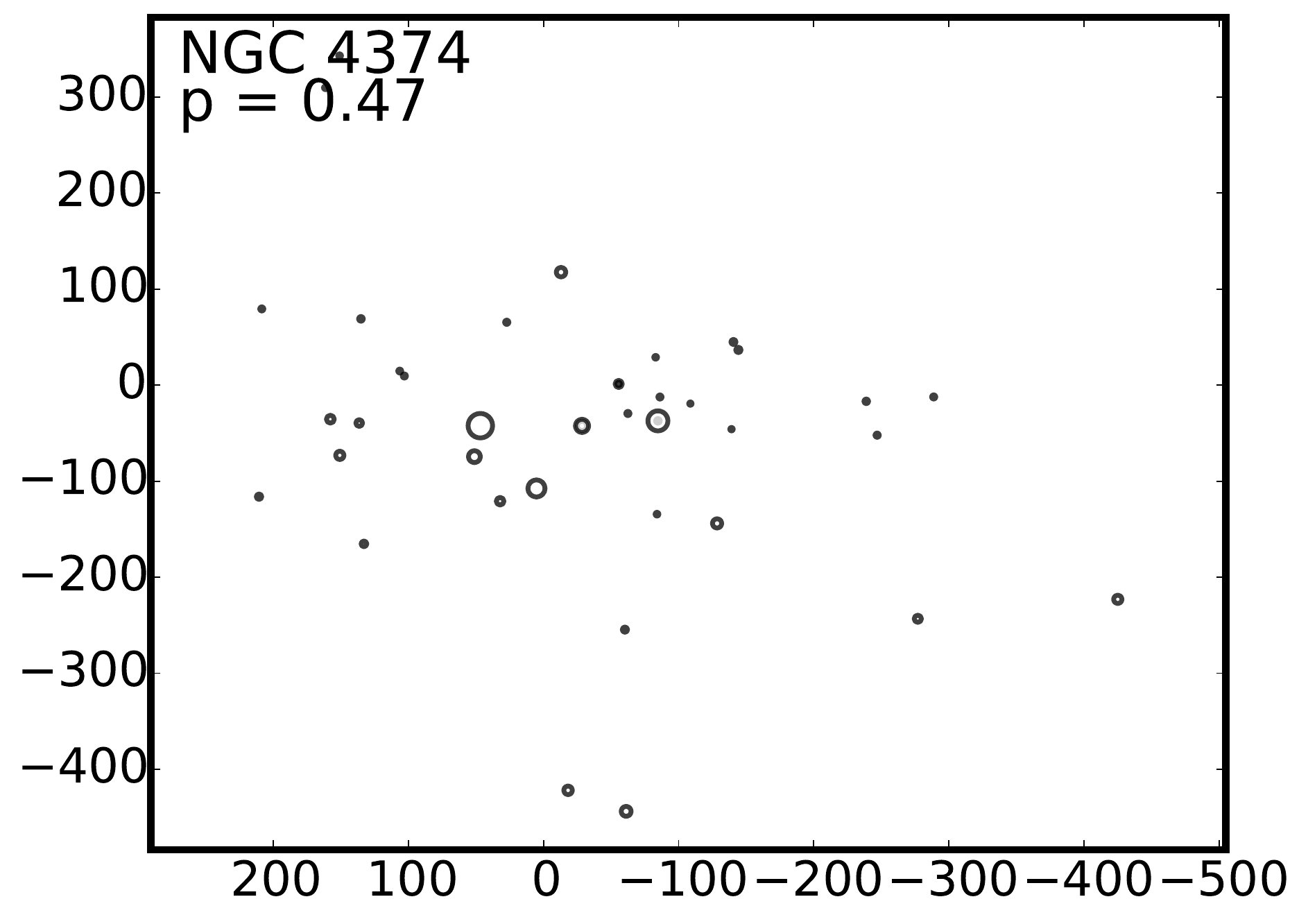}\\[0.05em]
\includegraphics[width=0.24\textwidth, height=0.19\textwidth]{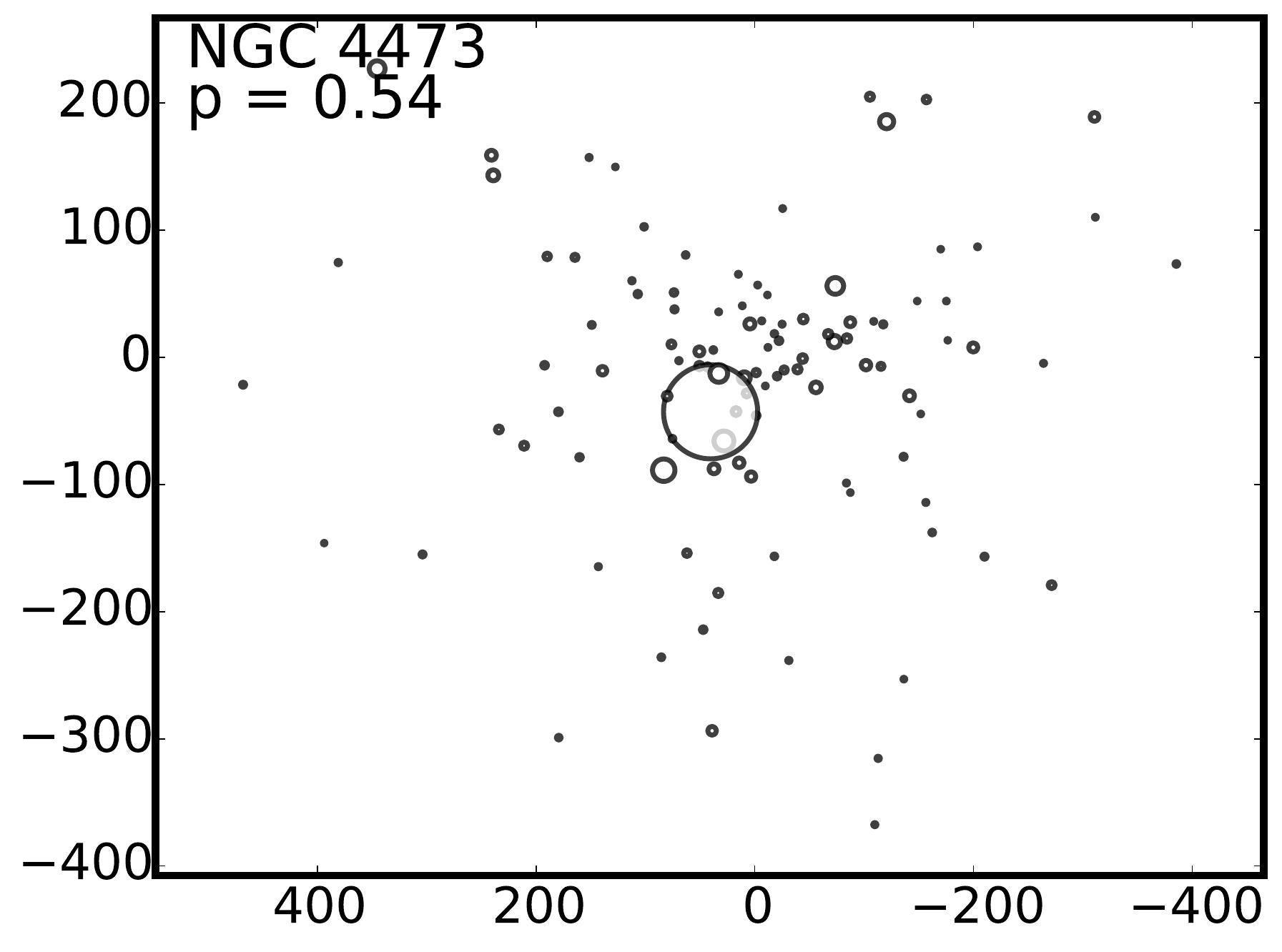}\hspace{0.001\textwidth}%
\includegraphics[width=0.24\textwidth, height=0.19\textwidth]{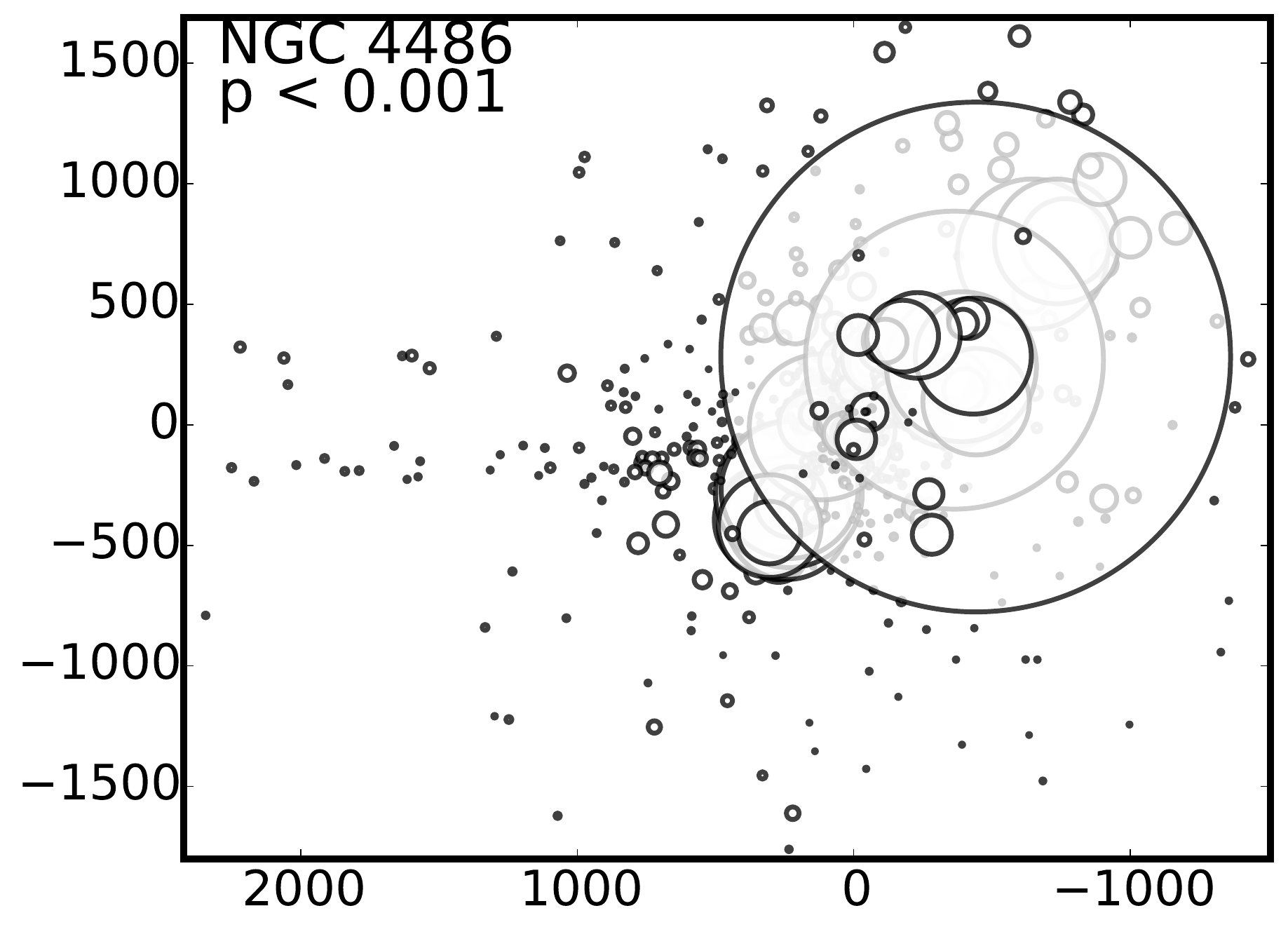}\hspace{0.001\textwidth}%
\includegraphics[width=0.24\textwidth, height=0.19\textwidth]{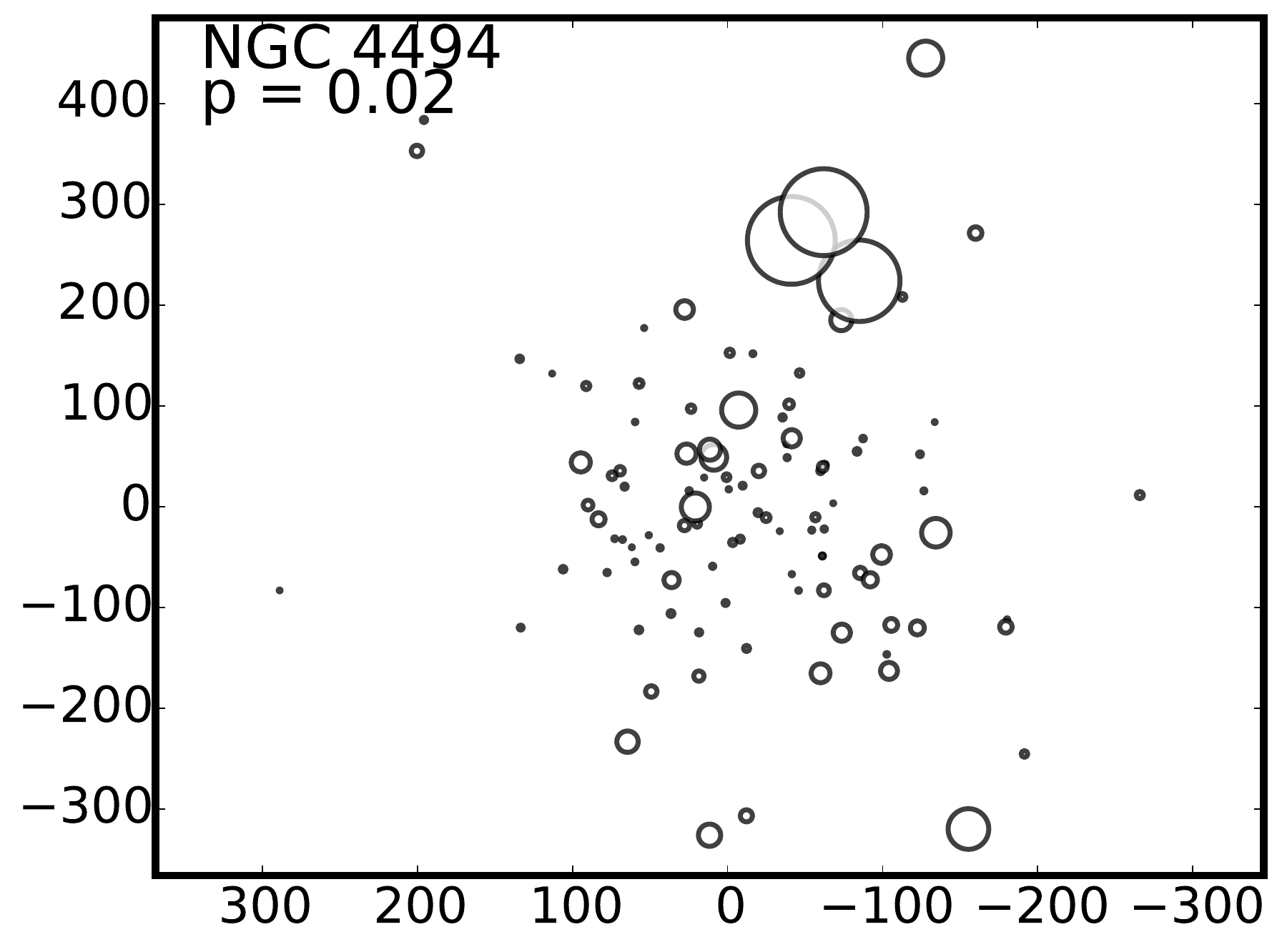}\hspace{0.001\textwidth}%
\includegraphics[width=0.24\textwidth, height=0.19\textwidth]{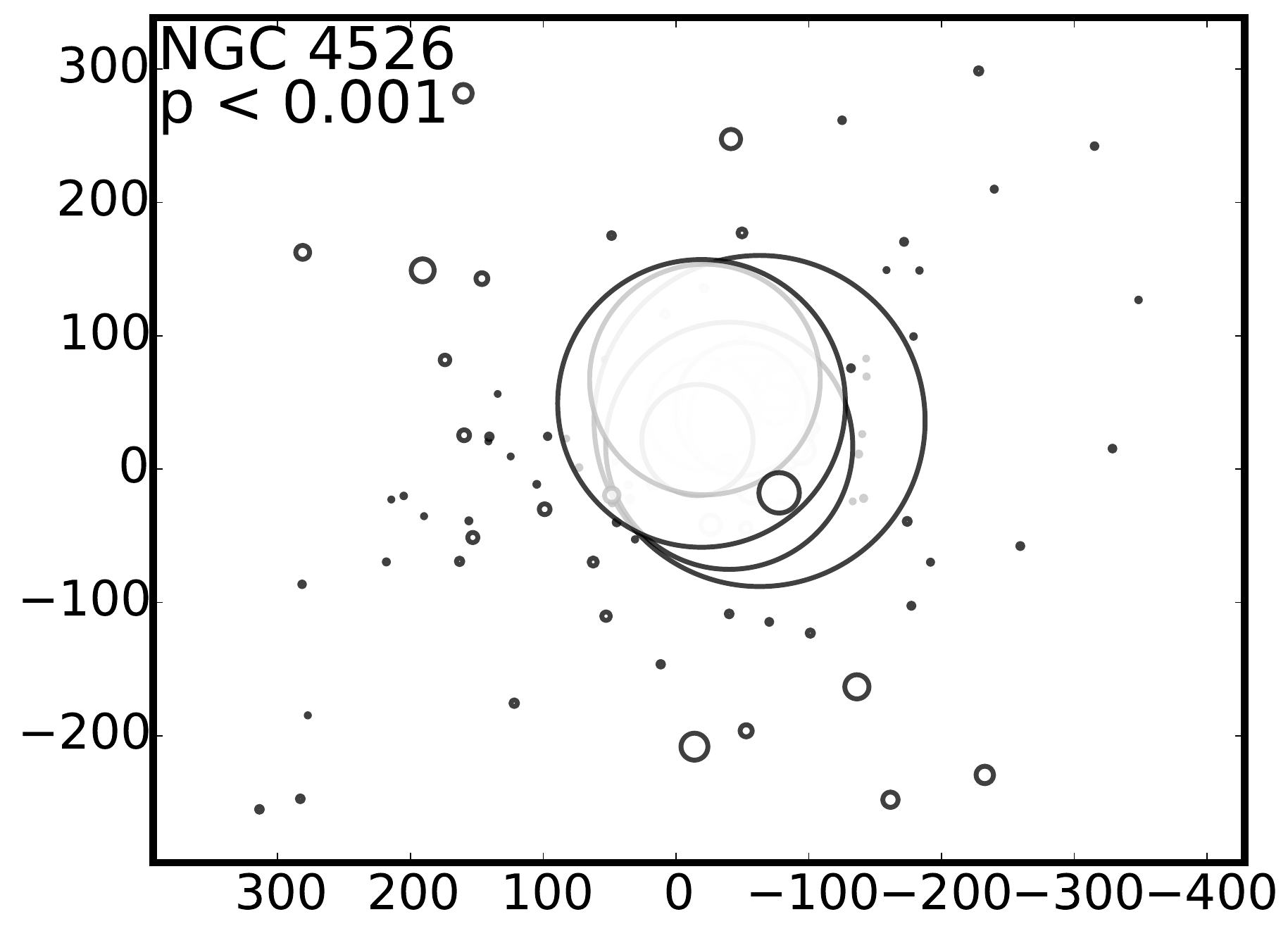}\\[0.05em]
\includegraphics[width=0.24\textwidth, height=0.19\textwidth]{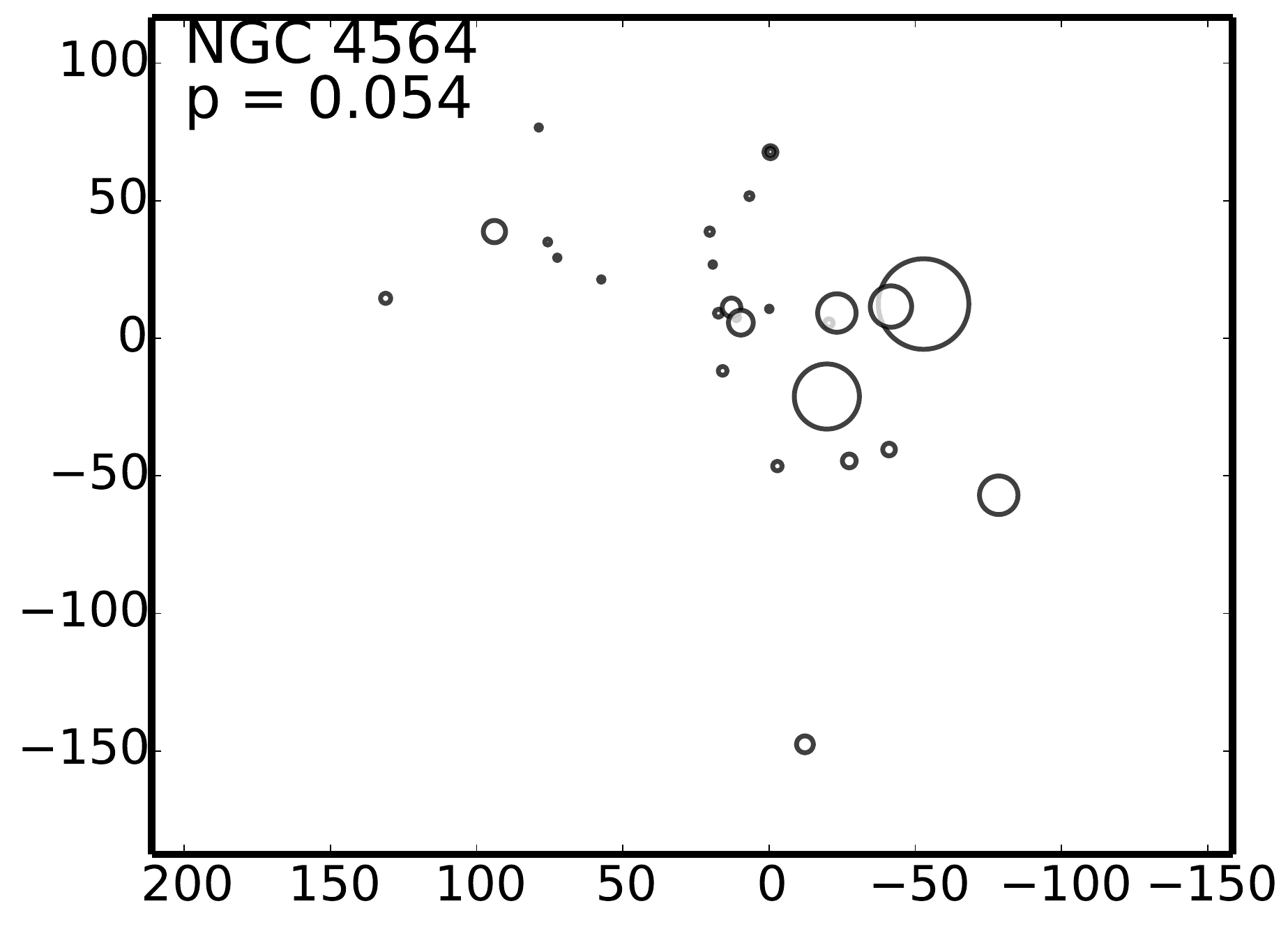}\hspace{0.001\textwidth}%
\includegraphics[width=0.24\textwidth, height=0.19\textwidth]{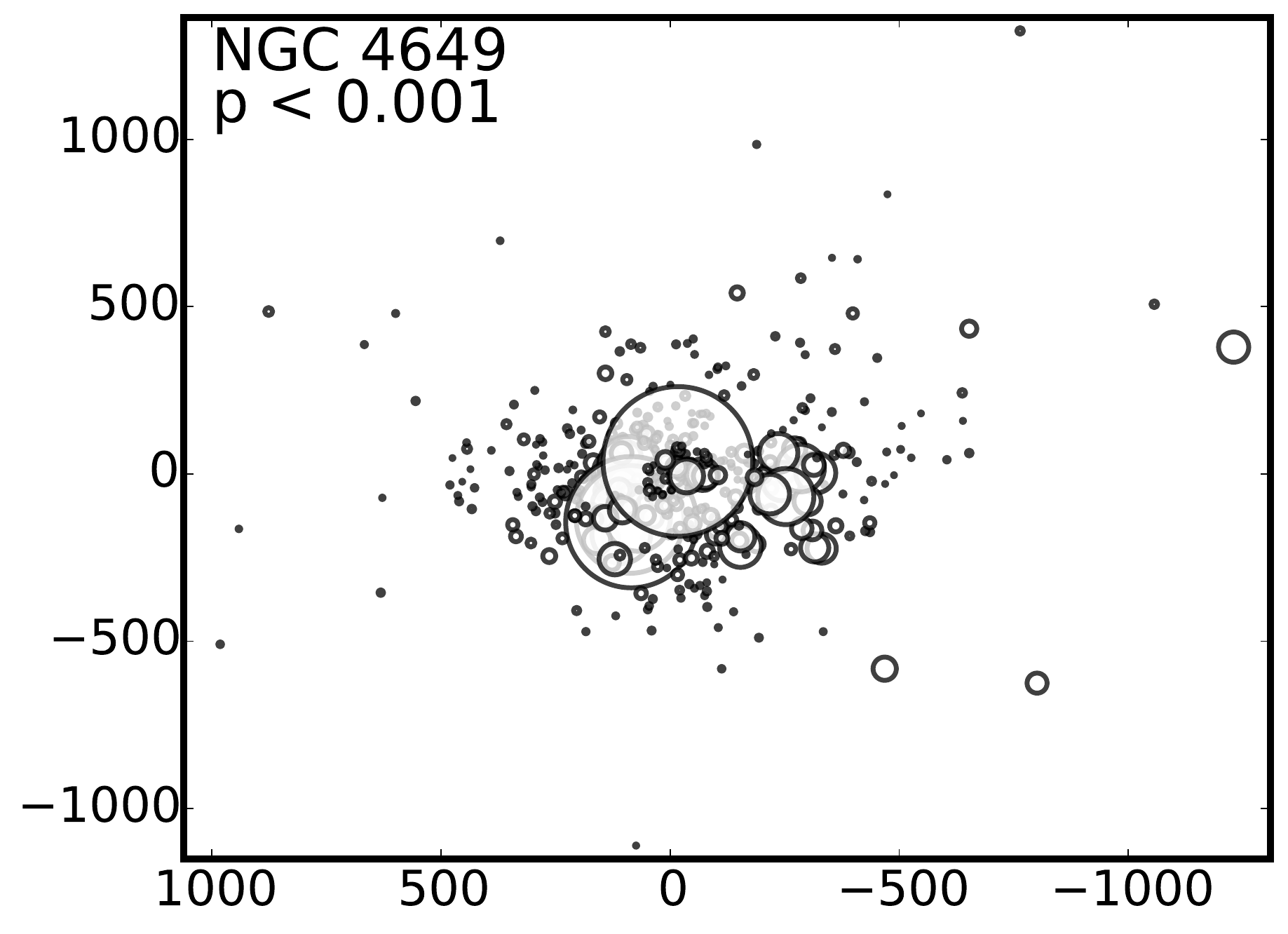}\hspace{0.001\textwidth}%
\includegraphics[width=0.24\textwidth, height=0.19\textwidth]{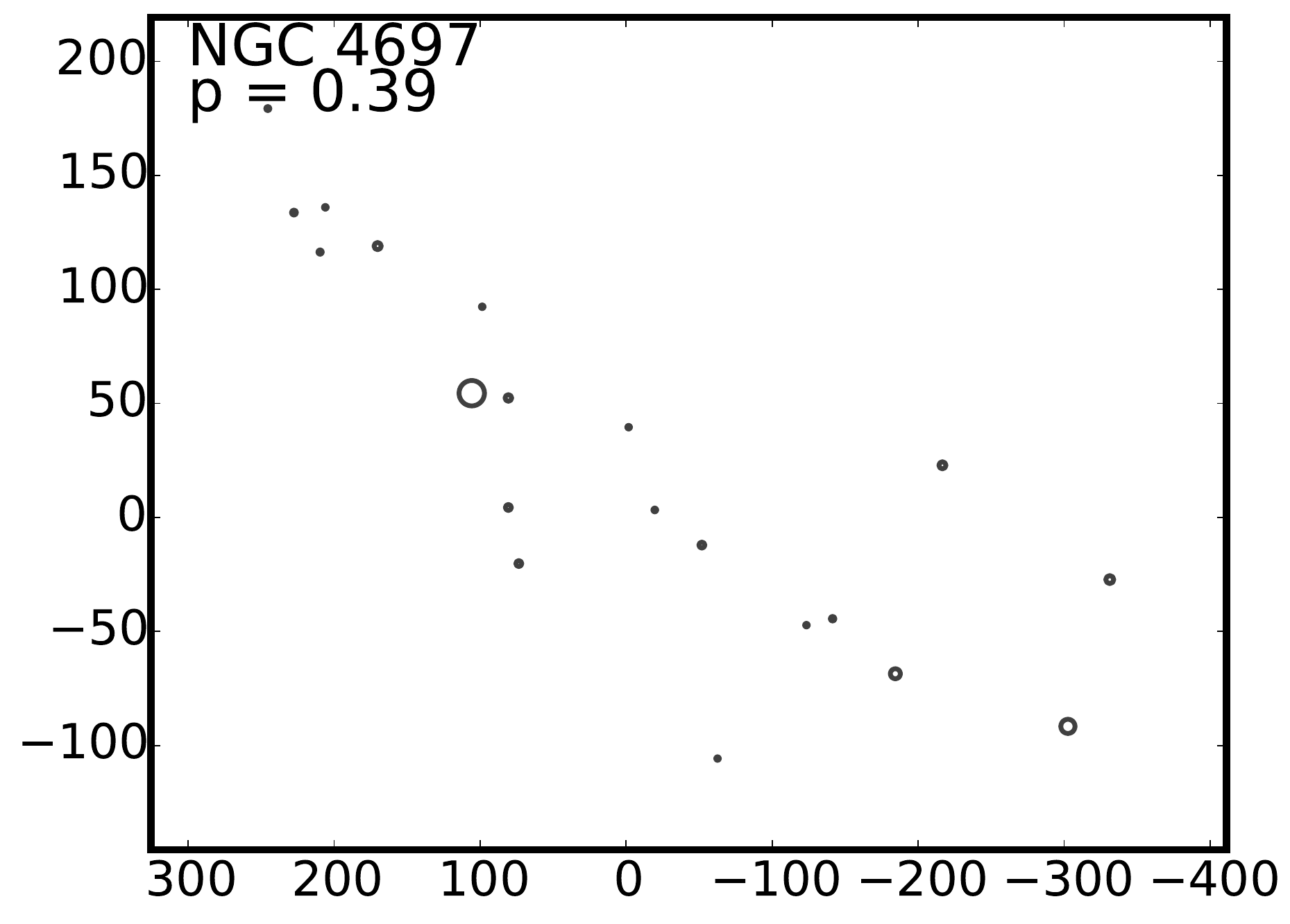}\hspace{0.001\textwidth}%
\includegraphics[width=0.24\textwidth, height=0.19\textwidth]{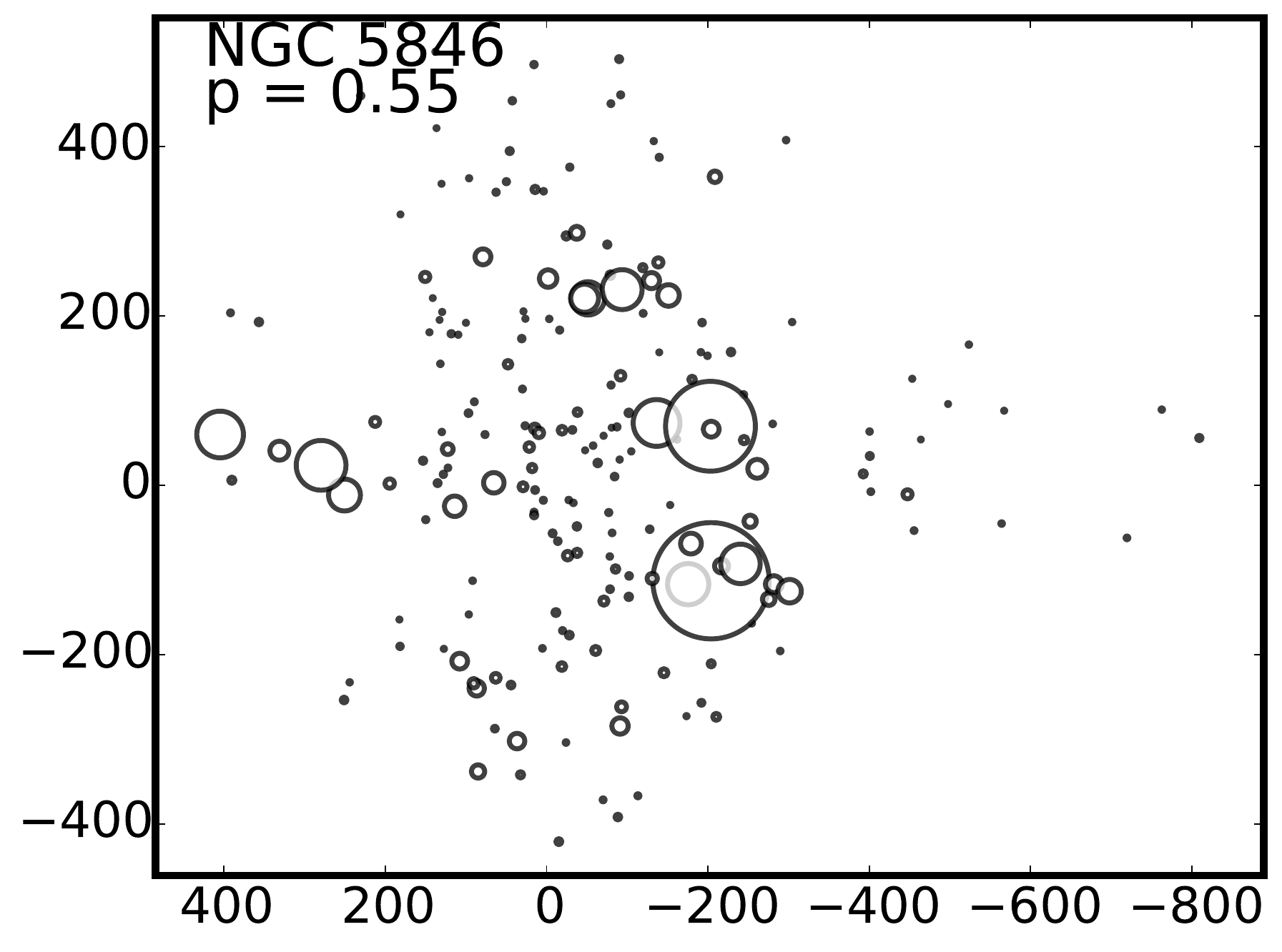}\\[0.05em]
\includegraphics[width=0.24\textwidth, height=0.19\textwidth]{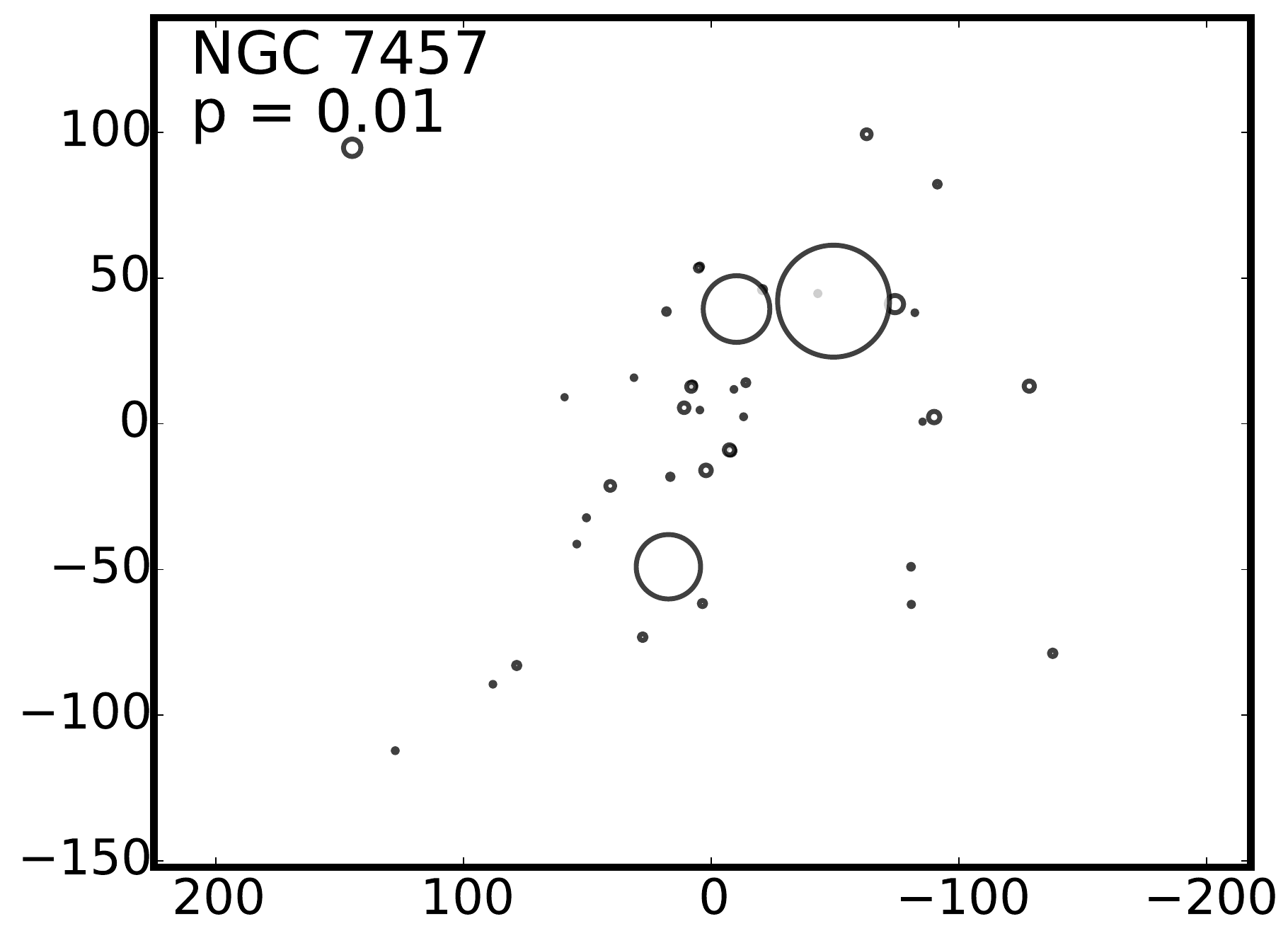}\hspace{0.001\textwidth}%
\includegraphics[width=0.24\textwidth, height=0.19\textwidth]{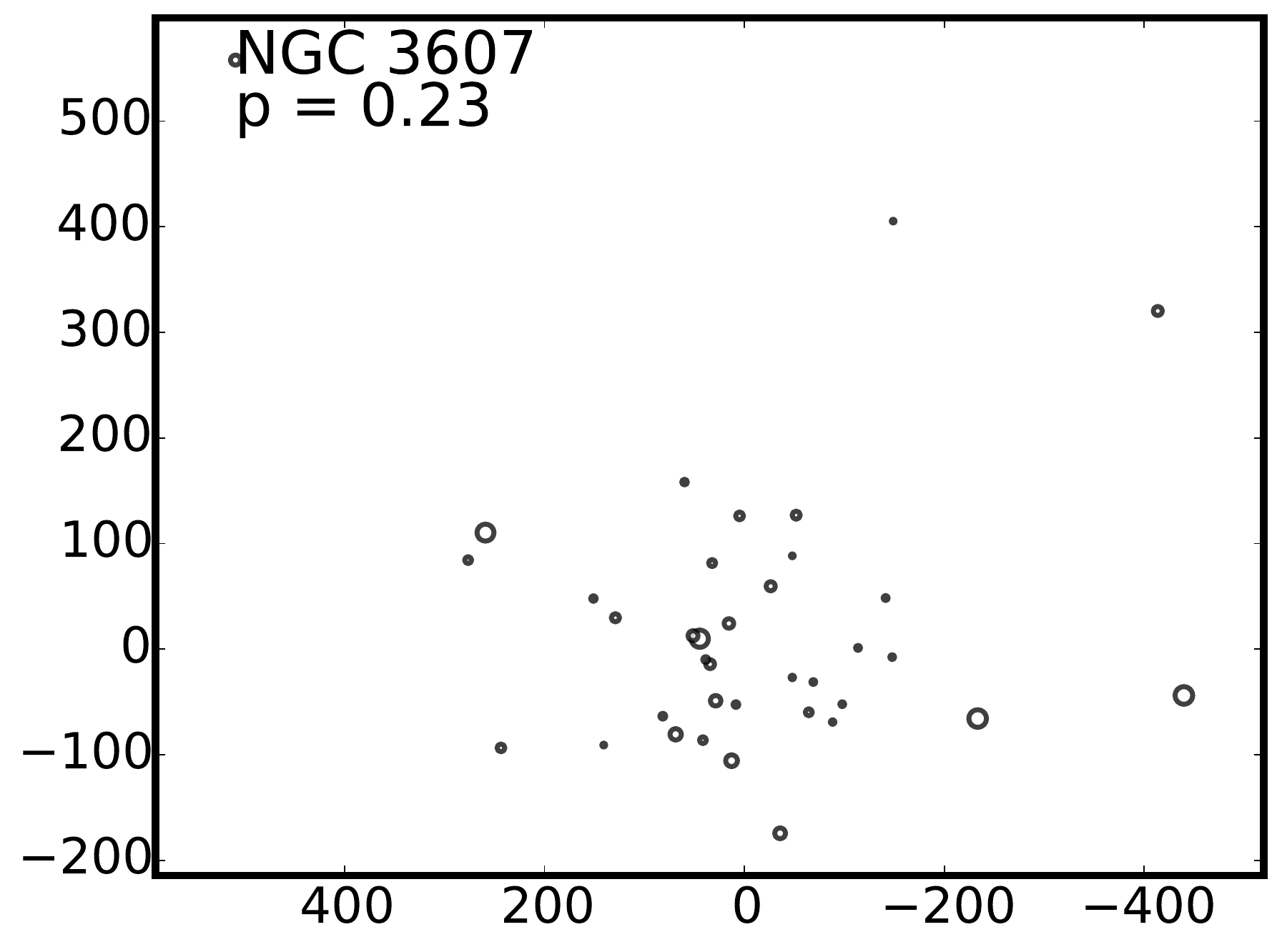}\hspace{0.001\textwidth}%
\includegraphics[width=0.24\textwidth, height=0.19\textwidth]{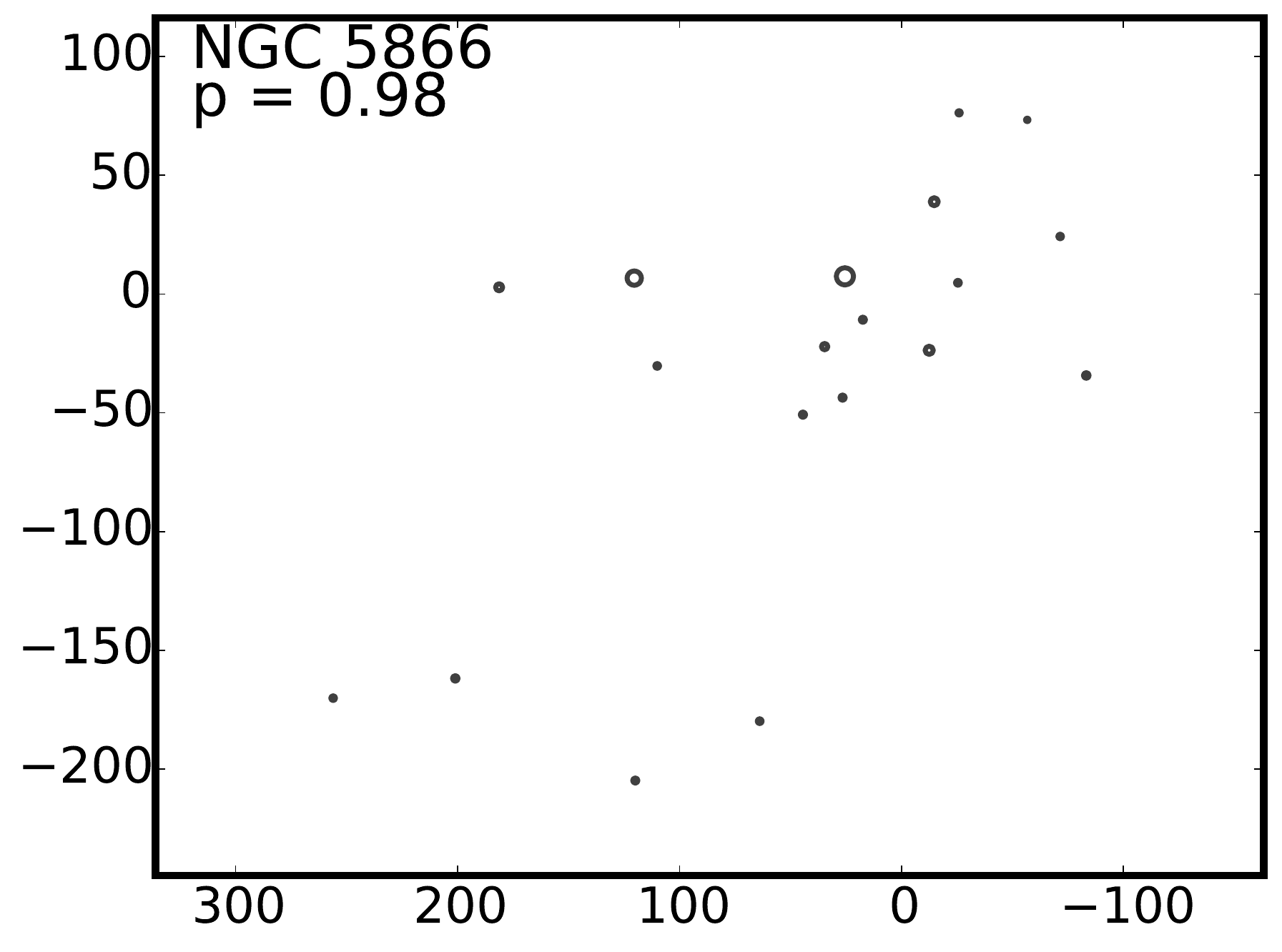}\hspace{0.001\textwidth}%
\color{white}
\includegraphics[draft, width=0.24\textwidth, height=0.19\textwidth]{plots/Bubble_plot_3607.pdf}\par
\color{black}
\leavevmode\smash{\makebox[0pt]{\hspace{1em}
  \rotatebox[origin=l]{90}{\hspace{35em}
    \large{$\rm \Delta DEC\ [arcsec]$}}%
}}\hspace{0pt plus 1filll}\null

\large{$\rm \Delta RA\ [arcsec]$}
\caption{\label{fig:DS_test} GC \textit{bubble} diagrams from the Dressler--Shectman substructure test. The circles represent the GCs and have been scaled to show the differences between local and global kinematics, such that bigger circles show higher probability of kinematic substructures. Galaxy ID and statistical significance of the identified substructures are shown on each plot (the smaller the $p$--value, the higher the significance of the substructure). North is up and East is left in all the plots.}
\end{figure*}
neighbours (as advised by \citealt{Pinkney_1996}). We then compare the local and global kinematics and sum over all the GCs to obtain $\rm \Delta$, the DS statistic, for the GC system using 
\begin{equation}
\Delta = \displaystyle\sum_{i}
\left \{\left(\frac{N_{\rm {nn}}+1}{\sigma_{\rm {global}}^2}\right)[(\bar{V}_{{\rm local},i}-\bar{V}_{\rm {global}})^2+(\sigma_{{\rm local},i}-\sigma_{\rm {global}})^2]\right \}^\frac{1}{2}.
\end{equation}
For a Gaussian--like $\vlos$ distribution, $\Delta$ is approximately of the order of $N_{\rm GC}$ and the larger its value, the more likely it is that the GC system has substructures. However, a non--Gaussian $\vlos$ distribution can also produce a $\Delta$ significantly different from $N_{\rm GC}$ even when there are no real substructures. Therefore, to properly identify substructures and statistically quantify their significance, we perform a Monte~Carlo experiment (repeated 5000 times) where we randomly shuffle the $\vlos$ of the GCs while keeping their positions fixed. This breaks any correlation between position and $\vlos$ while keeping the same velocity distribution and tests against the null hypothesis that there is no correlation between position and $\vlos$. The significance ($p-$value) is the number of times $\rm \Delta$ from the Monte~Carlo experiment is greater than that from the observed data divided by the total number of simulations, such that smaller $p-$values correspond to stronger substructure signatures. For GC systems with statistically significant substructures, i.e $p-val<0.05$, we identify and 
isolate the GCs with correlated kinematics and re--perform the DS test on the ``cleaned" dataset iteratively until $p-val>0.05$. The total numbers of GCs removed per globular cluster system are summarised in Table \ref{tab:observ}. Table \ref{tab:summary} contains the $p-$values for all the galaxies. We show the identified kinematic substructures from the DS test in Figure \ref{fig:DS_test}.

We ensure that our final samples are free of substructures as identified by the DS test. We further compare mass estimates with and without the identified substructures in Section \ref{subs:subx} to ascertain the effect of substructures on our mass estimation. However, we defer a detailed discussion of these substructures, within the context of hierarchical galaxy mass assembly, to a future paper.
	  
\section{Analysis}
\label{analy}
\subsection{Tracer mass estimators (TMEs)}
\label{TME}
The TMEs are generally expressed as 
	  \begin{equation}\label{eq:genTME}
	    M_{\rm p}(<\rout) = \frac{C}{GN}\sum_{i=1}^N {\vlos^2}_{,i} R^{\lambda}_i
\end{equation}
	  where $\rout$ is the de--projected radius of the outermost GC, $G$ is the gravitational constant and $M_{\rm p}$ is the pressure--supported mass, i.e. equation \ref{eq:genTME} assumes no rotation of the system. In practice $\rout$ is taken as the projected galactocentric radius of the outermost GC. The prefactor $C$ varies with TMEs but depends on the slope of the gravitational potential ($\alpha$, see Section \ref{alpha}), the orbital distribution of the GCs ($\beta$, see Section \ref{beta}) and the de--projected density profile of the GCs ($\gamma$, see Section \ref{gamma}). $C$ is defined with two choices as
\begin{equation}
\label{eq:C_eqn}
C = \begin{cases}\
	      \dfrac{16(\alpha+\gamma-2\beta)}{\pi (4-3\beta)}
	      \dfrac{4-\alpha-\gamma}{3-\gamma}
    \dfrac{1-(\rin/\rout)^{3-\gamma}}{1-(\rin/\rout)^{4-\alpha -\gamma}} &\hspace{5pt}  (i)
	  \medskip\\\
	      \dfrac{(\alpha+\gamma-2\beta)}{I_{\alpha,\beta}} \rout^{1-\alpha} &\hspace{5.5pt} (ii)
	  \end{cases}
\end{equation}
where $\rin$ is the de--projected radius of the innermost GC and  
\begin{equation}
I_{\alpha,\beta}
=\dfrac{\pi^{1/2}\Gamma(\tfrac\alpha2+1)}{4\Gamma(\tfrac\alpha2+\tfrac52)}
\left[\alpha+3-\beta(\alpha+2)\right]
\end{equation}
with $\Gamma(x)$ being the gamma function. Equations \ref{eq:C_eqn} \textit{(i)} and \textit{(ii)} are from \citet{Evans_2003} and \citet{Watkins_2010, An_2011}, respectively.
$\lambda\equiv1$ in the TME of \citet{Evans_2003} and $\lambda\equiv\alpha$ in those of \hyperlink{W+10}{W+10} and \citet{An_2011}.
%
%
Our kinematic data consist of $N$ line--of--sight velocity (${\vlos}_{,i}$) measurements at circularised galactocentric radii ($R_i$) defined as: 
\begin{equation}
	  {R=\sqrt{qX^2 + \frac{Y^2 }{q}}} 
	  \label{eq:radius}
\end{equation} 
where $q$ is the ratio of the galaxy photometric minor to major axis ($q=1-\epsilon$), with $X$ and $Y$ as the projected cartesian coordinates of individual GCs on the sky. Equation \ref{eq:radius} is from \citet{Romanowsky_2012}, and it ensures that $R_i$ is in a consistent format with the circularised effective radii \citep{Cappellari_2013b} we have used for our analysis.

The TME of \hyperlink{W+10}{W+10} has been shown to outperform that of Evans et al. (see \hyperlink{W+10}{W+10}), and that of An et al. is just a special case of \hyperlink{W+10}{W+10} where $\gamma\equiv3$. We therefore use the more general TME of \hyperlink{W+10}{W+10} for further analyses and hereafter refer to it as TME.

\subsection{Defining \texorpdfstring{$\alpha$, $\beta$ and $\gamma$}{}}
\label{subs:parameters}
\subsubsection{The power--law slope of the gravitational potential -- $\alpha$}
\label{alpha}
In the TME formalism, the gravitational potential is described mathematically by a
power--law function. This is assumed to be valid in the region probed and the slope is allowed to vary over $-1 \le \alpha \le 1$ such that 
\begin{equation}
\Phi(r)\propto\begin{cases}\
\displaystyle{		
\frac{v_0^2}{\alpha} \left( \frac{a}{r} \right)^{\alpha} }
&(\alpha \ne 0)
\medskip\\\
\displaystyle{		
v_0^2\, \log \left( \frac{a}{r} \right) } & (\alpha = 0).
\end{cases}
\label{eq:def_alpha}
\end{equation} 
$\alpha = 0$ corresponds to an isothermal potential with a flat circular velocity curve (CVC) and $\alpha = 1$ corresponds to a Keplerian potential around a point mass, characterised by a declining CVC. $v_0$ is the 
circular velocity at \textit{scale radius} $a$. 

The power--law slope of the gravitational potential is \textit{a priori} unknown and in the following we use different assumptions based on observations and/or theory to constrain our choice of $\alpha$.
The simplest clue about $\alpha$ is to be found from recent studies \citep[e.g.][]{Auger_2010, Thomas_2011, Cappellari_2015} where the total mass density of ETGs was found to be \textit{nearly} isothermal with a small intrinsic scatter i.e $\rho(r) \propto r^{-2}$.
These studies therefore suggest that $\alpha{\sim}0$. 
However, there are indications of a trend in the logarithmic slope of the total mass density profiles for ETGs with the more (less) massive 
ETGs having shallower (steeper) slopes both observationally (e.g. \citealt{Barnabe_2011},\hyperlink{D+12}{D+12},\citealt{Tortora_2014}) and from cosmological simulations \citep[e.g.][]{Remus_2013, Dutton_2013}. This implies that a variety of shapes would be seen in the CVCs at large radii.

\begin{figure}
    \includegraphics[width=0.48\textwidth]{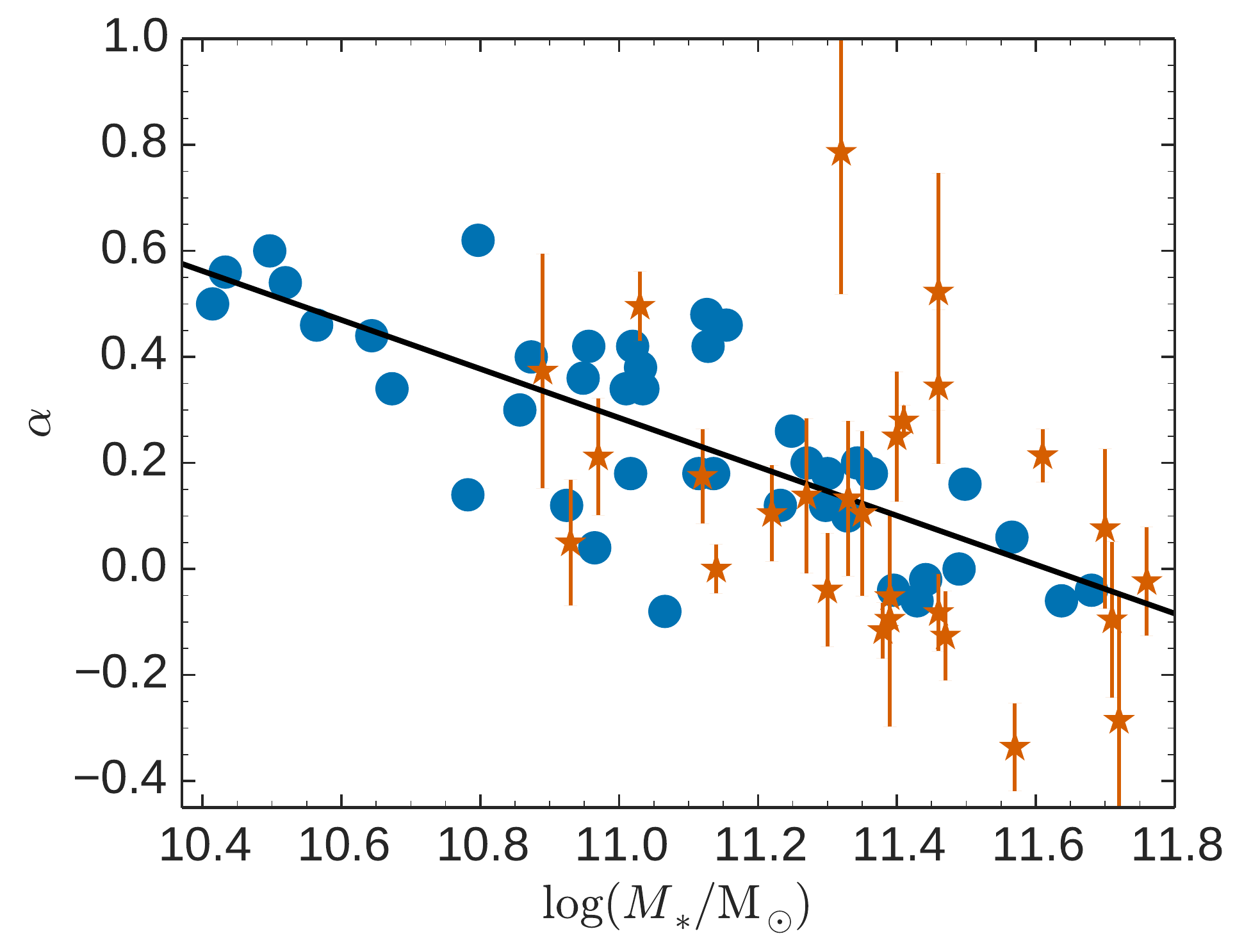}\hspace{0.01\textwidth}%
	\caption{\label{fig:fit_param1} The power--law slope of the gravitational potential, $\alpha$ vs. galaxy 
	stellar mass. The blue circles are from the simulated ETGs  
	in \citet{Wu_2014}, while the brown stars are from the observational data published in \citet{Trujillo_2011}. 
	The solid line is the best--fit to the predictions from \citet{Wu_2014}.} 
\end{figure}

Under the assumption of a power--law gravitational potential, $\alpha$ can be evaluated (see \citealt{Evans_1994a}) as the logarithmic slope of the CVC at large radii
\begin{equation}
\label{eq:det_alpha}
\alpha \equiv -\lim_{R \to \infty} \frac{\textrm{d~log\ }{V}_{\rm c}^2}{\textrm{d~log\ } R}.
\end{equation}
Using equation \ref{eq:det_alpha} we determine $\alpha$ given the CVCs from the cosmological hydrodynamical resimulations of \citet{Oser_2010, Oser_2012}. We use the logarithmic slopes of their CVCs as analysed by \hypertarget{Wu+14}{\citet{Wu_2014}}, hereafter \hyperlink{Wu+14}{Wu+14}, in 42 of these simulated ETGs.   
The simulated ETGs have stellar masses over the range $2.7 \times 10^{10} - 4.7 \times 10^{11}~\rm \Msun$, comparable to the stellar mass range in this study. The logarithmic slope is evaluated at 5~$\re$. 
We find an empirical relation between $\alpha$ and the logarithm of the stellar mass by fitting a linear function to the data (see Figure \ref{fig:fit_param1}). 
The best--fit linear function to the data is 
\begin{equation}\label{eq:fit_alpha}
\alpha = (-0.46\pm0.06) \times {\rm log}(M_*/\Msun) + (5.29\pm0.68)
\end{equation}
with a rms scatter of 0.13$\pm$0.01.
Using equation \ref{eq:det_alpha} and the radially extended CVC data (out to 20~kpc) for ETGs published in \citet{Trujillo_2011}, we confirm that the relation obtained above is consistent with observations in the region of overlap. Our best--fit function is similar to those reported in \citet{Tortora_2014} determined at much more central radii of 0.5 and 1~$\re$. When constrained this way, $\alpha$ reflects the shallower (steeper) total mass density profiles for more (less) massive ETGs. With equation \ref{eq:fit_alpha}, $\alpha\sim0.4$ for an arbitrary galaxy with MW--like stellar mass, consistent with the results for the MW potential in \citet{Yencho_2006} and \citet{Watkins_2010}. Table \ref{tab:summary} contains a summary of $\alpha$ adopted for the galaxies in this study, given their stellar mass. 


%
\subsubsection{The orbital anisotropy parameter -- $\beta$}
\label{beta}
The Binney anisotropy parameter, $\beta$,  \citep{Binney_1987} describes the orbital distribution of the GCs. It can be a major source of uncertainty in mass modelling of ETGs as it is poorly constrained. It is defined (assuming spherical symmetry) as 
\begin{equation}
	\beta = 1 - \frac{\sigma_{\theta}^2}{\sigma_{r}^2}
\end{equation}
where $\sigma_{\theta}$ and $\sigma_{r}$ are the tangential and radial velocity dispersions, respectively. The TMEs are based on the assumption of constant anisotropy with radius. We do not fit for $\beta$, but rather we derive
mass estimates assuming $\beta=0,0.5,-0.5$, corresponding to isotropic, strong radial and mild tangential anisotropies, respectively. Our choice of $\pm0.5$ is predicated on results from mass modelling where typical anisotropies are usually defined such that $-0.5\leq\beta\leq0.5$ \citep{Gerhard_2001, Cappellari_2007}. We show in Section \ref{subs:sensitivity} the sensitivity of our mass estimates to this parameter.  
%
%
\subsubsection{The power--law slope of the de--projected GC density profile -- $\gamma$}
\label{gamma}
\begin{figure}
    \includegraphics[width=0.48\textwidth]{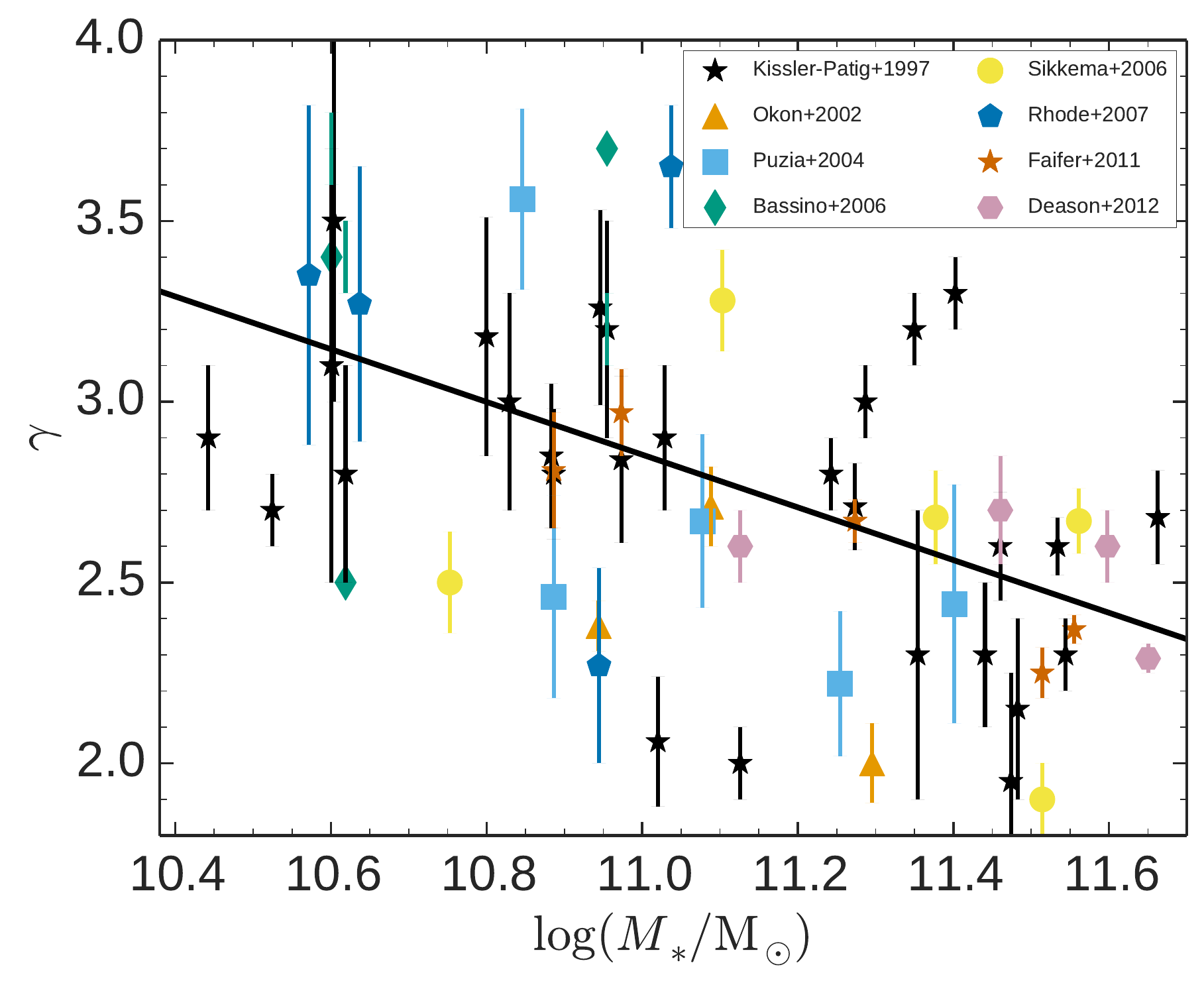}\hspace{0.01\textwidth}%
	\caption{\label{fig:fit_param2} The power--law 
	slope of the de--projected GC density profile, $\gamma$ vs. galaxy stellar mass. Data points are from 		\citealt{Kissler_1997, Okon_2002, Puzia_2004, Bassino_2006, Sikkema_2006, Rhode_2007, Faifer_2011} and \citealt{Deason_2012}, as summarized in the plot legend. The solid line is a linear fit to all of the data.}  
\end{figure}
We follow \citet{Harris_1976} and derive the de--projected GC density distribution $n(r) \sim r^{-\gamma}$ 
given the projected density profiles of photometric GCs in the plane of the sky i.e. $N(R) \sim R^{-(\gamma-1)}$. 
It is well known that the slope of the GC surface density profile varies with the galaxy luminosity
\citep[e.g.][]{Harris_1986, Kissler_1997, Dirsch_2005, Bekki_2006}. We therefore make a compilation of measured slopes of the GC density profiles (which we de--project) from wide--field photometric studies in the literature (\citealt{Kissler_1997, Okon_2002, Puzia_2004, Bassino_2006, Sikkema_2006, Rhode_2007, Faifer_2011}, \hyperlink{D+12}{D+12}) and the corresponding stellar mass of the host galaxy (using distance from \citealt{Tonry_2001, Blakeslee_2009, Brodie_2014}, the absolute $K$--band magnitude from 2MASS, the correction 
from \citealt{Scott_2013} and assuming a stellar $M/L_K=1$). Figure \ref{fig:fit_param2} shows the deprojected--power law slopes as a function of galaxy stellar mass. The best--fit linear function to the data is  
\begin{equation}\label{eq:fit_gamma}
\gamma = ( -0.63 \pm 0.17) \times {\rm log}(M_*/\Msun) + (9.81 \pm 1.94) 
\end{equation}
with a rms scatter of 0.29$\pm$0.04 in the data around the best--fit line.
With this linear relation we estimate the de--projected slope of the GC density profile of a galaxy given its stellar mass.
This is a useful tool when the photometric data are not readily available. Table \ref{tab:summary} contains a summary of  $\gamma$ for all the galaxies in this study. 

The power--law slope of the de--projected GC density profile is thus constrained to $2 \le \gamma \le 4$, with more massive ETGs having shallower profiles and lower--mass ETGs showing steeper profiles. For a galaxy with MW--like stellar mass, we find $\gamma=3.3$, similar to that of the Galaxy, ${\sim}3.5$ \citep{Harris_1976, Watkins_2010}. 

\subsection{Sensitivity of pressure--supported mass estimates to \texorpdfstring{$\alpha$, $\beta$ and $\gamma$}{}}
\label{subs:sensitivity}
\begin{figure*}
    \includegraphics[width=0.96\textwidth]{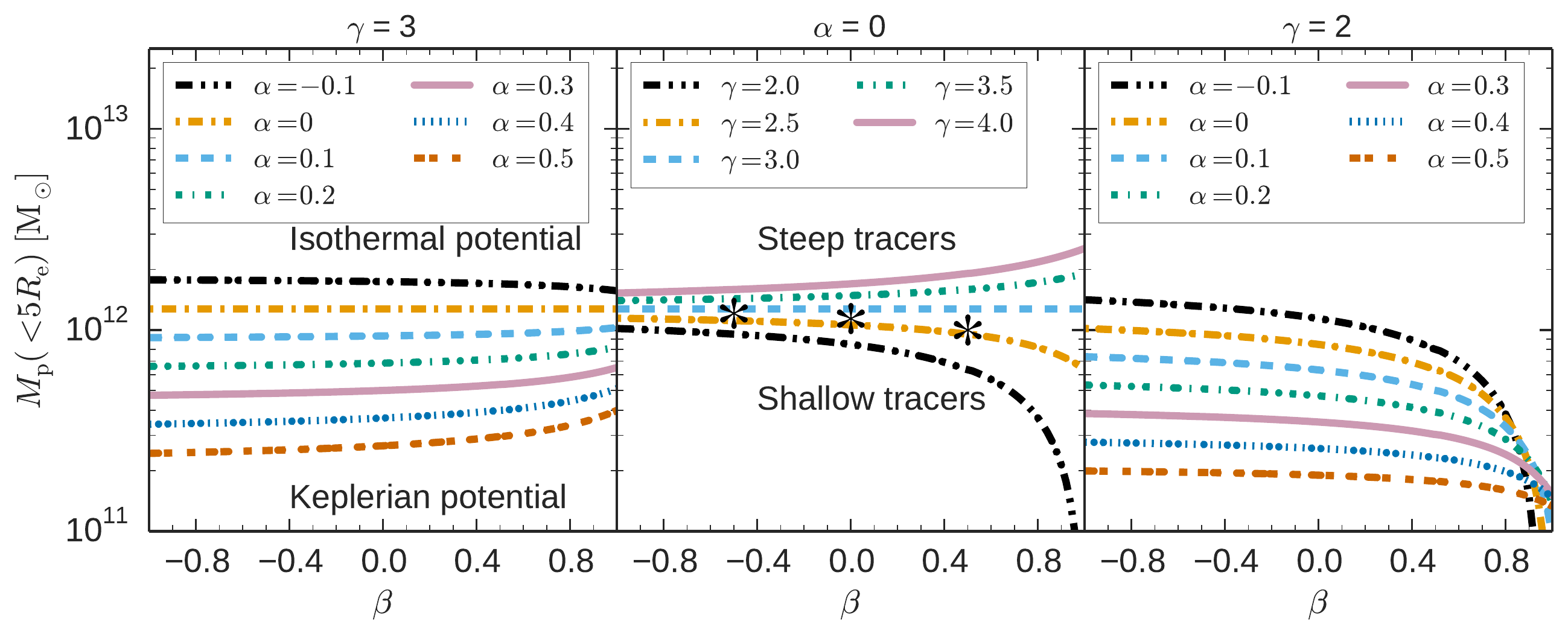}
	\caption{\label{fig:sensi} Sensitivity of pressure--supported mass estimate for NGC~1407 within 5 effective radii to parameters $\alpha$, $\beta$ and $\gamma$. \textit{Left panel}: Mass estimate when $\gamma\equiv 3$, with $-0.1 \leq \alpha \leq 0.5$ and $-1.0 \leq \beta \leq 1.0$. A 0.1 change in $\alpha$ at $\beta=0$ and $\gamma=3$ corresponds to a change in the mass estimate of ${\sim}$30 per cent. \textit{Middle panel}: Mass estimate when $\alpha \equiv 0$, with $2.0 \leq \gamma \leq 4.0$ and $-1.0 \leq \beta \leq 1.0$. A change of 0.5 in $\gamma$  at $\beta=0$ and $\alpha=0$ changes the mass estimate by ${\sim}$20 per cent. Mass estimates significantly diverge for $\beta=0.5$, with strongly radial orbital distributions producing extremely divergent mass estimates. Note that when $\alpha+\gamma=3$, mass estimates are very insensitive to $\beta$. The asterisks show the pressure--supported mass estimates for NGC~1407 when $\beta$ is $-0.5, 0, 0.5$, respectively. \textit{Right panel}: At the shallow limit of the power--law slope of the mass tracers (i.e. $\gamma=2$) and isothermal gravitational potential (i.e. $\alpha\sim0$), strongly radial orbits produce degenerate mass estimates.} 
\end{figure*}

We investigate the effects of the adopted values of $\alpha$, $\beta$ and $\gamma$ on the pressure--supported mass estimates, $M_{\rm p}$, from equation \ref{eq:genTME}, using NGC~1407 as a test case. \citet{diCintio_2012} showed that while the variations in $M_{\rm p}$ due to uncertainties in $\gamma$ and $\beta$ can be generalized, that due to changes in $\alpha$ is a complicated function (see their equation 19) that varies from galaxy to galaxy, depending on the radial distribution of the tracers. Figure \ref{fig:sensi} shows $M_{\rm p}$ within 5~$\re$ for NGC~1407. For our sensitivity tests, we extend the range of $\beta$ out to $\pm1$ to study mass--anisotropy dependencies at more extreme values. The \textit{left panel} shows $M_{\rm p}$ when $\gamma\equiv3$, $-0.1 \leq \alpha \leq 0.5$ and $-1.0 \leq \beta \leq 1.0$. In the \textit{middle panel}, $\alpha\equiv0$, $2 \leq \gamma \leq 4$ and $-1.0 \leq \beta \leq 1.0$. The two plots reveal that mass estimates are least sensitive to $\beta$ and most sensitive to the assumed potential slope, $\alpha$. For example, a 0.5 change in $\gamma$ when $\beta=0$ and $\alpha=0$ alters the mass estimate by ${\sim}$20 per cent while a change of 0.1 in $\alpha$ at $\beta=0$ and $\gamma=3$ changes the mass by ${\sim}$30 per cent. Ignorance of the nature of $\beta$ becomes an increasingly important issue only when the orbital distribution of the tracers is strongly radial, i.e., $\beta \geq 0.5$. We have also performed this test on all the other galaxies in the sample and confirm that in all cases $M_{\rm p}$ is most sensitive to $\alpha$ and least sensitive to $\beta$.

For galaxies with nearly isothermal gravitational potentials, radially biased orbital distributions increasingly lead to lower total mass estimates. However, in strongly Keplerian potentials, radially biased orbital distributions would lead to higher mass estimates. This implies that when $\alpha+\gamma<3$, the total mass obtained under the assumption of tangential anisotropy is greater than that obtained assuming radial anisotropy. In the same way, when $\alpha+\gamma>3$, the total mass obtained under the assumption of tangential anisotropy is less than that obtained assuming radial anisotropy. This is the classic situation from dynamical modelling studies with stars and PN in the far outer haloes e.g. \citep{Dekel_2005}. When $\alpha+\gamma\sim3$, the mass estimates are insensitive to $\beta$, similar to the result reported in \citet{Wolf_2010} for pressure supported galaxies. Also, when $\beta \to 1$ (see \textit{right panel}), particularly for galaxies with an isothermal gravitational potential i.e. more massive ellipticals, the mass estimates become degenerate \citep[see also][]{Wolf_2010}. The typical $\alpha$ and $\gamma$ pair adopted for the low--, intermediate-- and high-- mass galaxies in our sample are (0.4, 3.4), (0.2, 2.9) and (0, 2.6), respectively.
\subsection{Quantifying the effects of galaxy flattening and GC rotation on mass estimates}
\label{subs:flat_rot}
The tracer mass estimators are built on the assumption that galaxies are spherically symmetric and  pressure--supported. However, these assumptions are not always \textit{valid} and mass estimates thus need to properly account for other realities.
Edge--on and face--on galaxies, under the sphericity assumption, would have their masses over--estimated or under--estimated, respectively \citep[see][]{Bacon_1985, Bender_1994, Magorrian_2001}, closely mimicking the mass--anisotropy degeneracy.
A flattening--based mass correction of some sort is therefore necessary. Galaxies in our sample have 
been deliberately chosen with a bias towards edge--on inclinations to reduce confusion in mass estimates from projection effects, hence we are affected more by \textit{over--estimation}. 
We apply the normalizing factor from \citealt{Bacon_1985} to correct for the effect of galaxy flattening on our dynamical mass estimates (their eqn. 9), assuming that GC systems have the same ellipticity as the galaxy stars. 
We multiply our mass estimate $M_{\rm p}$ (obtained under the assumption of sphericity i.e. $q^{\prime}\sim1$), from equation \ref{eq:genTME}, by a factor $corr$ to normalise to mass when $q=1-\epsilon$. We use  
\begin{equation}
\label{eq:corr}
\begin{split}
corr(q^{\prime},q)&= \left(\frac{e^{\prime}}{e}\right)^{-3}\\
&\cdot\frac{(sin^{-1} e^{\prime} -e^{\prime}q^{\prime})(1-q^2)-2{q^{\prime}}^2(sin^{-1} e^{\prime} - e^{\prime}/q^{\prime})({q}^2-{q^{\prime}}^2)}{(1-{q^{\prime}}^2)(sin^{-1} e - eq)}
\end{split}
\end{equation}
where $e^{\prime} = (1-q^{\prime})^{1/2}$  and $e = (1-q)^{1/2}$.

\begin{figure}
    \includegraphics[width=0.48\textwidth]{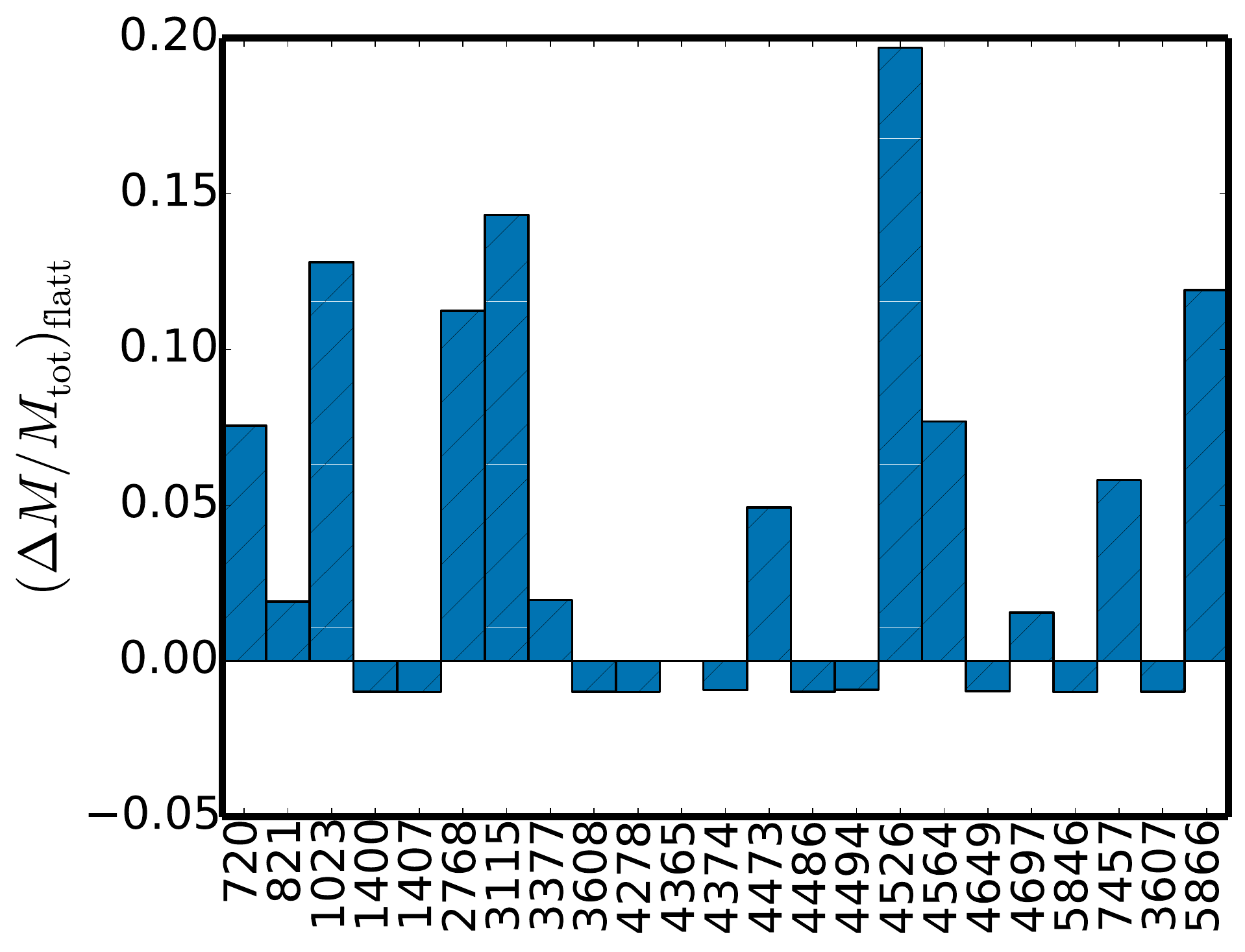}
	\caption{\label{fig:cmp_flatt} Bar chart showing effect of galaxy flattening on the total dynamical mass within 5~$\re$. For our galaxy sample, we are largely affected by mass over--estimation, with an average mass \textit{over--estimation} due to galaxy flattening of ${\sim}$5 per cent.} 
\end{figure}

Figure \ref{fig:cmp_flatt} shows the effects of galaxy flattening on the total mass estimates within 5~$\re$ after applying the correction from equation \ref{eq:corr}. The average difference in total mass estimates due to galaxy flattening is ${\sim}$5 per cent. This reflects the bias of our galaxy sample in favour of edge--on galaxies. We note that the severity of the over--estimation is highest for NGC~4526 (with $\epsilon=0.76$, where the total mass would have been over--estimated by ${\sim}$20 per cent). We list the correction factors so obtained for each galaxy in Table~\ref{tab:summary} and report dynamical mass estimates corrected for galaxy flattening in Table~\ref{tab:mass_summary}. 

Similarly, dynamical masses obtained under the assumption of non--rotating tracers would be largely \textit{under--estimated} for galaxies where the tracer population has kinematics dominated by rotation. Flattened (disky) ETGs have been shown to be mostly fast central rotators \citep{Krajnovic_2011}, with some of them observed to remain fast rotators even at large radii \citep[e.g.][]{Arnold_2014}. 
This result has been confirmed in studies that probed the kinematics of ETGs beyond 5~$\re$ \citep[e.g.][]{Coccato_2009, Pota_2013} with dynamical tracers often showing significant rotation in the outer haloes. 
Therefore, there is a non--negligible mass contribution from rotation, especially in the flattened ETGs, 
that needs to be accounted for. We obtain the best fit rotation amplitude ($V_{\rm rot}$), velocity dispersion ($\sigma$) and kinematic position angle ($PA_{\rm kin}$), respectively, for each galaxy by fitting
\begin{equation}\label{eq:model}
V_{\rm mod,\textit{i}}=V_{\rm sys} \pm \frac{V_{\rm rot}}{\sqrt{1+\left(\frac{\tan(PA_{i}-PA_{\rm kin})}{q_{\rm kin}}\right)^2}}
\end{equation}
to our GC data while we minimise
\begin{equation}\label{eq:model_min}
\chi^2\propto\sum^{}_{i} \left[\frac{(V_{i}-V_{\rm mod,\textit{i}})^2}{(\sigma^2+(\Delta V_{i})^2)}+\ln (\sigma^2+(\Delta V_{i})^2)\right].
\end{equation}
These equations are commonly used in studies of GC kinematics \citep[e.g.][]{Bergond_2006, Pota_2013}. In equations \ref{eq:model} and \ref{eq:model_min}, $V_{i}$, $\Delta V_{i}$ and $PA_{\rm i}$ are the measured radial velocities, uncertainties on the measured radial velocities and position angles of the GCs, respectively. $V_{\rm sys}$ is the galaxy recession velocity and we fix $q_{\rm kin}$ to the photometric axial ratio $q$. The uncertainties on the kinematic parameters are obtained through Monte Carlo simulations.

\begin{figure}
    \includegraphics[width=0.48\textwidth]{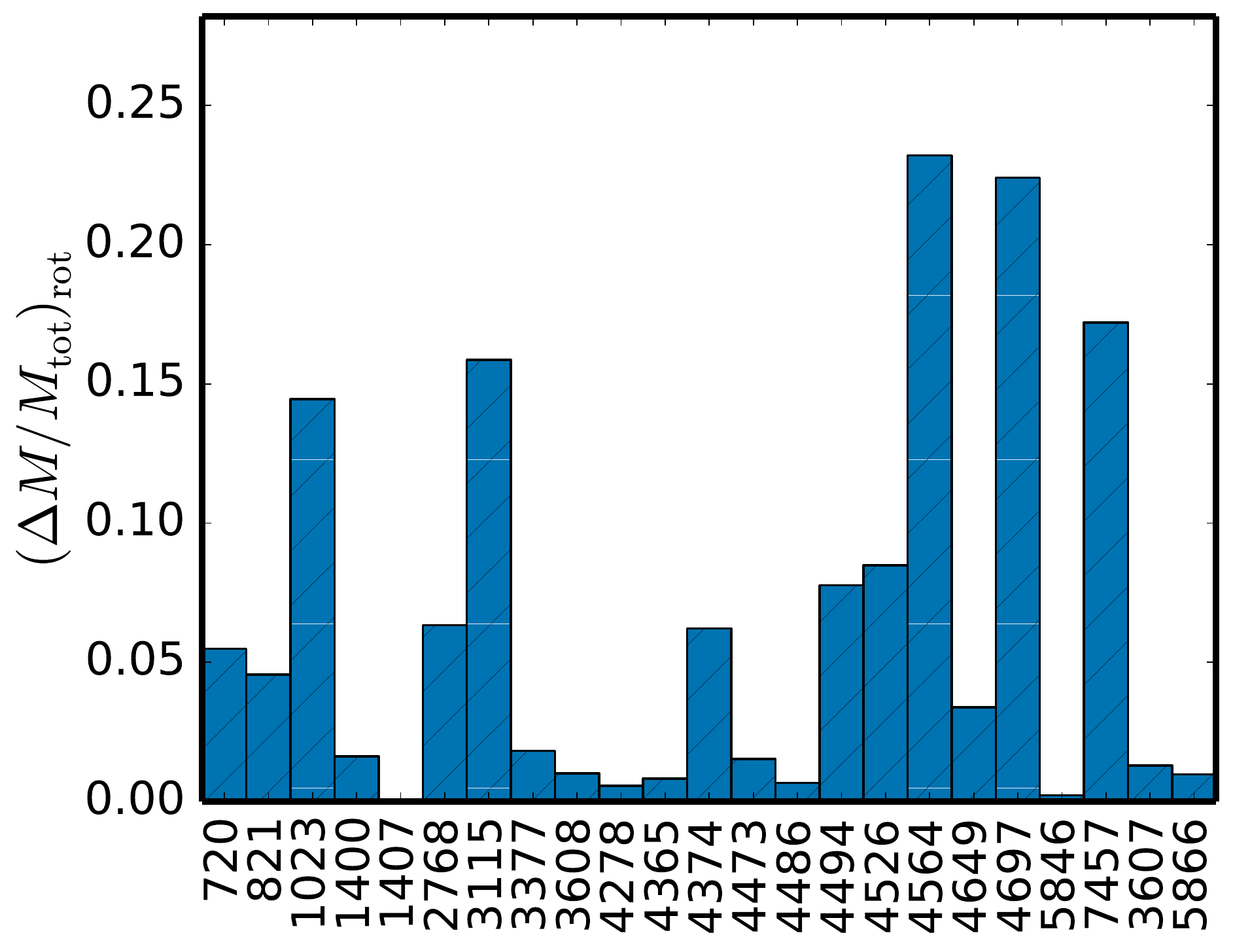}
	\caption{\label{fig:cmp_rot_sig} Bar chart showing effect of rotation in the tracer population on the total dynamical mass within 5~$\re$. For our galaxy sample, the average mass \textit{under--estimation} is ${\sim}$6 per cent.} 
\end{figure}

We summarise the significance of rotation by quantifying $V_{\rm rot}/\sigma$ for each GC system. We note that while few ETGs have GC systems that are rotation dominated with $V_{\rm rot}/\sigma > 1$, most of them show significant rotation ($V_{\rm rot}/\sigma \geq 0.4$). We therefore quantify the rotationally supported mass, $M_{\rm rot}$, enclosed within projected radius $R_{\rm out}$ using 
\begin{equation}\label{eq:mass_rot}
M_{\rm rot} = \frac{R_{\rm out}{V_{\rm rot}}^2}{G}
\end{equation}
Figure \ref{fig:cmp_rot_sig} shows the contribution from rotation to the total mass within 5~$\re$. For galaxies in our sample, the average contribution from rotation to the total mass is ${\sim}$6 per cent, with the maximum under--estimation of ${\sim}$20 per cent in NGC~4526 and NGC~4564. 

\subsection{Quantifying the effect of kinematic substructures on mass estimates}
\label{subs:subx}
For the galaxies with statistically significant kinematic substructures, identified in Section \ref{subs:iso_subx}, 
we obtain new mass estimates using the \textit{cleaned} catalogues and compare them with the mass estimates from the original catalogues, within 5~$\re$ and under different isotropy assumptions. 
\begin{figure}
    \includegraphics[width=0.48\textwidth]{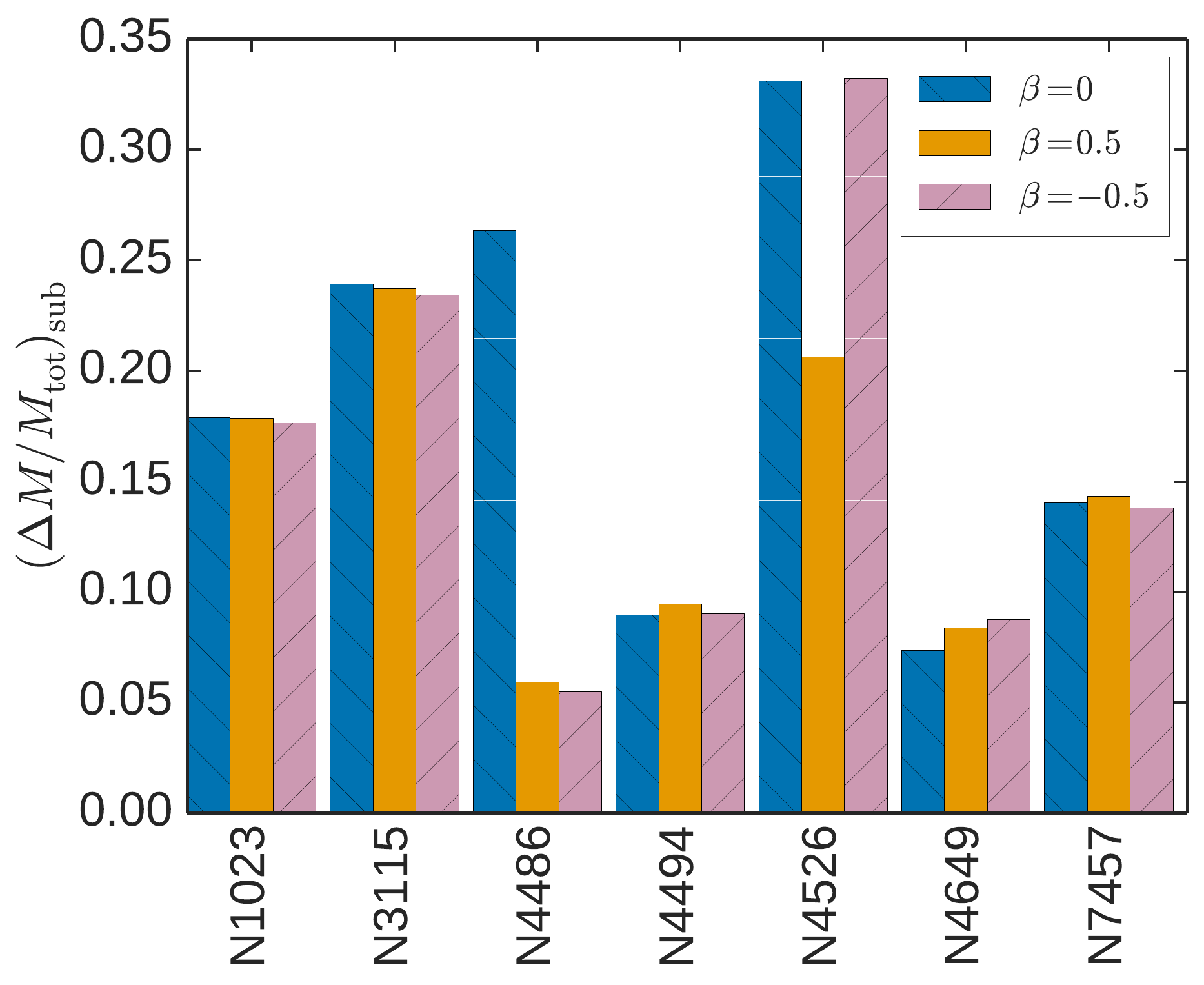}
	\caption{\label{fig:cmp_subx} Effect of kinematic substructures on mass estimates within 5~$\re$ for galaxies with statistically significant kinematic substructures. The bar chart shows the fractional mass \textit{over--estimation} due to kinematic substructures. It also shows how the mass estimate changes depending on the assumption made for the orbital anisotropy parameter, $\beta$. For our galaxy sample, kinematic substructures lead to ${\sim}$14--19 per cent \textit{over--estimation} in total mass, depending on
$\beta$.} 
\end{figure}
Figure \ref{fig:cmp_subx} shows the fractional change in the mass estimate due to the kinematic substructure $(\Delta M/M_{\rm tot})_{\rm sub}$ for isotropic, radial and tangential velocity distributions. Removing kinematic substructures lead to reduction in mass estimates and for our galaxy sample, the average \textit{over--estimation} varies from ${\sim}14-19$ per cent, depending on velocity anisotropy. This agrees with the study of \citet{Yencho_2006} where the effect of substructure on mass estimate of galaxies was found to be ${\sim}20$ per cent. The greatest fractional mass over--estimation is found in NGC~4526 (${\sim}$30 per cent). For the galaxies with identified kinematic substructures, the total mass estimates in Table \ref{tab:mass_summary} are from the \textit{cleaned} catalogues i.e. corrected for substructures.
\begin{table*}
\footnotesize
\captionsetup{width=.96\linewidth}
{\small
\caption{Summary of mass estimates and dark matter fractions assuming different anisotropy. The results shown here have been obtained using the tracer mass estimator of \citealt[][(see Section \ref{TME} for details]{Watkins_2010}) and assuming stellar $M/L_{\rm K}=1$. $M_{\rm p}$ is the pressure--supported mass and it has been corrected for the effect of galaxy flattening. $M_{\rm rot}$ is the rotationally--supported mass. 
$M_{\rm tot}$ is the total dynamical mass after correcting for galaxy flattening, rotation in the GC system and the presence of kinematic substructures (for galaxies with $p<0.05$). $f_{\rm DM}$ is the dark matter fraction. We list masses enclosed within spheres of radius 5~$\re$ and $R_{\rm max}$, the maximum galactocentric radius where we have GC kinematic data.} \label{tab:mass_summary}}
\begin{tabular}{lcccccccccl}
\hline\hline  
Galaxy & $\beta$ & $M_{\rm rot}(<5{\re})$ & $M_{\rm p}(<5{\re})$ & $M_{\rm tot}(<5{\re})$ & $f_{\rm DM} (<5{\re})$ & $R_{\rm max}$ & $M_{\rm rot}(<R_{\rm max})$ & $M_{\rm p}(<R_{\rm max})$ & $M_{\rm tot}(<R_{\rm max})$& $f_{\rm DM}(<R_{\rm max})$ \\
$\rm [NGC]$ &  & $[10^{10} \Msun]$ & $[10^{11} \Msun]$ & $[10^{11} \Msun]$ &  & [$\re$] & $[10^{10} \Msun]$ & $[10^{11} \Msun]$ & $[10^{11} \Msun]$ & 	\\ 
\hline
 720  &     0& 2.0 $\pm$0.4 & 3.4 $\pm$0.8 & 3.6 $\pm$0.7 & 0.46$\pm$0.16 & 19.05 & 7.6 $\pm$1.4 & 11.9 $\pm$2.4  & 12.7 $\pm$2.2 &  0.83$\pm$0.05 \\      
       &   0.5&              & 3.3 $\pm$0.7 & 3.5 $\pm$0.7 & 0.43$\pm$0.19 &       & 7.6 $\pm$1.4 & 11.3 $\pm$2.2  & 12.1 $\pm$2.1 &  0.82$\pm$0.05 \\      
       &  -0.5&              & 3.5 $\pm$0.8 & 3.7 $\pm$0.7 & 0.47$\pm$0.16 &       & 7.6 $\pm$1.4 & 12.1 $\pm$2.4  & 12.9 $\pm$2.2 &  0.83$\pm$0.05 \\ 
  \hline
  821  &     0& 1.9 $\pm$0.4 & 4.0 $\pm$0.8 & 4.2 $\pm$0.8 & 0.81$\pm$0.06 &  8.70 & 3.4 $\pm$0.7 & 5.6  $\pm$1.0  & 6.0  $\pm$1.0 &  0.85$\pm$0.04 \\      
       &   0.5&              & 4.2 $\pm$0.8 & 4.4 $\pm$0.8 & 0.81$\pm$0.06 &       & 3.4 $\pm$0.7 & 5.8  $\pm$1.1  & 6.2  $\pm$1.1 &  0.86$\pm$0.04 \\      
       &  -0.5&              & 4.0 $\pm$0.8 & 4.2 $\pm$0.8 & 0.8 $\pm$0.06 &       & 3.4 $\pm$0.8 & 5.5  $\pm$1.0  & 5.9  $\pm$1.0 &  0.85$\pm$0.04 \\ 
  \hline
  1023 &     0& 2.4 $\pm$0.4 & 1.4 $\pm$0.3 & 1.6 $\pm$0.2 & 0.48$\pm$0.13 & 16.15 & 7.7 $\pm$1.4 & 3.2  $\pm$0.6  & 4.0  $\pm$0.5 &  0.76$\pm$0.05 \\      
       &   0.5&              & 1.5 $\pm$0.3 & 1.7 $\pm$0.2 & 0.49$\pm$0.12 &       & 7.7 $\pm$1.4 & 3.3  $\pm$0.6  & 4.1  $\pm$0.5 &  0.77$\pm$0.05 \\      
       &  -0.5&              & 1.4 $\pm$0.3 & 1.6 $\pm$0.2 & 0.47$\pm$0.12 &       & 7.7 $\pm$1.4 & 3.1  $\pm$0.6  & 3.9  $\pm$0.5 &  0.76$\pm$0.05 \\      
 \hline
 1400 &     0& 0.4 $\pm$0.1 & 2.3 $\pm$0.5 & 2.3 $\pm$0.5 & 0.5 $\pm$0.19 & 20.62 & 1.6 $\pm$0.6 & 7.2  $\pm$1.3  & 7.4  $\pm$1.3 &  0.82$\pm$0.04 \\      
       &   0.5&              & 2.3 $\pm$0.5 & 2.4 $\pm$0.5 & 0.5 $\pm$0.18 &       & 1.6 $\pm$0.6 & 7.2  $\pm$1.3  & 7.4  $\pm$1.3 &  0.82$\pm$0.05 \\      
       &  -0.5&              & 2.3 $\pm$0.5 & 2.3 $\pm$0.6 & 0.49$\pm$0.23 &       & 1.6 $\pm$0.6 & 7.2  $\pm$1.3  & 7.3  $\pm$1.3 &  0.82$\pm$0.05 \\      
  \hline
  1407 &     0& 0.1 $\pm$0.0 & 11.5$\pm$1.1 & 11.5$\pm$1.1 & 0.71$\pm$0.06 & 14.14 & 0.2 $\pm$0.1 & 36.6 $\pm$2.7  & 36.6 $\pm$2.6 &  0.9 $\pm$0.02 \\      
       &   0.5&              & 10.2$\pm$1.0 & 10.2$\pm$1.0 & 0.67$\pm$0.07 &       & 0.2 $\pm$0.1 & 32.3 $\pm$2.4  & 32.3 $\pm$2.3 &  0.88$\pm$0.02 \\      
       &  -0.5&              & 12.2$\pm$1.2 & 12.2$\pm$1.2 & 0.72$\pm$0.06 &       & 0.2 $\pm$0.1 & 38.7 $\pm$2.8  & 38.7 $\pm$2.9 &  0.9 $\pm$0.02 \\      
  \hline
  2768 &     0& 4.5 $\pm$0.4 & 6.7 $\pm$1.1 & 7.1 $\pm$1.0 & 0.77$\pm$0.07 & 11.36 & 10.3$\pm$0.9 & 13.1 $\pm$2.1  & 14.1 $\pm$1.8 &  0.87$\pm$0.04 \\      
       &   0.5&              & 6.5 $\pm$1.1 & 6.9 $\pm$0.9 & 0.76$\pm$0.07 &       & 10.3$\pm$0.9 & 12.7 $\pm$2.0  & 13.7 $\pm$1.8 &  0.87$\pm$0.04 \\      
       &  -0.5&              & 6.8 $\pm$1.1 & 7.2 $\pm$1.0 & 0.77$\pm$0.07 &       & 10.3$\pm$0.8 & 13.3 $\pm$2.1  & 14.3 $\pm$1.8 &  0.87$\pm$0.04 \\      
  \hline
  3115 &     0& 3.2 $\pm$0.6 & 1.7 $\pm$0.3 & 2.0 $\pm$0.3 & 0.57$\pm$0.08 & 18.35 & 11.8$\pm$2.3 & 4.5  $\pm$0.7  & 5.7  $\pm$0.6 &  0.83$\pm$0.02 \\      
       &   0.5&              & 1.8 $\pm$0.3 & 2.1 $\pm$0.3 & 0.58$\pm$0.07 &       & 11.8$\pm$2.3 & 4.6  $\pm$0.7  & 5.8  $\pm$0.6 &  0.84$\pm$0.02 \\      
       &  -0.5&              & 1.7 $\pm$0.3 & 2.0 $\pm$0.3 & 0.56$\pm$0.09 &       & 11.8$\pm$2.2 & 4.4  $\pm$0.6  & 5.6  $\pm$0.6 &  0.83$\pm$0.02 \\      
  \hline
  3377 &     0& 0.1 $\pm$0.1 & 0.6 $\pm$0.1 & 0.6 $\pm$0.1 & 0.58$\pm$0.08 & 14.34 & 0.3 $\pm$0.2 & 1.2  $\pm$0.2  & 1.3  $\pm$0.2 &  0.79$\pm$0.04 \\      
       &   0.5&              & 0.7 $\pm$0.1 & 0.7 $\pm$0.1 & 0.63$\pm$0.07 &       & 0.3 $\pm$0.1 & 1.4  $\pm$0.2  & 1.5  $\pm$0.2 &  0.81$\pm$0.03 \\      
       &  -0.5&              & 0.6 $\pm$0.1 & 0.6 $\pm$0.1 & 0.55$\pm$0.1  &       & 0.3 $\pm$0.1 & 1.2  $\pm$0.2  & 1.2  $\pm$0.2 &  0.77$\pm$0.04 \\      
  \hline
  3608 &     0& 0.3 $\pm$0.2 & 3.3 $\pm$1.1 & 3.4 $\pm$1.1 & 0.82$\pm$0.18 &  9.75 & 0.7 $\pm$0.4 & 4.3  $\pm$1.2  & 4.4  $\pm$1.2 &  0.85$\pm$0.08 \\      
       &   0.5&              & 3.6 $\pm$1.1 & 3.6 $\pm$1.2 & 0.83$\pm$0.26 &       & 0.7 $\pm$0.4 & 4.6  $\pm$1.3  & 4.7  $\pm$1.3 &  0.86$\pm$0.07 \\      
       &  -0.5&              & 3.2 $\pm$1.0 & 3.3 $\pm$1.1 & 0.82$\pm$0.6  &       & 0.7 $\pm$0.4 & 4.2  $\pm$1.2  & 4.3  $\pm$1.1 &  0.85$\pm$0.06 \\      
  \hline
  4278 &     0& 0.2 $\pm$0.1 & 2.8 $\pm$0.3 & 2.8 $\pm$0.4 & 0.75$\pm$0.06 & 14.87 & 0.5 $\pm$0.2 & 6.5  $\pm$0.6  & 6.5  $\pm$0.6 &  0.88$\pm$0.02 \\      
       &   0.5&              & 2.9 $\pm$0.4 & 3.0 $\pm$0.4 & 0.76$\pm$0.06 &       & 0.5 $\pm$0.2 & 6.9  $\pm$0.6  & 6.9  $\pm$0.6 &  0.89$\pm$0.02 \\      
       &  -0.5&              & 2.7 $\pm$0.3 & 2.7 $\pm$0.3 & 0.74$\pm$0.06 &       & 0.5 $\pm$0.2 & 6.3  $\pm$0.5  & 6.4  $\pm$0.6 &  0.88$\pm$0.03 \\      
  \hline
  4365 &     0& 1.0 $\pm$0.2 & 12.0$\pm$1.3 &12.1 $\pm$1.3 & 0.78$\pm$0.05 & 12.90 & 2.6 $\pm$0.5 & 29.5 $\pm$2.6  & 29.8 $\pm$2.7 &  0.9 $\pm$0.02 \\      
       &   0.5&              & 11.0$\pm$1.2 & 11.1$\pm$1.2 & 0.76$\pm$0.05 &       & 2.6 $\pm$0.5 & 27.0 $\pm$2.4  & 27.3 $\pm$2.5 &  0.89$\pm$0.02 \\      
       &  -0.5&              & 12.5$\pm$1.3 & 12.6$\pm$1.4 & 0.79$\pm$0.05 &       & 2.6 $\pm$0.5 & 30.8 $\pm$2.8  & 31.1 $\pm$2.8 &  0.91$\pm$0.02 \\      
  \hline
  4374 &     0& 8.8 $\pm$2.0 & 13.2$\pm$3.4 & 14.1$\pm$3.3 & 0.82$\pm$0.06 &  9.22 & 16.2$\pm$3.7 & 21.0 $\pm$4.8  & 22.6 $\pm$4.8 &  0.88$\pm$0.04 \\      
       &   0.5&              & 12.2$\pm$3.1 & 13.1$\pm$3.1 & 0.81$\pm$0.07 &       & 16.2$\pm$3.9 & 19.4 $\pm$4.5  & 21.0 $\pm$4.5 &  0.87$\pm$0.04 \\      
       &  -0.5&              & 13.7$\pm$3.5 & 14.6$\pm$3.7 & 0.83$\pm$0.07 &       & 16.2$\pm$3.6 & 21.8 $\pm$5.0  & 23.4 $\pm$5.1 &  0.88$\pm$0.04 \\      
  \hline
  4473 &     0& 0.2 $\pm$0.1 & 1.4 $\pm$0.3 & 1.4 $\pm$0.3 & 0.52$\pm$0.12 & 17.35 & 0.8 $\pm$0.4 & 3.6  $\pm$0.5  & 3.6  $\pm$0.5 &  0.8 $\pm$0.04 \\      
       &   0.5&              & 1.5 $\pm$0.3 & 1.5 $\pm$0.3 & 0.55$\pm$0.12 &       & 0.8 $\pm$0.4 & 3.8  $\pm$0.6  & 3.9  $\pm$0.5 &  0.81$\pm$0.04 \\      
       &  -0.5&              & 1.4 $\pm$0.3 & 1.4 $\pm$0.3 & 0.51$\pm$0.12 &       & 0.8 $\pm$0.4 & 3.5  $\pm$0.5  & 3.5  $\pm$0.5 &  0.79$\pm$0.04 \\      
  \hline
  4486 &     0& 1.6 $\pm$0.2 & 24.0$\pm$1.7 & 24.2$\pm$1.8 & 0.88$\pm$0.01 & 30.52 & 9.8 $\pm$1.1 & 146.0$\pm$8.2  & 147.0$\pm$8.1 &  0.98$\pm$0.0  \\      
       &   0.5&              & 21.6$\pm$1.6 & 21.8$\pm$1.6 & 0.86$\pm$0.01 &       & 9.8 $\pm$1.1 & 131.0$\pm$7.4  & 132.0$\pm$7.2 &  0.97$\pm$0.0  \\      
       &  -0.5&              & 25.2$\pm$1.8 & 25.4$\pm$1.9 & 0.88$\pm$0.01 &       & 9.8 $\pm$1.1 & 153.0$\pm$8.6  & 154.0$\pm$8.6 &  0.98$\pm$0.0  \\      
  \hline
  4494 &     0& 1.2 $\pm$0.1 & 1.4 $\pm$0.2 & 1.5 $\pm$0.2 & 0.39$\pm$0.12 &  8.52 & 2.0 $\pm$0.3 & 1.9  $\pm$0.3  & 2.1  $\pm$0.3 &  0.53$\pm$0.09 \\      
       &   0.5&              & 1.4 $\pm$0.2 & 1.6 $\pm$0.2 & 0.41$\pm$0.12 &       & 2.0 $\pm$0.2 & 2.0  $\pm$0.3  & 2.2  $\pm$0.3 &  0.54$\pm$0.08 \\      
       &  -0.5&              & 1.4 $\pm$0.2 & 1.5 $\pm$0.2 & 0.38$\pm$0.13 &       & 2.0 $\pm$0.2 & 1.9  $\pm$0.3  & 2.1  $\pm$0.3 &  0.52$\pm$0.09 \\      
  \hline
  4526 &     0& 2.9 $\pm$0.6 & 3.1 $\pm$0.8 & 3.4 $\pm$0.6 & 0.56$\pm$0.14 & 12.06 & 7.0 $\pm$1.5 & 6.5  $\pm$1.3  & 7.2  $\pm$1.1 &  0.77$\pm$0.06 \\      
       &   0.5&              & 3.1 $\pm$0.8 & 3.4 $\pm$0.6 & 0.55$\pm$0.14 &       & 7.0 $\pm$1.4 & 6.4  $\pm$1.3  & 7.1  $\pm$1.0 &  0.77$\pm$0.06 \\      
       &  -0.5&              & 3.2 $\pm$0.8 & 3.5 $\pm$0.6 & 0.56$\pm$0.13 &       & 7.0 $\pm$1.4 & 6.6  $\pm$1.4  & 7.3  $\pm$1.1 &  0.77$\pm$0.06 \\ 
\hline
\end{tabular}
\end{table*}

\addtocounter{table}{-1}
\begin{table*}
\footnotesize
{\small \caption{Continued.}}
\begin{tabular}{lcccccccccl}
\hline\hline  
Galaxy & $\beta$ & $M_{\rm rot}(<5{\re})$ & $M_{\rm p}(<5{\re})$ & $M_{\rm tot}(<5{\re})$ & $f_{\rm DM} (<5{\re})$ & $R_{\rm max}$ & $M_{\rm rot}(<R_{\rm max})$ & $M_{\rm p}(<R_{\rm max})$ & $M_{\rm tot}(<R_{\rm max})$& $f_{\rm DM}(<R_{\rm max})$ \\
$\rm [NGC]$ &  & $[10^{10} \Msun]$ & $[10^{11} \Msun]$ & $[10^{11} \Msun]$ &  & [$\re$] & $[10^{10} \Msun]$ & $[10^{11} \Msun]$ & $[10^{11} \Msun]$ & 	\\ 
\hline
  4564 &     0& 2.4 $\pm$0.4 & 0.8 $\pm$0.3 & 1.0 $\pm$0.2 & 0.65$\pm$0.17 &  8.33 & 4.0 $\pm$0.7 & 0.9  $\pm$0.3  & 1.3  $\pm$0.3 &  0.72$\pm$0.1  \\      
       &   0.5&              & 0.9 $\pm$0.3 & 1.1 $\pm$0.3 & 0.68$\pm$0.13 &       & 4.0 $\pm$0.6 & 1.0  $\pm$0.3  & 1.4  $\pm$0.3 &  0.74$\pm$0.08 \\      
       &  -0.5&              & 0.7 $\pm$0.2 & 1.0 $\pm$0.2 & 0.64$\pm$0.16 &       & 4.0 $\pm$0.6 & 0.9  $\pm$0.3  & 1.3  $\pm$0.3 &  0.71$\pm$0.09 \\      
  \hline
  4649 &     0& 3.8 $\pm$0.3 & 10.9$\pm$0.9 & 11.3$\pm$0.9 & 0.72$\pm$0.05 & 24.25 & 18.6$\pm$1.6 & 53.8 $\pm$3.8  & 55.7 $\pm$3.7 &  0.94$\pm$0.01 \\      
       &   0.5&              & 9.8 $\pm$0.8 & 10.2$\pm$0.8 & 0.69$\pm$0.06 &       & 18.6$\pm$1.6 & 48.2 $\pm$3.4  & 50.1 $\pm$3.4 &  0.93$\pm$0.01 \\      
       &  -0.5&              & 11.5$\pm$0.9 & 11.9$\pm$1.0 & 0.74$\pm$0.05 &       & 18.6$\pm$1.6 & 56.6 $\pm$4.0  & 58.5 $\pm$4.1 &  0.94$\pm$0.01 \\      
  \hline
  4697 &     0& 20.3$\pm$3.4 & 7.0 $\pm$2.4 & 9.1 $\pm$2.4 & 0.9 $\pm$0.04 &  4.66 & -- & --  & -- &  -- \\      
       &   0.5&              & 7.2 $\pm$2.4 & 9.3 $\pm$2.4 & 0.9 $\pm$0.05 &       & -- & --  & -- &  -- \\      
       &  -0.5&              & 6.9 $\pm$2.3 & 9.0 $\pm$2.3 & 0.9 $\pm$0.42 &       & -- & --  & -- &  -- \\      
  \hline
  5846 &     0& 0.3 $\pm$0.1 & 12.4$\pm$1.6 & 12.4$\pm$1.7 & 0.83$\pm$0.05 & 13.68 & 0.8 $\pm$0.2 & 32.6 $\pm$3.5  & 32.7 $\pm$3.5 &  0.93$\pm$0.02 \\      
       &   0.5&              & 11.6$\pm$1.5 & 11.6$\pm$1.5 & 0.81$\pm$0.05 &       & 0.8 $\pm$0.2 & 30.6 $\pm$3.3  & 30.7 $\pm$3.2 &  0.92$\pm$0.02 \\      
       &  -0.5&              & 12.8$\pm$1.7 & 12.8$\pm$1.7 & 0.83$\pm$0.04 &       & 0.8 $\pm$0.2 & 33.6 $\pm$3.6  & 33.7 $\pm$3.5 &  0.93$\pm$0.02 \\      
  \hline
  7457 &     0& 1.9 $\pm$0.2 & 0.9 $\pm$0.3 & 1.1 $\pm$0.2 & 0.84$\pm$0.05 &  6.26 & 2.4 $\pm$0.3 & 1.0  $\pm$0.3  & 1.2  $\pm$0.2 &  0.85$\pm$0.05 \\      
       &   0.5&              & 1.1 $\pm$0.3 & 1.3 $\pm$0.3 & 0.86$\pm$0.05 &       & 2.4 $\pm$0.3 & 1.1  $\pm$0.3  & 1.3  $\pm$0.3 &  0.87$\pm$0.05 \\      
       &  -0.5&              & 0.9 $\pm$0.2 & 1.0 $\pm$0.2 & 0.83$\pm$0.06 &       & 2.4 $\pm$0.3 & 0.9  $\pm$0.2  & 1.1  $\pm$0.2 &  0.84$\pm$0.05 \\      
  \hline
  3607 &     0& 0.3 $\pm$0.1 & 2.5 $\pm$0.7 & 2.5 $\pm$0.7 & 0.3 $\pm$0.45 & 20.72 & 1.3 $\pm$0.6 & 10.0 $\pm$2.4  & 10.1 $\pm$2.4 &  0.81$\pm$0.1  \\      
       &   0.5&              & 2.4 $\pm$0.6 & 2.4 $\pm$0.6 & 0.28$\pm$0.6  &       & 1.3 $\pm$0.6 & 9.6  $\pm$2.3  & 9.7  $\pm$2.4 &  0.8 $\pm$0.09 \\      
       &  -0.5&              & 2.5 $\pm$0.7 & 2.5 $\pm$0.7 & 0.31$\pm$0.32 &       & 1.3 $\pm$0.6 & 10.1 $\pm$2.4  & 10.2 $\pm$2.4 &  0.81$\pm$0.07 \\      
  \hline
  5866 &     0& 0.1 $\pm$0.3 & 1.2 $\pm$0.6 & 1.3 $\pm$0.5 & 0.33$\pm$0.45 &  5.75 & -- & --  & -- &  -- \\      
       &   0.5&              & 1.3 $\pm$0.6 & 1.3 $\pm$0.5 & 0.35$\pm$1.01 &       & -- & --  & -- &  -- \\      
       &  -0.5&              & 1.2 $\pm$0.6 & 1.2 $\pm$0.5 & 0.31$\pm$1.93 &       & -- & --  & -- &  -- \\      
\hline
\end{tabular}
\end{table*}

\subsection{Total mass estimates}
\label{subs:tot_mass}
Table \ref{tab:mass_summary} contains a summary of the mass contribution from rotation, $M_{\rm rot}$ and pressure support, $M_{\rm p}$ (corrected for galaxy flattening) for all the galaxies studied in this paper.
$M_{\rm p}$ is calculated with $V_{\rm rot}$ subtracted from $V_{\rm los,i}$ as prescribed in \citet{Evans_2003}.
We account for the effects of galaxy flattening and rotation in our total dynamical mass estimate, $M_{\rm tot}(<R_{\rm out})$ using
\begin{equation}\label{eq:mass_tot}
M_{\rm tot}(<R_{\rm out}) = corr \times M_{\rm p}(<R_{\rm out}) + M_{\rm rot} 
\end{equation}
Again, we note that for galaxies with identified kinematic substructures, the mass estimates in Table \ref{tab:mass_summary} are from the \textit{cleaned} catalogues. For example, in NGC~3115, we account for the mass \textit{over--estimation} due to galaxy flattening by applying a ${\sim}$15 per cent reduction to $M_{\rm p}$, and it is this corrected value that we show in Table \ref{tab:mass_summary}.  

The mass estimates we list in Table \ref{tab:mass_summary} have been obtained assuming $\beta\equiv-0.5, 0, 0.5$. The uncertainty on the total mass varies with the total number of GCs, $N_{\rm GC}$, such that when $N_{\rm GC} \ge 100$ and after accounting for individual $\vlos$ error via Monte Carlo simulations, the typical uncertainty is ${\sim}0.12$ dex. For galaxies with $N_{\rm GC} \sim 70$ and $\le 40$, typical uncertainties on total mass are ${\sim}0.20$ and ${\sim}0.25$ dex, respectively. 

\begin{figure}
    \includegraphics[width=0.48\textwidth]{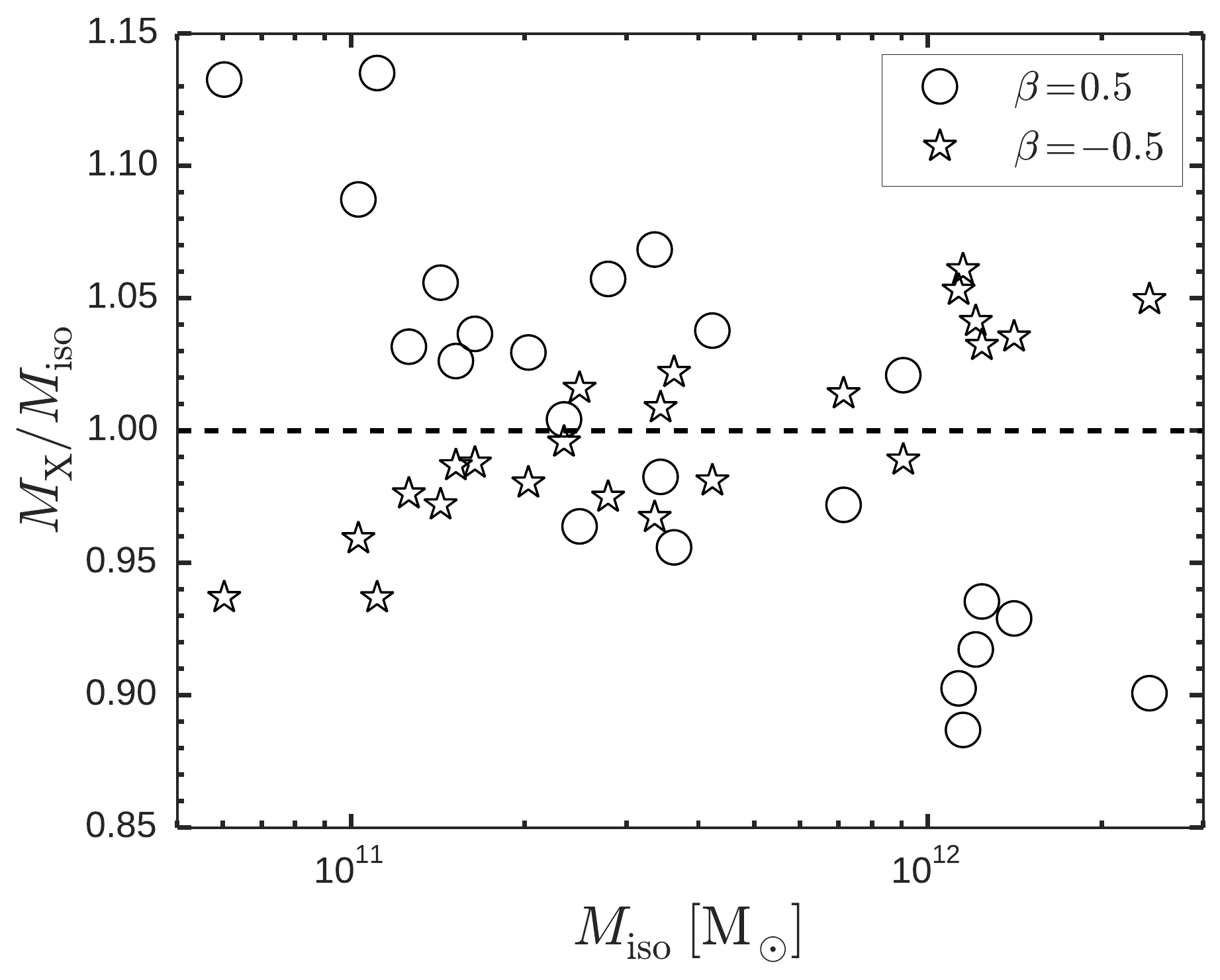}
	\caption{\label{fig:dev_ani} Deviation of mass estimate for radial or tangential anisotropy compared to the isotropic mass estimate. The circles are the mass estimates when orbital anisotropy is radially biased ($\beta=0.5$), while the stars are mass estimates when tangential anisotropy ($\beta=-0.5$) is assumed.The total mass estimates within 5~$\re$ in most of the galaxies are insensitive to anisotropy, with the average variation of the mass estimates being $\leq5$ per cent.} 
\end{figure}

We show how the total mass estimates, assuming mildly tangential and strong radial anisotropies, deviate from that obtained under isotropy condition in Figure \ref{fig:dev_ani}. Mass estimates are largely insensitive to our choice of $\beta$: only NGC~3377, NGC~7457 and NGC~1407 show deviations larger than 10 per cent, in agreement with the findings of \citet{Bacon_1985}. In what follows, we adopt the mass estimates obtained under isotropy conditions, bearing in mind the potential deviations for each galaxy. 

\subsection{Comparison of mass estimates with results from the literature}
\label{subs:cmp_lit}
\begin{table}
\centering
{\small \caption{Mass estimates from the literature obtained using different mass tracers and modelling techniques, and comparison with results from this work, obtained assuming $\beta=0$ and allowing the slope of the gravitational potential to vary with galaxy stellar mass. $M_{\rm lit.}$ and $M_{\rm TME}$ are the total masses from the literature and this work, respectively, within projected radial distance $R$.} \label{tab:comparison} }
\begin{tabular}{@{}r c l l c}
\hline
\hline
Galaxy & $R$	& $M_{\rm lit.}$ & $M_{\rm TME}$	 & Tracer\\
$\rm [NGC]$	   & [kpc] & [$10^{11} \Msun$] &	[$10^{11} \Msun$]	 &\\
\hline
 720 &	  20 &	 5.1$\pm$ 0.4 &	2.8$\pm$0.6	& X--ray$^j$\\
 821 &	  22 &	 2.3$\pm$ 0.6 &	 4.3$\pm$0.1	    & PNe$^a$\\
1023 &	  10 &	 1.7$\pm$ 0.6 &	 1.6$\pm$0.3	& PNe$^m$\\
1407 &	  68 &	 30.6$\pm$ 3.9 &	 20.9$\pm$1.7	& GC$^b$\\
	 &	  29 &	 9.4$\pm$ 1.3 &	 8.2$\pm$0.9	& GC$^a$\\
	 &	  25 &	 21.6$\pm$ 6.9 &	 7.1$\pm$0.7  & X--ray$^d$\\
	 &	 100 &	 100.0			&	 30.5$\pm$2.2	& X--ray$^k$\\
2768 &	  14 &	 3.2$\pm$ 1.5 &	 3.7$\pm$0.6	& PNe$^m$\\
3115 &	   7 &	 1.1$\pm$ 0.5 &	 2.1$\pm$0.3	& PNe$^m$\\
3377 &	  10 &	 0.7$\pm$ 0.2 &	 0.6$\pm$0.1 	& PNe$^a$\\
4365 &	  15 &	 3.9$\pm$ 0.6 & 6.2$\pm$0.6	& X--ray$^c$\\
4374 &	  32 &	 11.5$\pm$ 1.2 &	 17.7$\pm$3.9	& PNe$^g$\\
	 &	  30 &	 15.9$\pm$ 1.9 &	 17.1$\pm$3.8	& PNe$^a$\\
	 &	  29 &	 19.2$\pm$ 1.8 &	 16.5$\pm$3.7	& PNe$^l$\\
4486 &	  46 &	 33.3$\pm$ 3.3 &	 35.4$\pm$2.3	& GC$^a$\\
	 &	 135 &	 85.2$\pm$ 10.1 &	 97.2$\pm$5.4	& GC$^f$\\
	 &	 180 &	 149.6$\pm$20.0 &	 130.6$\pm$7.2	& GC$^l$\\
	 &   180 &   192.0$\pm$ 66.0 &    130.6$\pm$7.2 & GC$^r$\\     
	 &	  47 &	 57.0$\pm$ 11.0 &	 36.3$\pm$2.3	& GC$^o$\\
	 &	 120 &	 125.0$\pm$ 7.0  &	 86.7$\pm$4.9  & X--ray$^d$\\
4494 &	  20 &	 1.6$\pm$ 0.3 &	 1.4$\pm$0.2		& PNe$^n$\\
	 &	  20 &	 1.2$\pm$ 0.2 &	 1.4$\pm$0.2		& PNe$^a$\\
	 &	  19 &	 2.1$\pm$ 0.1 &	 1.3$\pm$0.2		& PNe$^h$\\
4564 &	   7 &	 0.4$\pm$ 0.1 &   1.0$\pm$0.2	& PNe$^a$\\
4649 &	  46 &	 8.7$\pm$ 1.3 &	 19.1$\pm$1.3	& PNe$^a$\\
	 &	  25 &	 16.3$\pm$ 4.3 &	 10.5$\pm$0.8  & X--ray$^d$\\
	 &    45 &   34 		    &	 18.6$\pm$1.3  & GC$^p$\\
	 &    45 &   22 		    &	 18.6$\pm$1.3  & GC$^s$\\
4697 &	  17 &	 1.4$\pm$ 0.2 &	  8.9$\pm$2.2	& PNe$^a$\\
	 &	  15 &	 1.9$\pm$ 0.3 &	  8.3$\pm$2.2	& PNe$^i$\\
5846 &	  56 &	 17.0$\pm$ 3.0 &	 19.8$\pm$2.1	& PNe,GC$^q$\\
     &	  45 &	 16.0$\pm$ 3.3 &	 16.2$\pm$1.8	& GC$^g$\\
	 &	  45 &	 11.2$\pm$ 2.7 &	 16.2$\pm$1.8	& PNe$^a$\\
	 &	  25 &	 12.5$\pm$ 1.9 &	 9.9$\pm$1.5  & X--ray$^d$\\
7457 &	   5 &	 0.2$\pm$ 0.1 &	 0.6$\pm$0.1	& PNe$^m$\\
\hline
3607 &	  20 &	 3.3$\pm$ 0.7 &	 2.0$\pm$0.5	& X--ray$^c$\\
\hline
\end{tabular}
\begin{flushleft}
{\small 
References : \textit{a}. \citet{Deason_2012}, \textit{b}. \citet{Pota_2015}, \textit{c}. \citet{Nagino_2009, Trujillo_2011}, \textit{d}. \citet{Das_2010}, \textit{e}. \citet{Napolitano_2011}, \textit{f}. \citet{Agnello_2014}, \textit{g}. \citet{Napolitano_2014}, \textit{h}. \citet{Morganti_2013}, \textit{i}. \citet{DeLorenzi_2008},  \textit{j}. \citet{Humphrey_2010, Trujillo_2011}, \textit{k}. \citet{Su_2014}, \textit{l}. \citet{Zhu_2014}, \textit{m}. \citet{Cortesi_2013}, \textit{n}. \citet{Napolitano_2009}, \textit{o}. \citet{Murphy_2011}, \textit{p}. \citet{Shen_2010}, \textit{q}. Zhu et al. (in prep.), \textit{r}. \citet{Oldham_2016}, \textit{s}. \citet{Das_2011}}\end{flushleft}
\end{table}
\begin{figure*}
    		\includegraphics[width=0.48\textwidth]{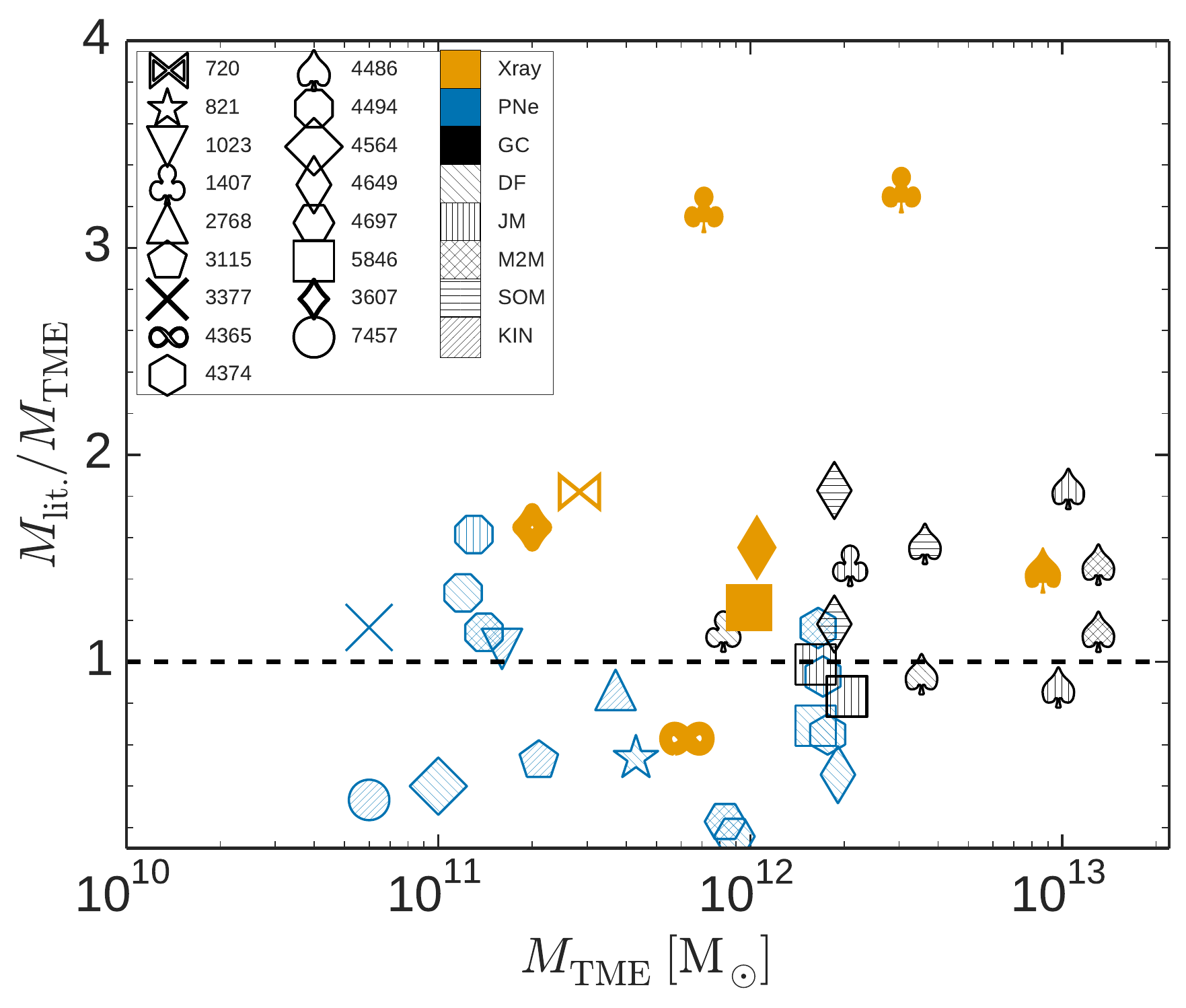}\hspace{0.01\textwidth}%
    		\includegraphics[width=0.51\textwidth]{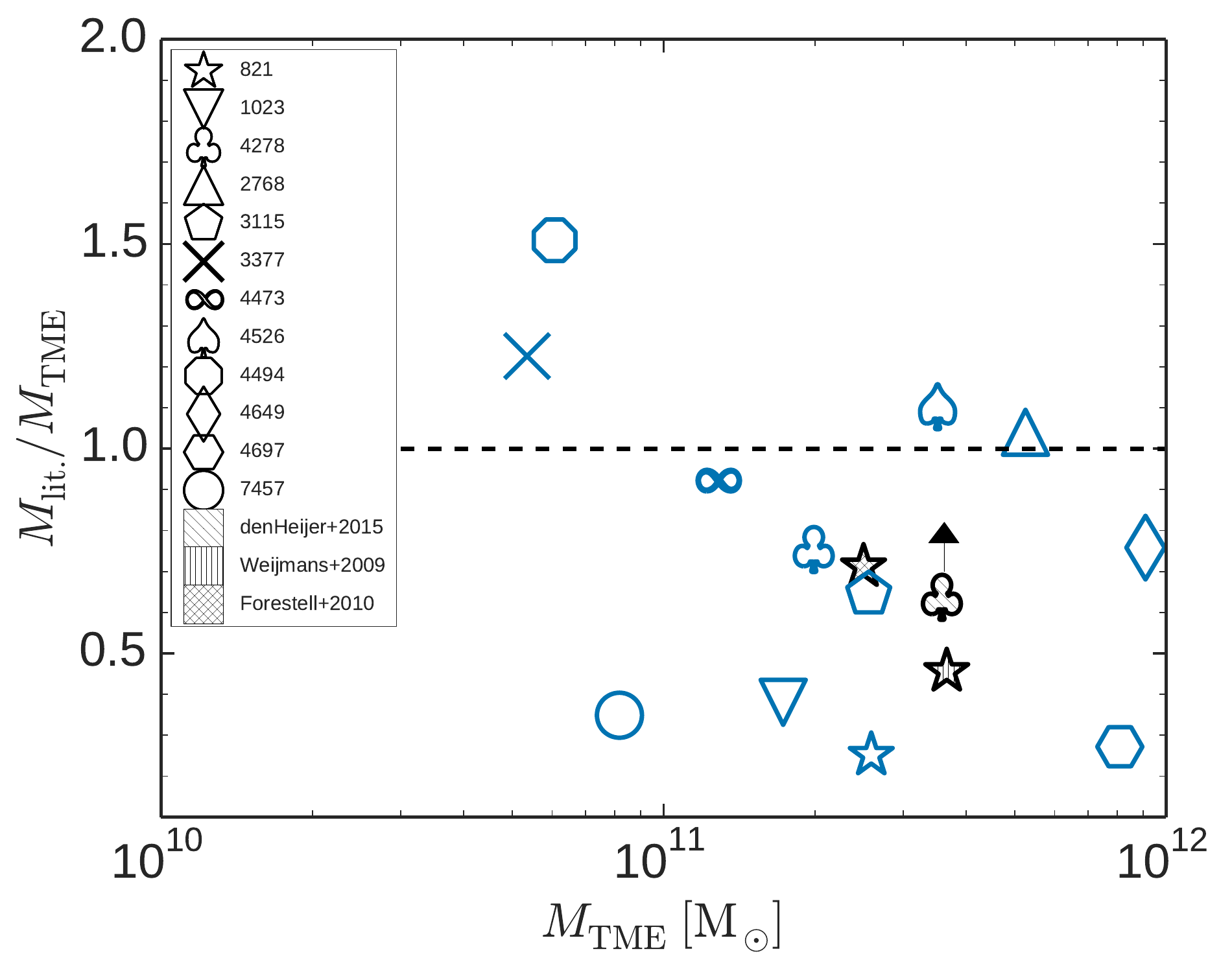}\\
	\caption{\label{fig:cmp_tracers} Comparison of mass estimates for galaxies in our sample with results from the literature obtained using different mass tracers and modelling techniques. \textit{Left panel} Galaxies are identified as shown in the plot legend, with brown, blue and black symbols highlighting the mass tracers used. The hatch marks differentiate galaxies according to the modelling technique employed. DF indicates the phase--space distribution function technique used in \citet{Deason_2012}, JM is the traditional Jeans mass modelling technique, e.g. \citet{Pota_2015}, M2M is the made--to--measure mass model, e.g. \citet{DeLorenzi_2008}, SOM is the Schwarzschild orbit--based modelling technique used in \citet{Murphy_2011} while KIN is the asymmetric drift method employed by \citet{Cortesi_2013} to extract circular velocities from PNe kinematics, respectively. \textit{Right panel} Comparison of mass estimates with results from the literature based on  extended stellar kinematics. Most of the datapoints are from the extended stellar kinematics study of \citet{Cappellari_2015} which we supplement with results from \citet{Weijmans_2009} and \citet{Forestell_2010} for NGC~821. We also include the mass estimate for NGC~4278 from the cold--gas study of \citet{denHeijer_2015}, assuming their result was measured at 15 kpc (the arrow shows how the mass estimates compare towards 28 kpc). Combining all the data (i.e., PNe, GCs and stars, without the X--ray data) and assuming comparable errors, we observe a 1~$\sigma$ scatter of 0.2 dex between our mass estimates and literature values.} 
\end{figure*}
In Table \ref{tab:comparison}, we compare mass estimates for our galaxies to the literature. We show the comparison in Figure \ref{fig:cmp_tracers}. The literature sample include studies with PNe, GCs and X--rays as the mass tracers and a variety of generally more sophisticated mass modelling techniques. For example, 9 galaxies from our sample were studied homogeneously by \hyperlink{D+12}{D+12} using PNe and/or GC kinematic data out to 5~$\re$. They did not account for galaxy flattening, rotation of the tracers and kinematic substructures in their mass estimates, even though they modelled the velocity anisotropy. The comparisons are done at the same galactocentric radii (not always at 5~$\re$) as reported in the literature.
We have excluded mass estimates from the X--ray study of \citet{Churazov_2010} from our comparison since those results are simple power--law fits to their data which under--estimates the total mass. 

The most deviant results are for NGC~1407 from the X--ray studies of \citet{Das_2010} and \citet{Su_2014} which both suggest a greater total mass (by a factor of ${\sim}3$) compared to what is obtained from GCs. This could be due to their assumption of hydrostatic equilibrium for the Eridanus A group, which may be wrong as the ripples in the X--ray maps \citep{Su_2014} seem to suggest. Mass estimates obtained by studying the phase--space distribution function of tracers (\hyperlink{D+12}{D+12}) tend to be systematically lower than those from other methods. This is most likely related to the modelling assumptions made in the study e.g., they restricted $\alpha > 0$. For mass estimates obtained from GC kinematics, the two galaxies with the greatest offsets from our results are NGC~4486 \citep{Murphy_2011} and NGC~4649 \citep{Shen_2010}. The mass overestimation for NGC~4486 has been attributed in the literature to the problematic data used in the study (see \citealt{Strader_2011, Zhu_2014}). For NGC~4649, there is a wide spread in the mass estimates obtained from GC \citep{Shen_2010}, X--ray \citep{Das_2010} and PN (\hyperlink{D+12}{D+12}) data. The GC data used in \citet{Shen_2010} comes in part from the catalogue of \citet{Lee_2008}, in which \citet{Pota_2015} identified some extreme velocity objects. This, combined with the kinematic substructures we have identified in this galaxy, could be the source of the differences in the mass estimates for NGC~4649. Another interesting case is NGC~4494, where results using essentially the same dataset but different methods give mass estimates that vary by a factor of ${\sim}$2.

Lastly, we compare our mass estimates to results from the extended stellar kinematics in the right panel of Figure \ref{fig:cmp_tracers}. Most of the results are from \hypertarget{C+15}{\citet{Cappellari_2015}}, hereafter \hyperlink{C+15}{C+15}, where we obtain the total mass by integrating their total mass density profiles. We have also added results from the extended stellar kinematics of \citet{Weijmans_2009} and \citet{Forestell_2010} for NGC~821 and the cold--gas study of \citet{denHeijer_2015} for NGC~4278. The agreement between our mass estimates and literature mass measurements from stellar kinematics is similar to that in the \textit{right panel} of Figure \ref{fig:cmp_tracers}, though with some individual discrepancies. For example, we find the largest deviation in NGC~821, where our mass estimate differs from that of \hyperlink{C+15}{C+15} by a factor of 4, while being more consistent with the results from \citet{Weijmans_2009} and \citet{Forestell_2010}. Also, the mass estimate for NGC~4494 from \hyperlink{C+15}{C+15} is a factor of ${\sim}2$ higher than what we have found. 

From Figure \ref{fig:cmp_tracers}, mass estimates from PNe appear to be systematically lower compared to those from GCs and X--ray data, especially for the more massive galaxies. Our masses also appear to be systematically lower than literature values obtained using GCs. If we assume that all the mass measurements (stars, GCs, PN and X--rays) have comparable errors, then the observed 1-$\sigma$ scatter about the one--to--one relation between the literature values and our mass estimates is 0.3 dex. If we exclude the X--ray data, the scatter is reduced to 0.2 dex. These rms scatters are however upper limits since we only consider total mass estimates obtained assuming isotropy, $\alpha$ varying with stellar mass and stellar $M/L_K=1$ for the comparison. On a galaxy by galaxy basis, the scatter can be reduced significantly by considering specific combinations of these parameters. Our mass estimates therefore compare well with results from more sophisticated modelling techniques, and from different mass tracers over a wide radial range that extends out to 180~kpc. 
\subsection{Dark matter fraction}
\label{subs:DMfrac}
The dark matter fraction is a useful parameter in understanding the mass distribution as a function of radius in galaxies. We define the DM fraction, $f_{\rm DM}$, as: 
\begin{equation}\label{eq:DMfrac}
	f_{\rm DM}(<R) = 1 - M_{*}(<R)/M_{\rm tot}(<R),
\end{equation}
where $M_{*}(<R)$ and $M_{\rm tot}(<R)$ are the enclosed stellar and total dynamical mass, respectively,  within the projected radial distance $R$. Equation \ref{eq:DMfrac} assumes that gas and dust do not contribute significantly to the baryonic mass. The total stellar mass enclosed within $R$ is described by the projected \Sersic/ mass profile \citep{Sersic_1968, Terzic_2005} and it depends on the \Sersic/ index, $n$. We use the $\re$--$n$ relation from \citet{Graham_2013} to obtain unique \Sersic/ indices for our galaxies and summarize $f_{\rm DM}$ in Table \ref{tab:mass_summary}. Similar results are obtained if a luminosity--concentration relation or a de Vaucouleurs' profile ($n=4$) is assumed for our galaxy sample.

\begin{figure*}
    \includegraphics[width=1.0\textwidth]{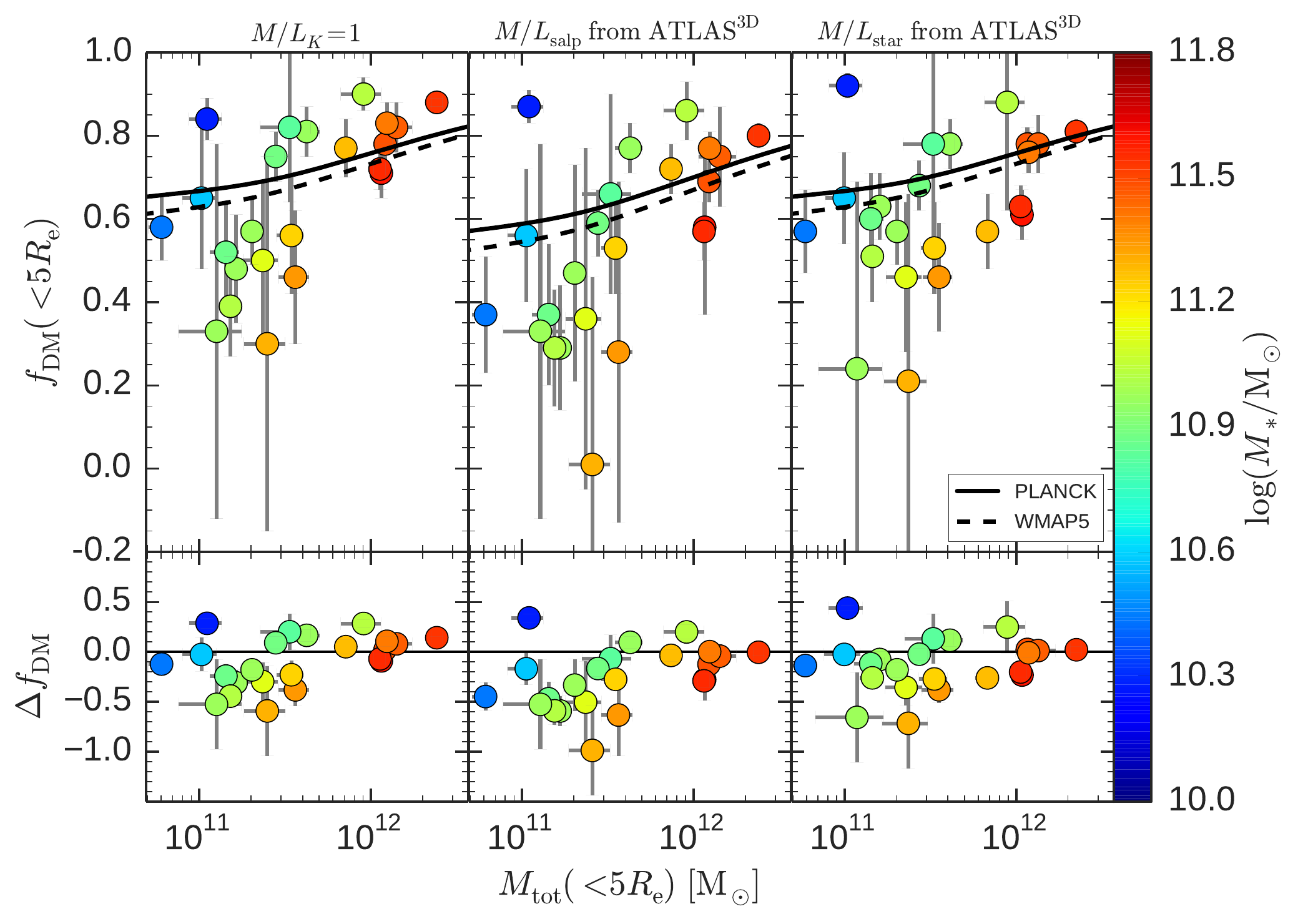}\hspace{0.01\textwidth}\\
	\caption{\label{fig:predict_fdm1} Measured dark matter fraction, $f_{\rm DM}$, versus the total mass, $M_{\rm tot}$, within 5 effective radii ($\re$). \textit{Top panels}: In all the top panels, the solid lines show the predicted dark matter fractions within 5~$\re$ assuming \textit{Planck} cosmology. The dashed lines show the same but assuming \textit{WMAP5} cosmology (see text for details). The marker colour shows the stellar mass of the galaxies. Minimum $f_{\rm DM}$ is observed at stellar mass ${\sim}10^{11}\Msun$. In the left panel, we assume a stellar $M/L_K=1$ while in the middle and right panels, we use stellar mass--to--light ratios from the ATLAS$^{\rm 3D}$ survey based on a Salpeter~IMF and best--fit stellar mass--to--light ratios from dynamical modelling (total dynamical mass minus dark matter mass), respectively \citep{Cappellari_2013, Cappellari_2013b}. Regardless of the adopted stellar--mass--to light ratios, galaxies in the intermediate stellar mass bin have $f_{\rm DM}$ significantly different from what is predicted. Also, low-- and high--stellar mass galaxies have higher measured dark matter fractions than intermediate stellar mass galaxies. \textit{Bottom panels}: These panels show residuals between predictions (with Planck cosmology) and observations, calculated as (observed-predicted)/predicted.}
\end{figure*}

\subsection{Total mass and dark matter fraction beyond 5~$\re$}
\label{max_mass}
We extend our mass estimation method to GC kinematic data beyond 5~$\re$ and obtain the total mass and dark matter fraction enclosed within the maximum probed radius ($R_{\max}$). We summarise our results in Table \ref{tab:mass_summary}, where we have assumed stellar $M/L_K=1$. NGC~4697 and NGC~5866 have been excluded from this analysis due to the limited radial extent of their GC kinematic data.

\begin{figure}
        \includegraphics[width=0.48\textwidth]{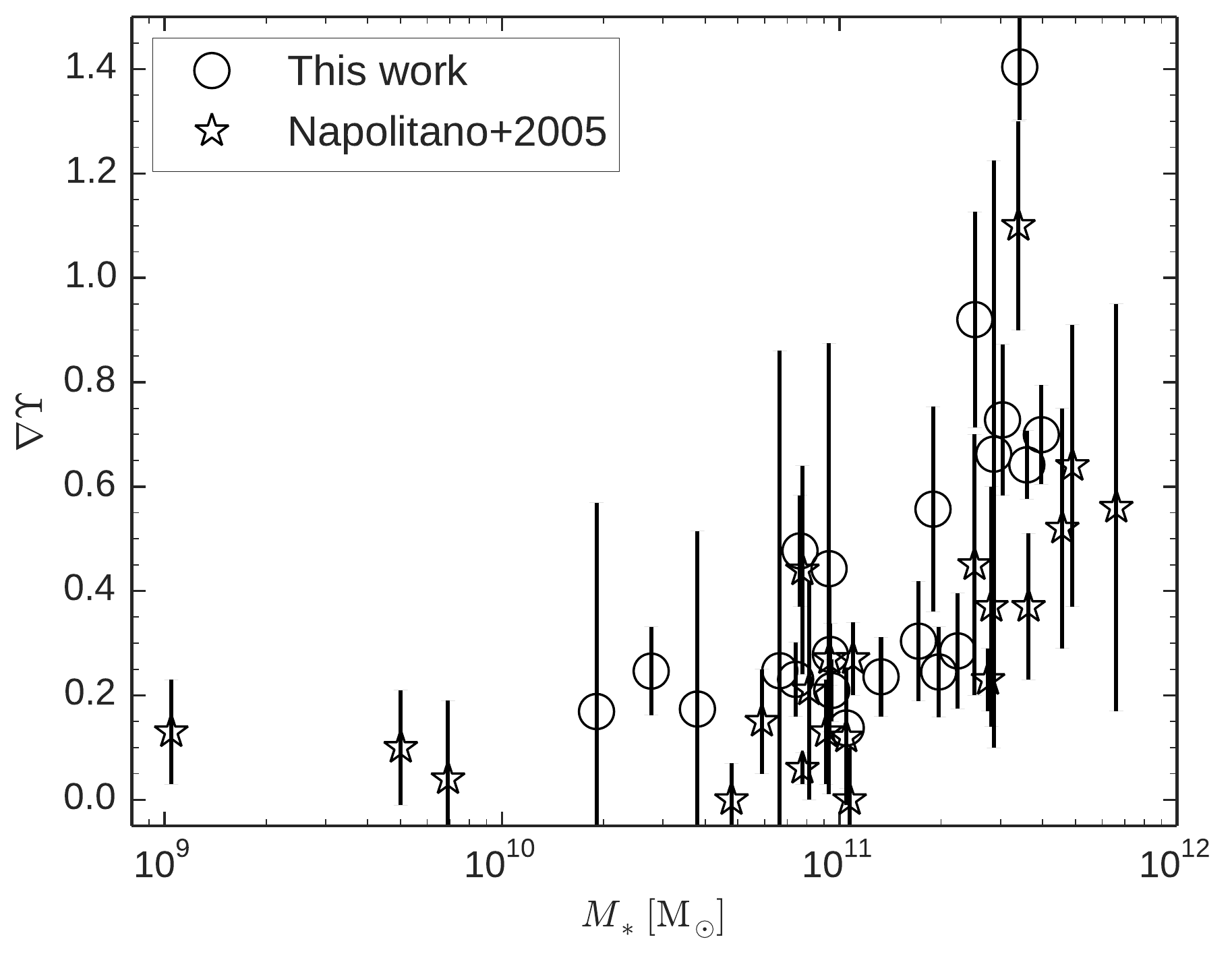}
	\caption{\label{fig:cmp_grad_lit} Mass--to--light gradient (between 5~$\re$ and the maximum radial limit) versus total stellar mass. The circles are the results from this work, while the stars are from \citet{Napolitano_2005}. Note that \citet{Napolitano_2005} obtained their gradients over the range 0.1--4~$\re$, while the gradients in this work have been obtained between 5~$\re$ and larger radii. The systematic offset between the trend in our data and that of \citet{Napolitano_2005} is because we probe radial regions that are more dark matter dominated (see their figure 3). The low and intermediate stellar mass galaxies have shallow total mass gradients and the more massive galaxies show much steeper gradients. NGC~4697 and NGC~5866 are not shown in this plot due to the limited radial extent of their GC kinematic data.} 
\end{figure}

To properly understand how the total mass changes with galactocentric radius, we use the method of \citet{Napolitano_2005} to obtain the mass--to--light gradient between 5~$\re$ and the maximum radius. We use 
\begin{equation}
\nabla\Upsilon \equiv \dfrac{\re}{\Delta R}\left[\left(\dfrac{M_{\rm DM}}{M_*}\right)_{\rm out} - \left(\dfrac{M_{\rm DM}}{M_*}\right)_{\rm in}\right]
\end{equation}
where $\nabla\Upsilon$ is the mass--to--light gradient, $M_{\rm DM}$ and $M_*$ are the enclosed DM and stellar mass, respectively.
Figure \ref{fig:cmp_grad_lit} shows $\nabla\Upsilon$ versus the total stellar mass of our galaxies. For comparison, we have added datapoints from \citet{Napolitano_2005}, where a similar analysis was done using data extending out to ${\sim}4$~$\re$ (they compiled results from the literature from dynamical studies based on discrete tracers and extended integrated stellar light). The systematic offset between the trend in our data and that of \citet{Napolitano_2005} is because we probe radial regions that are more dark matter dominated (see their figure 3). We note that similar results are obtained when $\alpha=0$ or  an outer radius beyond 5~$\re$ is used.

The gradient is shallow for galaxies with stellar mass below ${\sim}10^{11.2}\Msun$, however beyond this transition stellar mass, a sharp upturn in the gradient is observed, with the more luminous galaxies showing a wide variety of gradients. This dichotomy is the direct effect of the difference in the relative radial distribution of stellar mass and DM in ETGs. The transition stellar mass coincides with the upturn in the galaxy $M_* - \re$ relation, such that in the lower mass galaxies, where $\re$ varies slowly with $M_*$, the scale radius of the DM halo also varies slowly with $M_*$, hence the flat gradients. For the more massive galaxies, as $\re$ increases rapidly with $M_*$, we are able to probe more DM.

\section{Discussion}
\label{discu}
In the previous section, we homogeneously obtained total mass estimates and dark matter fraction within 5~$\re$ and beyond, and showed that our mass estimates are consistent with previous studies in the literature, with an observed rms scatter (upper limit) of 0.2 dex. We also used the dark matter fraction, $f_{\rm DM}$, to describe the relative radial distribution of the stellar and dark matter in our sample.

\subsection{Dark matter fractions and galaxy models}
\label{models}
To properly understand these results within the $\Lambda$CDM framework, we compare the $f_{\rm DM}$ within 5~$\re$ with predictions from a simple galaxy model where we assume that the DM content follows a NFW profile \citep{Navarro_1996}, with the stellar content described by a \citet{Sersic_1968} mass profile. Starting with the galaxy stellar mass, we use the $\re$--$M_*$ relation from \citet{Lange_2015} to obtain model galaxy sizes over our stellar mass range. Next, the non--linear $M_*$--halo mass relation for ETGs from \citet{Dutton_2010} gives the galaxy halo mass, $M_{\rm 200}$, for a given total stellar mass. The halo is then completely parametrised by obtaining the halo concentration parameter, $c_{\rm 200}$, using the $M_{\rm 200}$--$c_{\rm 200}$ relation from \citet{Dutton_2014} based on the \textit{Planck} cosmology. We note that at a fixed halo mass, \textit{Planck} cosmology yields higher halo concentration than the \textit{WMAP5} cosmology, but only slightly alters the $f_{\rm DM}$. We then obtain the scale radius, $r_{\rm s}$, of the galaxy halo using 
$M_{\rm 200}=4\pi\Delta_{\rm vir}\rho_{\rm cri}{r_{\rm 200}}^3/3$ and $r_s \equiv r_{\rm 200}/c_{\rm 200}$. Armed with the $r_{\rm s}$, $c_{\rm 200}$ for any given $\re$--$M_*$ pair plus a universal baryon fraction of 0.17 \citep{Spergel_2007}, we then produce the cumulative NFW DM--only radial profiles out to large radii. Likewise, for each $\re$--$M_*$ pair, we use the $\re$--$n$ relation from \citet{Graham_2013} and describe the cumulative stellar mass radial profile as defined in \citet{Terzic_2005}.

Our total stellar masses have been obtained assuming a global stellar mass--to--light ratio of $M/L_{\rm K}=1$. This assumption does not reflect differences in the stellar population parameters (e.g. age, metallicity, stellar initial mass function) of ETGs, especially in their central regions. However, we note that our SLUGGS galaxies are generally dominated by very old (8--14 Gyrs) stellar populations and have a small range in mean metallicity \citep{Mcdermid_2015}. The $M/L_{\rm K}$ is largely insensitive to metallicity variations \citep{Forbes_2008, Conroy_2012}. For example, figure 10 from \citet{Forbes_2008} shows that the stellar $M/L_{\rm K}$ can vary by ${\sim}0.15$ dex within the metallicity range of our sample ($-0.2\leq \rm{[Fe/H]} \leq 0.1$), which is comparable to the uncertainties on our stellar mass estimates.
A similar uncertainty is associated with the observed age variation, i.e. 8--14 Gyrs, of our sample. 

To test how adopting a stellar $M/L_{\rm K}=1$ (corresponding to a Kroupa IMF \citealt{Kroupa_1993}) may affect our $f_{\rm DM}$, we also obtain $f_{\rm DM}$ using stellar masses from the ATLAS$^{\rm 3D}$ survey. We first use their $(M/L_r)_{\rm Salp}$, obtained from stellar population synthesis models which assumed a Salpeter IMF \citep[table 1, column 5 in][]{Cappellari_2013b} and the galaxy luminosity in the SDSS $r$--band \citep[table 1, column 15 in][]{Cappellari_2013} to estimate individual stellar masses for the galaxies we have in common. For the four galaxies in our sample that are not in the ATLAS$^{\rm 3D}$ survey, we use the best--fit function to the $K$--band magnitude and stellar mass data of their 260 galaxies to infer the stellar masses. We also use their best--fit $(M/L_r)_{\rm stars}$ \citep[table 1, column 4 in][]{Cappellari_2013b}, obtained from dynamical modelling as total mass minus DM mass, to obtain the stellar mass. This method avoids the potential issue of a non--universal stellar $M/L_{\rm K}$ for our sample when deriving the stellar masses, since recent results suggest that stellar $M/L$ systematically varies with galaxy mass \citep[e.g.][]{Cappellari_2012, Conroy_2012, Spiniello_2014, Pastorello_2014}. On average, these stellar masses are consistent with those listed in Table \ref{tab:summary} within ${\sim}0.3$ dex.

In Figure \ref{fig:predict_fdm1}, we compare the \textit{predicted} $f_{\rm DM}$ for our galaxy sample with the \textit{measured} $f_{\rm DM}$ within 5~$\re$. The average $f_{\rm DM}$ for our sample is $0.6\pm0.2$, varying from $0.3$ in NGC~3607 to $0.9$ in NGC~4486. $f_{\rm DM}$ is predicted to increase with galaxy stellar mass while low and high stellar mass galaxies are seen to be DM dominated within 5~$\re$, consistent with the model predictions. However, our $f_{\rm DM}$ measurements reveal discrepancies between predicted and measured $f_{\rm DM}$ for galaxies in the intermediate stellar mass bin ($\sim10^{11}\Msun$). To ascertain if this trend is driven by our stellar $M/L_{\rm K}=1$ assumption, we repeat the entire analyses, adopting the stellar masses obtained earlier with alternative $M/L_{\rm K}$ assumptions. This is an important exercise, bearing in mind the uncertain contribution from the stellar mass to the total mass estimate. \textit{The trend in the measured $f_{\rm DM}$ within 5~$\re$ persists for a variety of stellar $M/L$ assumptions}. We note again for clarity that while the results we show in Figure \ref{fig:predict_fdm1} were obtained under the additional assumption of isotropy, the trends are the same regardless of orbital anisotropies. For some of our galaxies, especially in the intermediate stellar mass bin, the Salpeter IMF (\citealt{Salpeter_1955}) gives stellar mass greater than the total dynamical mass estimate.
The tension between predictions and measurements is however reduced when a Kroupa IMF (\citealt{Kroupa_1993}) is assumed. This is not surprising as a Kroupa IMF implies ${\sim}$40 per cent less stellar mass compared to a Salpeter IMF. The low-- and high--stellar mass galaxies are however consistent with both Salpeter and Kroupa IMF. 

We have also checked to see if the corrections we applied for galaxy flattening and inclinations alter our results. We performed our entire analysis assuming that all our galaxies are spherical and observed edge--on i.e. $q=1$ and $i=90$ degrees. This implies that in Table \ref{tab:mass_summary} and in Equations \ref{eq:radius} and \ref{eq:mass_tot}, $q=1$ and $corr=1$, respectively. A 2--sided KS test of $M_{\rm tot}$ and $f_{\rm DM}$ thus obtained with our earlier results shows that they are identical, i.e. one cannot rule out that they are drawn from the same distribution.

\begin{figure}
        \includegraphics[width=0.48\textwidth]{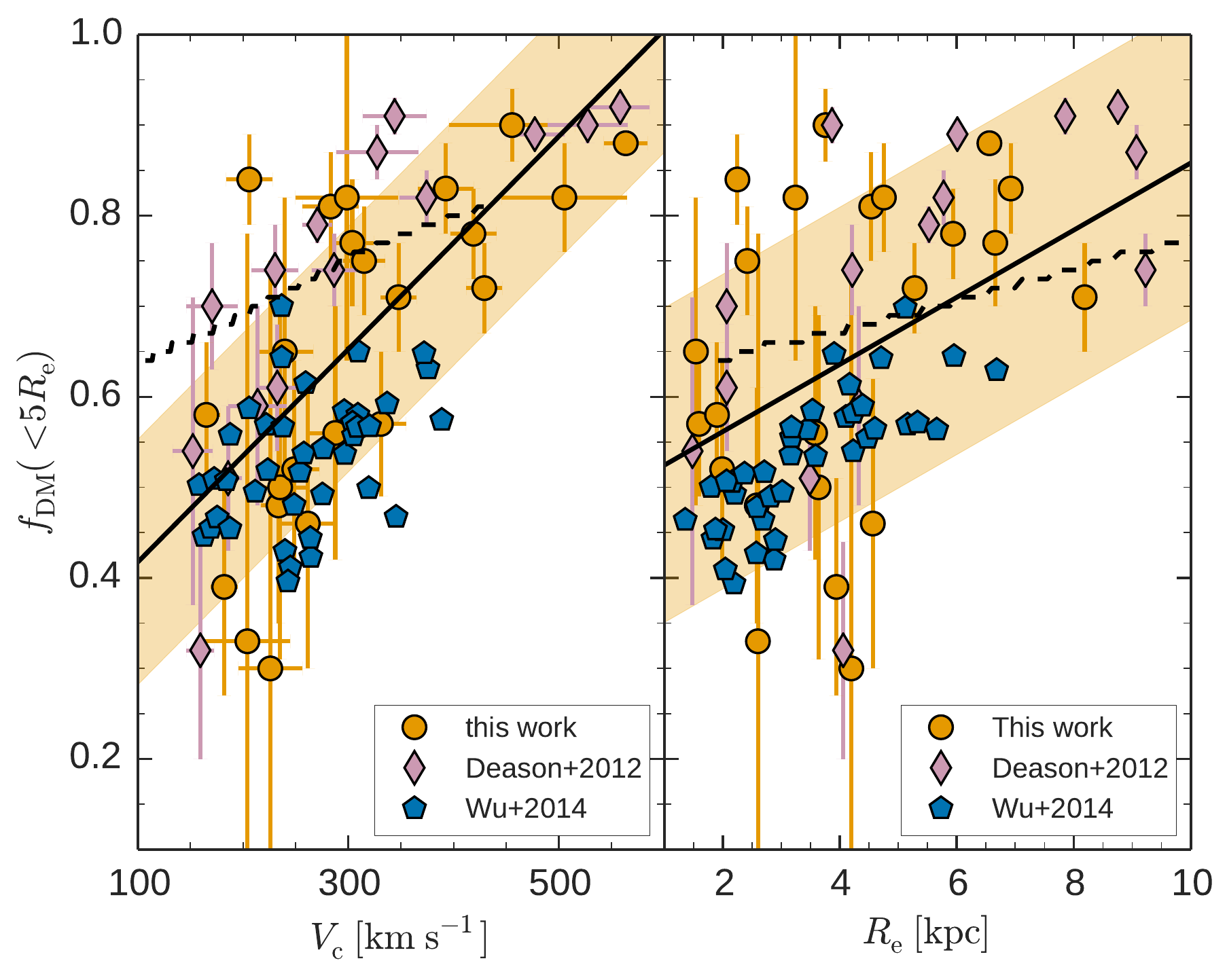}
	\caption{\label{fig:cmp_Wu} Dark matter fraction within 5~$\re$ versus circular velocity 
	(at 5~$\re$) and galaxy size. \textit{Left panel}: Dark matter fraction within 5~$\re$ versus 
	circular velocity. Data from our sample, \citet{Deason_2012} and from the simulations of 
	\citet{Wu_2014} are shown as indicated in the plot legend. The best linear fit and intrinsic 
	1~$\sigma$ scatter to our data are shown by the line and the shaded band, respectively. 
	\textit{Right panel}: Same as in \textit{left panel}, but now showing dark matter fraction 
	versus galaxy size. Our study, as well as that of \citet{Deason_2012} finds a wider range of 
	dark matter fraction than in the simulations of \citet{Wu_2014}. The simulations yield dark 
	matter fractions more consistent with the low measurements we have for some of our intermediate 
	mass galaxies. The dashed lines in both panels are from the simple galaxy model with a pristine 
	NFW DM density distribution as discussed in the text.}
\end{figure}

The galaxy model above is simplistic and does not explicitly account for processes which may alter the distribution of DM during galaxy assembly. In Figure \ref{fig:cmp_Wu}, we therefore compare the observed $f_{\rm DM}$ within 5~$\re$ with results from the simulation of \hyperlink{W+14}{W+14}, where both the observed and simulated galaxies covered a comparable stellar mass range. In their simulations, they allowed the DM density distribution to be modified during galaxy assembly via processes like adiabatic halo contraction and halo expansion, such that the inner DM density is different from the NFW DM density we adopted in our simple galaxy model. The \hyperlink{W+14}{W+14} simulations however did not account for AGN and$/$or Supernovae feedback processes, therefore, their haloes host galaxies with efficient star formation histories and stellar masses a factor of 2--3 above the expectations from a typical galaxy $M_*$--halo mass relation. At any given halo mass, their simulations yield significantly lower 
$f_{\rm DM}$ than our vanilla model predicts, but in better agreement with our measurements for the intermediate mass galaxies with lowered $f_{\rm DM}$. While it is obvious that processes which maximize the stellar mass would result in lower $f_{\rm DM}$, it is however not clear from the simulation if the low 
$f_{\rm DM}$ is exclusively driven by the baryon--DM interaction or by the feedback processes. 

 \begin{figure*}
    \includegraphics[width=1.0\textwidth]{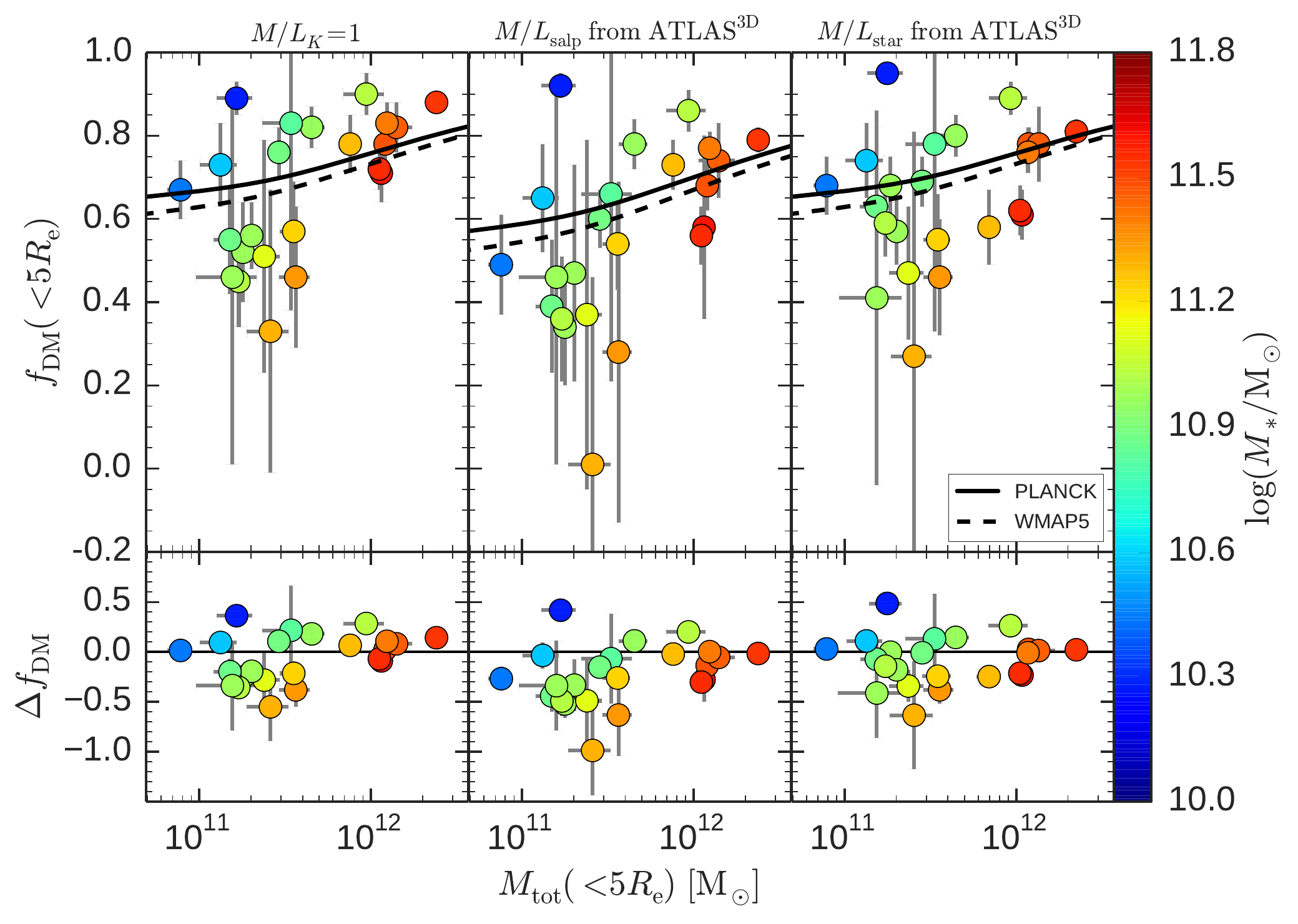}\hspace{0.01\textwidth}\\
	\caption{\label{fig:predict_fdm2} Same as in Figure \ref{fig:predict_fdm1}, but assuming a logarithmic potential for our sample, i.e. $\alpha\equiv0$ in equation \ref{eq:def_alpha}. The intermediate stellar mass galaxies still have the lowest $f_{\rm DM}$ in our sample, same as in Figure \ref{fig:predict_fdm1}. } 
\end{figure*}

\subsection{Total mass and dark matter fraction, with $\alpha\equiv0$}
\label{subs:alp_0}
In our total mass estimation, we used $\alpha$ derived from the slopes of the circular velocity profiles in the simulation of \hyperlink{Wu+14}{Wu+14} (see Section \ref{subs:parameters}). This allowed $\alpha$ to vary freely between the extremes of Keplerian and logarithmic potentials depending on the galaxy stellar mass. For the most massive galaxies, $\alpha\sim0$ (see Table \ref{tab:summary}). However, it is plausible that the low and intermediate stellar mass ETGs reside in isothermal gravitational potential, such that they are better described by $\alpha\equiv0$. For example, \citealt{Cappellari_2015} found $\alpha = 0.27 \pm 0.23$, on 1--4~$\re$ scales for galaxies with a wide range of stellar mass. Also, \citet{Thomas_2009} showed that ETGs in the Coma cluster are better described by logarithmic DM haloes rather than NFW DM haloes. This would mean that in our earlier analysis, the total mass and $f_{\rm DM}$ especially for these galaxies would be \textit{under--estimated}, depending on how much their $\alpha$ parameter deviates from 0. 
Since our mass estimator is most sensitive to the $\alpha$ parameter on a galaxy by galaxy basis, it is imperative that we check if the earlier trend we found in the distribution $f_{\rm DM}$ with stellar mass is robust to the value of $\alpha$.

We therefore re--perform our mass estimation assuming a logarithmic gravitational potential, i.e. $\alpha\equiv0$, for our sample, and show the result in Figure \ref{fig:predict_fdm2}. \textit{The earlier--observed trends in $f_{\rm DM}$ persist, and they are therefore independent of the assumed slope of the gravitational potential, $\alpha$, as well as the adopted stellar $M/L$ and the orbital anisotropy of the tracers}. NGC~3607 has the least $f_{\rm DM}$ within 5~$\re$ in our sample regardless of the adopted stellar $M/L$ ratio. The total mass for NGC~7457 and NGC~4494 are also increased by $\sim$45 and 35~per~cent, respectively. We summarise these mass estimates and $f_{\rm DM}$ in Tables \ref{tab:appendix_2} and \ref{tab:appendix_3}. 
\subsection{Tension between observations and predictions}
\label{tension}
The results in Figures \ref{fig:predict_fdm1} and \ref{fig:predict_fdm2} show that the mismatch between observations and predictions of $f_{\rm DM}$ is systematic. Intermediate stellar mass galaxies with 
$M_* \sim 10^{11} \Msun$ (NGC~4494, NGC~3607, and NGC~5866) show the greatest deviation from the predicted $f_{\rm DM}$, all with low $f_{\rm DM}$ within 5~$\re$. It is helpful to note that this stellar mass range coincides with the sharp upturn in the galaxy $\re$--$M_*$ relation 
\citep[e.g.][]{Hyde_2009, Lange_2015} and galaxy peak star formation efficiency 
\citep[e.g.][]{Shankar_2006, Conroy_2009, Sparre_2015} beyond which halo quenching 
prevents massive galaxies from accretion of cold gas \citep[e.g.][]{White_1978, Dutton_2014}. From our simple galaxy model (see Section \ref{subs:DMfrac}), it is also the stellar mass beyond which 
$\re/r_{\rm s}$, the ratio of the galaxy size to the scale radius of the DM halo, starts to fall sharply. While the low $f_{\rm DM}$ of these galaxies can be directly linked to a more efficient star formation history, it is interesting to explore why they show more scatter in their $f_{\rm DM}$ compared to $\Lambda$CDM predictions. \citet{Dutton_2011} showed that intermediate mass galaxies are consistent with Salpeter IMF only when their $V_{\rm c}(\re)/\sigma \geq 1.6$. In the top middle panel of Figure \ref{fig:predict_fdm1}, the intermediate mass galaxies that are consistent with a Salpeter IMF are NGC~3608, NGC~821, NGC~4697, NGC~2768 and NGC~4278. We find that these galaxies have $V_{\rm c}(5\re)/\sigma \geq 1.3$. NGC~3607, which has the lowest 
$V_{\rm c}(5\re)/\sigma \sim 0.9$, has a negative $f_{\rm DM}$ when a Salpeter IMF is assumed. A simple experiment in which we vary the stellar $M/L_{K}$ ratio (a proxy for the IMF) reveals that the maximum stellar $M/L_{K}$ that gives positive $f_{\rm DM}$ for all the galaxies in our sample is ${\sim}1.4$. This is shallower than the Salpeter IMF, but steeper than a Kroupa/Chabrier IMF (at a fixed age and metallicity). 

One of the intermediate mass galaxies studied here, NGC~4494, has been notoriously difficult to model in the literature, with results ranging from a low DM content \citep{Romanowsky_2003, Napolitano_2009, Deason_2012} to a high DM content \citep{Morganti_2013}. Here, using GC kinematic data that extend far out into the halo, we find that NGC~4494 is DM poor, i.e., $f_{\rm DM}\leq0.5$ at $R_{\max}\sim9\ \re$ regardless of the adopted stellar $M/L_{\rm K}$ and the GC orbital anisotropy when $\alpha$, the slope of the gravitational potential, is assumed to be 0.2. Our $R_{\max}$ for NGC~4494 is close to the scale radius of the NFW DM halo, where the mass distribution is expected to be DM dominated, such that in a typical $10^{13} \Msun$ halo, $f_{\rm DM}\sim0.9$ at the scale radius. However, when $\alpha\equiv0$ in equation \ref{eq:def_alpha}, we obtain $f_{\rm DM}\sim0.3-0.6$ within 5~$\re$ and $f_{\rm DM}\sim0.5-0.7$ within $R_{\max}$, for varying stellar $M/L_{\rm K}$. This is similar to the result from \citet{Morganti_2013}, obtained also by assuming a logarithmic DM halo. A more detailed dynamical mass modelling of NGC~4494 that combines the existing literature data and the GC data we have studied here would be desirable. Such a study should explore a wide suite of gravitational potentials, galaxy shapes and orbital distributions while incorporating stellar population models.

Galaxies with marginally low $f_{\rm DM}$ within 5~$\re$ e.g. NGC~720, NGC~4526 and NGC~1023, they can be seen to rapidly increase their $f_{\rm DM}$ between 5~$\re$ and their respective $R_{\rm max}$, showing that they are dark matter dominated. Our study also includes the two most dominant members of the Leo II group (NGC~3607 and NGC~3608) with intriguing $f_{\rm DM}$ measurements. The most luminous member of the group, NGC~3607 ($M_{\rm K}=-24.96$) has $f_{\rm DM}\sim0.3$ within 5~$\re$. NGC~3607 has the lowest $f_{\rm DM}$ within 5~$\re$ in our sample even when the DM content is maximised with a logarithmic potential, regardless of the adopted stellar $M/L_{\rm K}$. However, beyond 5~$\re$, the $f_{\rm DM}$ in NGC~3607 increases steeply up to ${\sim}0.8$, again showing that the outer halo is dominated by DM. The next most luminous member of the group with $M_{\rm K}=-23.78$, NGC~3608, however, has a higher $f_{\rm DM}$ of ${\sim}0.8$ within 5~$\re$. Within 5~$\re$, NGC~3607 has an average dark matter density of $\log \left\langle\rho_{\rm DM}\right\rangle$ $\sim6.2\ {\rm \Msun kpc^{-3}}$, the lowest in our sample, unlike NGC~3608 with a denser DM halo with $\log \left\langle\rho_{\rm DM}\right\rangle$ $\sim7.2\ {\rm \Msun kpc^{-3}}$. This suggests that both galaxies have DM haloes that are structurally different, with implications for their assembly time, such that the galaxy with the denser DM halo assembled earlier \citep{Navarro_1996, Bullock_2001, Thomas_2009}. As a group, the intermediate stellar mass galaxies with low $f_{\rm DM}$ in our galaxy sample also have the lowest average DM densities. This mirrors the results from \citet{Romanowsky_2003} and \citet{Napolitano_2005, Napolitano_2009} where some discy, fast rotating, intermediate stellar mass galaxies showed more diffuse DM haloes than expected. A more detailed investigation of the structural parameters of the DM haloes (with adiabatic halo contraction) is however beyond the scope of this paper.

\subsection{Correlations between dark matter fraction and galaxy properties}
\label{DM_cor}
\begin{figure}
    \includegraphics[width=0.48\textwidth]{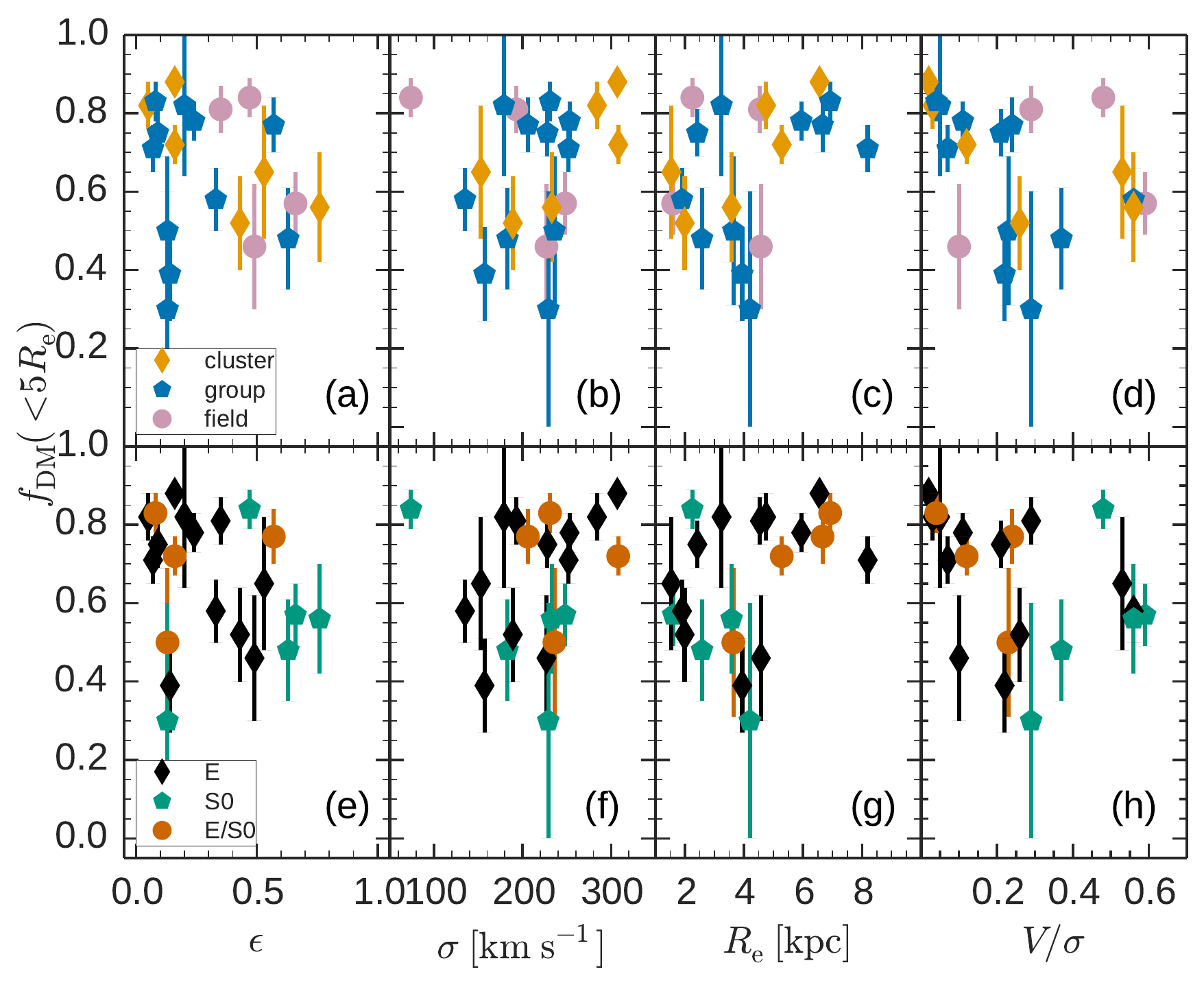}
	\caption{\label{fig:corr_fdm1} Dark matter fraction within $5~\re$ versus galaxy parameters. \textit{Top panels} are color--coded according to galaxy environment as shown in \textit{panel a} and the \textit{bottom panels} according to galaxy morphology as shown in \textit{panel e}. (\textit{Panels a, e}) galaxy ellipticity -- $\epsilon$, (\textit{panels b, f}) central velocity dispersion, (\textit{panel c, g}) effective radius and (\textit{panels d, h}) rotation dominance parameter. There is no clear trend either as a function of galaxy morphology or environment and the trends with galaxy properties are generally weak.} 
\end{figure}

In this section, we look for trends in $f_{\rm DM}$ as a function of other galaxy properties. Figure \ref{fig:corr_fdm1} shows how the measured $f_{\rm DM}$ within 5~$\re$ varies with galaxy ellipticity, central velocity dispersion, galaxy size and galaxy rotation dominance parameter, and we also highlight the environment and morphology of the galaxies (see Table \ref{tab:summary}). The rotational dominance parameters are from \citet{Arnold_2011}. Table \ref{tab:spearman} shows the Spearman rank correlation coefficient and statistical significance of the correlation between the $f_{\rm DM}$ and galaxy properties. The correlations are generally weak, mainly due to the huge scatter introduced by the intermediate stellar mass galaxies identified and discussed in Section \ref{tension}. There is a visible trend in $f_{\rm DM}$ with $\epsilon$.
Here, we find that $f_{\rm DM}$ within 5~$\re$ decreases with $\epsilon$. However, there are notable outliers to this trend. The trends with $\re$, $\sigma$ and $V/\sigma$ are weak. While the slow--rotators in our sample generally have high $f_{\rm DM}$, there is no clear pattern in the fast--rotators. The S0 galaxies, however, show a decreasing $f_{\rm DM}$ with $\sigma$ and $\re$. We do not see any strong trend as a function of environment or morphology. 

\begin{figure}
        \includegraphics[width=0.48\textwidth]{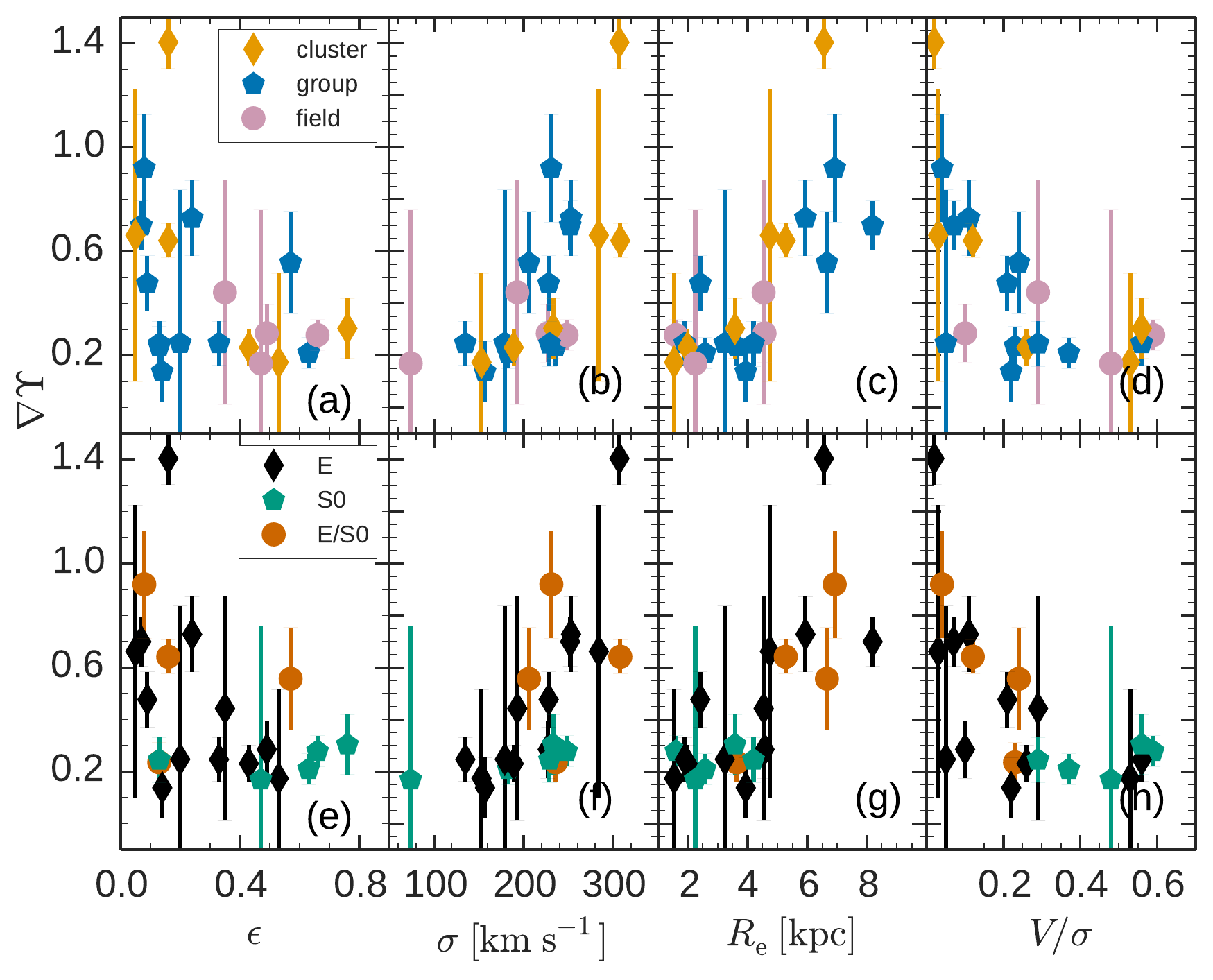}
	\caption{\label{fig:corr_grad_env} Mass--to--light gradient, $\nabla\Upsilon$,  versus galaxy parameters. Color coding is same as in Figure \ref{fig:corr_fdm1}. (\textit{Panels a, e}) $\nabla\Upsilon$ vs. galaxy ellipticity -- $\epsilon$, (\textit{panels b, f}) $\nabla\Upsilon$ vs. central velocity dispersion, (\textit{panels c, g}) $\nabla\Upsilon$ vs. effective radius and (\textit{panels d, h}) $\nabla\Upsilon$ vs. rotation dominance parameter. Galaxies with $\epsilon \sim 0$, large $\sigma$ and larger $\re$ have steeper gradients. An interesting trend is seen in \textit{panel h} where the S0 galaxies have shallow $\nabla\Upsilon$ regardless of galaxy ellipticity, size or mass.}
\end{figure}
%
%
In Figure \ref{fig:corr_grad_env}, we show how $\nabla\Upsilon$ varies with galaxy properties. The correlations are now stronger and statistically significant (see Table \ref{tab:spearman}). The gradients are shallower for flattened galaxies, with the more spherical galaxies showing a great variety of $\nabla\Upsilon$. Larger and more massive galaxies have steeper $\nabla\Upsilon$. There is no clear trend with galaxy environment. However, when the different galaxy morphologies are highlighted, the S0 galaxies are seen to have shallow $\nabla\Upsilon$ regardless of galaxy ellipticity, size, total mass or rotational dominance parameter (the same trend is also evident from the result in \citealt{Napolitano_2005}). The net effect for massive S0 galaxies is to reduce their $\re/r_{\rm s}$ compared to similar stellar mass ellipticals, hence their flattened gradients i.e. lower $\nabla\Upsilon$. 
\begin{table}
{\small \caption{Spearman correlation test and statistical significance of the correlation between the $f_{\rm DM}$ and $\nabla\Upsilon$ and galaxy properties.} \label{tab:spearman}}
\begin{tabular}{l c c | l c c }
\hline
\hline
Parameters & coeff & p-val & Parameters & coeff & p-val\\
\hline
$f_{\rm DM}-\epsilon$ & -0.22 & 0.32  & $\nabla\Upsilon-\epsilon$ & -0.37 & 0.09 \\
$f_{\rm DM}-\sigma$   & 0.18  & 0.44  & $\nabla\Upsilon-\sigma$   & 0.76  & 0.001 \\
$f_{\rm DM}-\re$      & 0.32  & 0.16 & $\nabla\Upsilon-\re$      & 0.75  & 0.001 \\
$f_{\rm DM}-V/\sigma$ & -0.44  & 0.04  & $\nabla\Upsilon-V/\sigma$ & -0.63  & 0.001 \\
\hline
\end{tabular}
\end{table}

\section{Conclusions}
\label{concl}
We have employed a tracer mass estimator to homogeneously obtain mass estimates of 23 ETGs out to 
5~$\re$ and beyond, using their GC kinematic data. The galaxies we have studied cover a wide range of total galaxy stellar mass and include galaxies from the field, group and cluster environments. The GC kinematic data have been obtained using the Keck/DEIMOS multi--object spectrograph as part of the SLUGGS survey. We accounted for kinematic substructures, galaxy flattening and rotation in the GC system in our mass estimates. We have done an extensive comparison of our mass estimates with results from the literature obtained using various mass tracers and more sophisticated modelling techniques. 

From the mass profiles, we have obtained the dark matter fraction enclosed within 5~$\re$ and compared our results with predictions from a simple galaxy model (NFW profile for dark matter plus \Sersic/ mass profile for the stars). We have also studied the effect of varying the stellar mass--to--light ratio (consistent with either a Salpeter or a Kroupa--like IMF or one that varies with galaxy stellar mass) on our results. Since our GC data extends well beyond 5~$\re$, we have quantified the gradient of the dark matter fraction between 5~$\re$ and the maximum probed radius. Lastly, we studied trends in the dark matter fraction as a function of galaxy properties.

The salient results are:
\begin{itemize}
\item	Mass estimates obtained using GC kinematic data and the tracer mass estimator are consistent with those obtained from more sophisticated modelling techniques and with various mass tracers over a radial range that extends out to ${\sim}13\ \re$. Using the tracer mass estimator, we are able to obtain mass estimates out to ${\sim}10\ \re$ in low--mass galaxies with relatively sparse dynamical tracers. We find an upper limit of 0.2 dex in the observed 1-$\sigma$ scatter around the one--to--one comparison line between our mass estimates and those from the literature. 
\item	On average in our sample, kinematic substructures in GC systems leads to mass over--estimation by ${\sim}19$ per cent. Not accounting for GC system rotation leads to mass under--estimation by ${\sim}6$ per cent, while galaxy flattening is responsible for a ${\sim}5$ per cent mass over--estimation with the caveat that our galaxies are mostly edge--on.
\item 	By comparing the total mass enclosed within 5~$\re$ under various assumptions of velocity anisotropy, we are able to establish that total mass estimates are largely insensitive to GC orbital anisotropy. Only NGC~3377, NGC~7457 and NGC~1407 show mass deviations greater than $10$ per cent when mildly tangential or radial anisotropies are assumed rather than isotropy conditions.
\item	The dark matter fraction within 5~$\re$, $f_{\rm DM}$, generally increases with galaxy stellar mass. It increases from $f_{\rm DM}{\sim}0.6$ in low mass ETGs to $f_{\rm DM}{\sim}0.8$ in high mass ETGs, in line with $\Lambda$CDM predictions. However, some intermediate mass galaxies (${\sim}10^{11}\ \Msun$), i.e., NGC~4494, NGC~3607 and NGC~5866, have $f_{\rm DM}$ that are significantly lower than what a vanilla galaxy model would predict. This is independent of the assumed stellar $M/L_{\rm K}$ ratio, the orbital anisotropy of the mass tracers or the shape of the gravitational potential. These lower $f_{\rm DM}$ measurements are consistent with results from the cosmological simulations of \citet{Wu_2014} where the pristine DM density distribution has been modified via baryon--DM interactions during galaxy assembly. The widely reported dearth of dark matter in the outer halo of NGC~4494 is alleviated by assuming a logarithmic gravitational potential.
\item	Using total mass estimates within 5~$\re$ and larger radii (usually comparable to the scale radii of the dark matter haloes), low and intermediate stellar mass galaxies in our sample have shallow mass--to--light gradients, with the more massive galaxies generally having steeper gradients. This reflects the relative difference in the radial scale of baryons and dark matter in ETGs. However, lenticular galaxies, regardless of galaxy stellar mass, ellipticity, size and rotational dominance parameter, have shallow gradients. 
\item	We find hints that intermediate stellar mass galaxies with low dark matter fractions have halo structural parameters that are not typical i.e., they possess very diffuse dark matter haloes and they assembled late. This result is interesting and calls for a systematic study of the structural parameters of the haloes of ETGs. 
\end{itemize}

\section*{Acknowledgements}
We wish to thank the anonymous referee for the useful feedback. The data presented herein were obtained at the W.M. Keck Observatory, which is operated as a scientific partnership among the California
Institute of Technology, the University of California and the National Aeronautics and Space Administration. 
The Observatory was made possible by the generous financial support of the W.M. Keck Foundation. The authors wish to
recognize and acknowledge the very significant cultural role and reverence that the summit of Mauna Kea has always had within 
the indigenous Hawaiian community.
The analysis pipeline used to reduce the DEIMOS data was developed at UC Berkeley with support from NSF grant AST--0071048. JPB acknowledges support from NSF grant AST--1211995. DAF and JJ thank the ARC for financial support via DP 130100388. We thank Cappellari M. for making his mass profiles available. This research made use of TOPCAT \citep{Taylor_2005} and \textsc{hyper.fit} \citep{Robotham_2015}.

\bibliographystyle{mnras}
\bibliography{mass_est}

\begin{thebibliography}{}
\makeatletter
\relax
\def\mn@urlcharsother{\let\do\@makeother \do\$\do\&\do\#\do\^\do\_\do\%\do\~}
\def\mn@doi{\begingroup\mn@urlcharsother \@ifnextchar [ {\mn@doi@}
  {\mn@doi@[]}}
\def\mn@doi@[#1]#2{\def\@tempa{#1}\ifx\@tempa\@empty \href
  {http://dx.doi.org/#2} {doi:#2}\else \href {http://dx.doi.org/#2} {#1}\fi
  \endgroup}
\def\mn@eprint#1#2{\mn@eprint@#1:#2::\@nil}
\def\mn@eprint@arXiv#1{\href {http://arxiv.org/abs/#1} {{\tt arXiv:#1}}}
\def\mn@eprint@dblp#1{\href {http://dblp.uni-trier.de/rec/bibtex/#1.xml}
  {dblp:#1}}
\def\mn@eprint@#1:#2:#3:#4\@nil{\def\@tempa {#1}\def\@tempb {#2}\def\@tempc
  {#3}\ifx \@tempc \@empty \let \@tempc \@tempb \let \@tempb \@tempa \fi \ifx
  \@tempb \@empty \def\@tempb {arXiv}\fi \@ifundefined
  {mn@eprint@\@tempb}{\@tempb:\@tempc}{\expandafter \expandafter \csname
  mn@eprint@\@tempb\endcsname \expandafter{\@tempc}}}

\bibitem[\protect\citeauthoryear{{Agnello}, {Evans}, {Romanowsky}  \&
  {Brodie}}{{Agnello} et~al.}{2014}]{Agnello_2014}
{Agnello} A.,  {Evans} N.~W.,  {Romanowsky} A.~J.,   {Brodie} J.~P.,  2014,
  \mn@doi [\mnras] {10.1093/mnras/stu960}, \href
  {http://adsabs.harvard.edu/abs/2014MNRAS.442.3299A} {442, 3299}

\bibitem[\protect\citeauthoryear{{An} \& {Evans}}{{An} \&
  {Evans}}{2011}]{An_2011}
{An} J.~H.,  {Evans} N.~W.,  2011, \mn@doi [\mnras]
  {10.1111/j.1365-2966.2011.18251.x}, \href
  {http://adsabs.harvard.edu/abs/2011MNRAS.413.1744A} {413, 1744}

\bibitem[\protect\citeauthoryear{{Arnold}, {Romanowsky}, {Brodie}, {Chomiuk},
  {Spitler}, {Strader}, {Benson}  \& {Forbes}}{{Arnold}
  et~al.}{2011}]{Arnold_2011}
{Arnold} J.~A.,  {Romanowsky} A.~J.,  {Brodie} J.~P.,  {Chomiuk} L.,  {Spitler}
  L.~R.,  {Strader} J.,  {Benson} A.~J.,   {Forbes} D.~A.,  2011, \mn@doi
  [\apjl] {10.1088/2041-8205/736/2/L26}, \href
  {http://adsabs.harvard.edu/abs/2011ApJ...736L..26A} {736, L26}

\bibitem[\protect\citeauthoryear{{Arnold} et~al.,}{{Arnold}
  et~al.}{2014}]{Arnold_2014}
{Arnold} J.~A.,  et~al., 2014, \mn@doi [\apj] {10.1088/0004-637X/791/2/80},
  \href {http://adsabs.harvard.edu/abs/2014ApJ...791...80A} {791, 80}

\bibitem[\protect\citeauthoryear{{Ashman} \& {Bird}}{{Ashman} \&
  {Bird}}{1993}]{Ashman_1993}
{Ashman} K.~M.,  {Bird} C.~M.,  1993, \mn@doi [\aj] {10.1086/116799}, \href
  {http://adsabs.harvard.edu/abs/1993AJ....106.2281A} {106, 2281}

\bibitem[\protect\citeauthoryear{{Auger}, {Treu}, {Bolton}, {Gavazzi},
  {Koopmans}, {Marshall}, {Moustakas}  \& {Burles}}{{Auger}
  et~al.}{2010}]{Auger_2010}
{Auger} M.~W.,  {Treu} T.,  {Bolton} A.~S.,  {Gavazzi} R.,  {Koopmans}
  L.~V.~E.,  {Marshall} P.~J.,  {Moustakas} L.~A.,   {Burles} S.,  2010,
  \mn@doi [\apj] {10.1088/0004-637X/724/1/511}, \href
  {http://adsabs.harvard.edu/abs/2010ApJ...724..511A} {724, 511}

\bibitem[\protect\citeauthoryear{{Bacon}}{{Bacon}}{1985}]{Bacon_1985}
{Bacon} R.,  1985, \aap, \href
  {http://adsabs.harvard.edu/abs/1985A%26A...143...84B} {143, 84}

\bibitem[\protect\citeauthoryear{{Bahcall} \& {Tremaine}}{{Bahcall} \&
  {Tremaine}}{1981}]{Bahcall_1981}
{Bahcall} J.~N.,  {Tremaine} S.,  1981, \mn@doi [\apj] {10.1086/158756}, \href
  {http://adsabs.harvard.edu/abs/1981ApJ...244..805B} {244, 805}

\bibitem[\protect\citeauthoryear{{Barnab{\`e}}, {Czoske}, {Koopmans}, {Treu}
  \& {Bolton}}{{Barnab{\`e}} et~al.}{2011}]{Barnabe_2011}
{Barnab{\`e}} M.,  {Czoske} O.,  {Koopmans} L.~V.~E.,  {Treu} T.,   {Bolton}
  A.~S.,  2011, \mn@doi [\mnras] {10.1111/j.1365-2966.2011.18842.x}, \href
  {http://adsabs.harvard.edu/abs/2011MNRAS.415.2215B} {415, 2215}

\bibitem[\protect\citeauthoryear{{Bassino}, {Richtler}  \& {Dirsch}}{{Bassino}
  et~al.}{2006}]{Bassino_2006}
{Bassino} L.~P.,  {Richtler} T.,   {Dirsch} B.,  2006, \mn@doi [\mnras]
  {10.1111/j.1365-2966.2005.09919.x}, \href
  {http://adsabs.harvard.edu/abs/2006MNRAS.367..156B} {367, 156}

\bibitem[\protect\citeauthoryear{{Bekki} \& {Forbes}}{{Bekki} \&
  {Forbes}}{2006}]{Bekki_2006}
{Bekki} K.,  {Forbes} D.~A.,  2006, \mn@doi [\aap]
  {10.1051/0004-6361:20053686}, \href
  {http://adsabs.harvard.edu/abs/2006A%26A...445..485B} {445, 485}

\bibitem[\protect\citeauthoryear{{Bender}, {Saglia}  \& {Gerhard}}{{Bender}
  et~al.}{1994}]{Bender_1994}
{Bender} R.,  {Saglia} R.~P.,   {Gerhard} O.~E.,  1994, \mnras, \href
  {http://adsabs.harvard.edu/abs/1994MNRAS.269..785B} {269, 785}

\bibitem[\protect\citeauthoryear{{Bergond}, {Zepf}, {Romanowsky}, {Sharples}
  \& {Rhode}}{{Bergond} et~al.}{2006}]{Bergond_2006}
{Bergond} G.,  {Zepf} S.~E.,  {Romanowsky} A.~J.,  {Sharples} R.~M.,   {Rhode}
  K.~L.,  2006, \mn@doi [\aap] {10.1051/0004-6361:20053697}, \href
  {http://adsabs.harvard.edu/abs/2006A%26A...448..155B} {448, 155}

\bibitem[\protect\citeauthoryear{{Binney} \& {Mamon}}{{Binney} \&
  {Mamon}}{1982}]{Binney_1982}
{Binney} J.,  {Mamon} G.~A.,  1982, \mnras, \href
  {http://adsabs.harvard.edu/abs/1982MNRAS.200..361B} {200, 361}

\bibitem[\protect\citeauthoryear{{Binney} \& {Tremaine}}{{Binney} \&
  {Tremaine}}{1987}]{Binney_1987}
{Binney} J.,  {Tremaine} S.,  1987, {Galactic dynamics}

\bibitem[\protect\citeauthoryear{{Blakeslee} et~al.,}{{Blakeslee}
  et~al.}{2009}]{Blakeslee_2009}
{Blakeslee} J.~P.,  et~al., 2009, \mn@doi [\apj] {10.1088/0004-637X/694/1/556},
  \href {http://cdsads.u-strasbg.fr/abs/2009ApJ...694..556B} {694, 556}

\bibitem[\protect\citeauthoryear{{Brodie} et~al.,}{{Brodie}
  et~al.}{2014}]{Brodie_2014}
{Brodie} J.~P.,  et~al., 2014, \mn@doi [\apj] {10.1088/0004-637X/796/1/52},
  \href {http://adsabs.harvard.edu/abs/2014ApJ...796...52B} {796, 52}

\bibitem[\protect\citeauthoryear{{Bullock} \& {Johnston}}{{Bullock} \&
  {Johnston}}{2005}]{Bullock_2005}
{Bullock} J.~S.,  {Johnston} K.~V.,  2005, \mn@doi [\apj] {10.1086/497422},
  \href {http://adsabs.harvard.edu/abs/2005ApJ...635..931B} {635, 931}

\bibitem[\protect\citeauthoryear{{Bullock}, {Kolatt}, {Sigad}, {Somerville},
  {Kravtsov}, {Klypin}, {Primack}  \& {Dekel}}{{Bullock}
  et~al.}{2001}]{Bullock_2001}
{Bullock} J.~S.,  {Kolatt} T.~S.,  {Sigad} Y.,  {Somerville} R.~S.,  {Kravtsov}
  A.~V.,  {Klypin} A.~A.,  {Primack} J.~R.,   {Dekel} A.,  2001, \mn@doi
  [\mnras] {10.1046/j.1365-8711.2001.04068.x}, \href
  {http://adsabs.harvard.edu/abs/2001MNRAS.321..559B} {321, 559}

\bibitem[\protect\citeauthoryear{{Cappellari} et~al.,}{{Cappellari}
  et~al.}{2007}]{Cappellari_2007}
{Cappellari} M.,  et~al., 2007, \mn@doi [\mnras]
  {10.1111/j.1365-2966.2007.11963.x}, \href
  {http://adsabs.harvard.edu/abs/2007MNRAS.379..418C} {379, 418}

\bibitem[\protect\citeauthoryear{{Cappellari} et~al.,}{{Cappellari}
  et~al.}{2012}]{Cappellari_2012}
{Cappellari} M.,  et~al., 2012, \mn@doi [\nat] {10.1038/nature10972}, \href
  {http://adsabs.harvard.edu/abs/2012Natur.484..485C} {484, 485}

\bibitem[\protect\citeauthoryear{{Cappellari} et~al.,}{{Cappellari}
  et~al.}{2013a}]{Cappellari_2013}
{Cappellari} M.,  et~al., 2013a, \mn@doi [\mnras] {10.1093/mnras/stt562}, \href
  {http://adsabs.harvard.edu/abs/2013MNRAS.432.1709C} {432, 1709}

\bibitem[\protect\citeauthoryear{{Cappellari} et~al.,}{{Cappellari}
  et~al.}{2013b}]{Cappellari_2013b}
{Cappellari} M.,  et~al., 2013b, \mn@doi [\mnras] {10.1093/mnras/stt644}, \href
  {http://adsabs.harvard.edu/abs/2013MNRAS.432.1862C} {432, 1862}

\bibitem[\protect\citeauthoryear{{Cappellari} et~al.,}{{Cappellari}
  et~al.}{2015}]{Cappellari_2015}
{Cappellari} M.,  et~al., 2015, \mn@doi [\apjl] {10.1088/2041-8205/804/1/L21},
  \href {http://adsabs.harvard.edu/abs/2015ApJ...804L..21C} {804, L21}

\bibitem[\protect\citeauthoryear{{Churazov} et~al.,}{{Churazov}
  et~al.}{2010}]{Churazov_2010}
{Churazov} E.,  et~al., 2010, \mn@doi [\mnras]
  {10.1111/j.1365-2966.2010.16377.x}, \href
  {http://adsabs.harvard.edu/abs/2010MNRAS.404.1165C} {404, 1165}

\bibitem[\protect\citeauthoryear{{Coccato} et~al.,}{{Coccato}
  et~al.}{2009}]{Coccato_2009}
{Coccato} L.,  et~al., 2009, \mn@doi [\mnras]
  {10.1111/j.1365-2966.2009.14417.x}, \href
  {http://adsabs.harvard.edu/abs/2009MNRAS.394.1249C} {394, 1249}

\bibitem[\protect\citeauthoryear{{Conroy} \& {Wechsler}}{{Conroy} \&
  {Wechsler}}{2009}]{Conroy_2009}
{Conroy} C.,  {Wechsler} R.~H.,  2009, \mn@doi [\apj]
  {10.1088/0004-637X/696/1/620}, \href
  {http://adsabs.harvard.edu/abs/2009ApJ...696..620C} {696, 620}

\bibitem[\protect\citeauthoryear{{Conroy} \& {van Dokkum}}{{Conroy} \& {van
  Dokkum}}{2012}]{Conroy_2012}
{Conroy} C.,  {van Dokkum} P.~G.,  2012, \mn@doi [\apj]
  {10.1088/0004-637X/760/1/71}, \href
  {http://adsabs.harvard.edu/abs/2012ApJ...760...71C} {760, 71}

\bibitem[\protect\citeauthoryear{{Cooper}, {Newman}, {Davis}, {Finkbeiner}  \&
  {Gerke}}{{Cooper} et~al.}{2012}]{Cooper_2012}
{Cooper} M.~C.,  {Newman} J.~A.,  {Davis} M.,  {Finkbeiner} D.~P.,   {Gerke}
  B.~F.,  2012, {spec2d: DEEP2 DEIMOS Spectral Pipeline}, Astrophysics Source
  Code Library (\mn@eprint {ascl} {1203.003})

\bibitem[\protect\citeauthoryear{{Cooper}, {D'Souza}, {Kauffmann}, {Wang},
  {Boylan-Kolchin}, {Guo}, {Frenk}  \& {White}}{{Cooper}
  et~al.}{2013}]{Cooper_2013}
{Cooper} A.~P.,  {D'Souza} R.,  {Kauffmann} G.,  {Wang} J.,  {Boylan-Kolchin}
  M.,  {Guo} Q.,  {Frenk} C.~S.,   {White} S.~D.~M.,  2013, \mn@doi [\mnras]
  {10.1093/mnras/stt1245}, \href
  {http://adsabs.harvard.edu/abs/2013MNRAS.434.3348C} {434, 3348}

\bibitem[\protect\citeauthoryear{{Cortesi} et~al.,}{{Cortesi}
  et~al.}{2013}]{Cortesi_2013}
{Cortesi} A.,  et~al., 2013, \mn@doi [\mnras] {10.1093/mnras/stt529}, \href
  {http://adsabs.harvard.edu/abs/2013MNRAS.432.1010C} {432, 1010}

\bibitem[\protect\citeauthoryear{{Das}, {Gerhard}, {Churazov}  \&
  {Zhuravleva}}{{Das} et~al.}{2010}]{Das_2010}
{Das} P.,  {Gerhard} O.,  {Churazov} E.,   {Zhuravleva} I.,  2010, \mn@doi
  [\mnras] {10.1111/j.1365-2966.2010.17417.x}, \href
  {http://adsabs.harvard.edu/abs/2010MNRAS.409.1362D} {409, 1362}

\bibitem[\protect\citeauthoryear{{Das}, {Gerhard}, {Mendez}, {Teodorescu}  \&
  {de Lorenzi}}{{Das} et~al.}{2011}]{Das_2011}
{Das} P.,  {Gerhard} O.,  {Mendez} R.~H.,  {Teodorescu} A.~M.,   {de Lorenzi}
  F.,  2011, \mn@doi [\mnras] {10.1111/j.1365-2966.2011.18771.x}, \href
  {http://adsabs.harvard.edu/abs/2011MNRAS.415.1244D} {415, 1244}

\bibitem[\protect\citeauthoryear{{DeLorenzi}, {Gerhard}, {Saglia}, {Sambhus},
  {Debattista}, {Pannella}  \& {M{\'e}ndez}}{{DeLorenzi}
  et~al.}{2008}]{DeLorenzi_2008}
{DeLorenzi} F.,  {Gerhard} O.,  {Saglia} R.~P.,  {Sambhus} N.,  {Debattista}
  V.~P.,  {Pannella} M.,   {M{\'e}ndez} R.~H.,  2008, \mn@doi [\mnras]
  {10.1111/j.1365-2966.2008.12905.x}, \href
  {http://adsabs.harvard.edu/abs/2008MNRAS.385.1729D} {385, 1729}

\bibitem[\protect\citeauthoryear{{Deason}, {Belokurov}, {Evans}  \&
  {McCarthy}}{{Deason} et~al.}{2012}]{Deason_2012}
{Deason} A.~J.,  {Belokurov} V.,  {Evans} N.~W.,   {McCarthy} I.~G.,  2012,
  \mn@doi [\apj] {10.1088/0004-637X/748/1/2}, \href
  {http://adsabs.harvard.edu/abs/2012ApJ...748....2D} {748, 2}

\bibitem[\protect\citeauthoryear{{Dekel}, {Stoehr}, {Mamon}, {Cox}, {Novak}  \&
  {Primack}}{{Dekel} et~al.}{2005}]{Dekel_2005}
{Dekel} A.,  {Stoehr} F.,  {Mamon} G.~A.,  {Cox} T.~J.,  {Novak} G.~S.,
  {Primack} J.~R.,  2005, \mn@doi [\nat] {10.1038/nature03970}, \href
  {http://adsabs.harvard.edu/abs/2005Natur.437..707D} {437, 707}

\bibitem[\protect\citeauthoryear{{Di Cintio}, {Knebe}, {Libeskind}, {Hoffman},
  {Yepes}  \& {Gottl{\"o}ber}}{{Di Cintio} et~al.}{2012}]{diCintio_2012}
{Di Cintio} A.,  {Knebe} A.,  {Libeskind} N.~I.,  {Hoffman} Y.,  {Yepes} G.,
  {Gottl{\"o}ber} S.,  2012, \mn@doi [\mnras]
  {10.1111/j.1365-2966.2012.21013.x}, \href
  {http://adsabs.harvard.edu/abs/2012MNRAS.423.1883D} {423, 1883}

\bibitem[\protect\citeauthoryear{{Dirsch}, {Schuberth}  \& {Richtler}}{{Dirsch}
  et~al.}{2005}]{Dirsch_2005}
{Dirsch} B.,  {Schuberth} Y.,   {Richtler} T.,  2005, \mn@doi [\aap]
  {10.1051/0004-6361:20035737}, \href
  {http://adsabs.harvard.edu/abs/2005A%26A...433...43D} {433, 43}

\bibitem[\protect\citeauthoryear{{Dressler} \& {Shectman}}{{Dressler} \&
  {Shectman}}{1988}]{Dressler_1988}
{Dressler} A.,  {Shectman} S.~A.,  1988, \mn@doi [\aj] {10.1086/114694}, \href
  {http://adsabs.harvard.edu/abs/1988AJ.....95..985D} {95, 985}

\bibitem[\protect\citeauthoryear{{Dutton} \& {Macci{\`o}}}{{Dutton} \&
  {Macci{\`o}}}{2014}]{Dutton_2014}
{Dutton} A.~A.,  {Macci{\`o}} A.~V.,  2014, \mn@doi [\mnras]
  {10.1093/mnras/stu742}, \href
  {http://adsabs.harvard.edu/abs/2014MNRAS.441.3359D} {441, 3359}

\bibitem[\protect\citeauthoryear{{Dutton}, {Conroy}, {van den Bosch}, {Prada}
  \& {More}}{{Dutton} et~al.}{2010}]{Dutton_2010}
{Dutton} A.~A.,  {Conroy} C.,  {van den Bosch} F.~C.,  {Prada} F.,   {More} S.,
   2010, \mn@doi [\mnras] {10.1111/j.1365-2966.2010.16911.x}, \href
  {http://adsabs.harvard.edu/abs/2010MNRAS.407....2D} {407, 2}

\bibitem[\protect\citeauthoryear{{Dutton} et~al.,}{{Dutton}
  et~al.}{2011}]{Dutton_2011}
{Dutton} A.~A.,  et~al., 2011, \mn@doi [\mnras]
  {10.1111/j.1365-2966.2011.19038.x}, \href
  {http://adsabs.harvard.edu/abs/2011MNRAS.416..322D} {416, 322}

\bibitem[\protect\citeauthoryear{{Dutton}, {Macci{\`o}}, {Mendel}  \&
  {Simard}}{{Dutton} et~al.}{2013}]{Dutton_2013}
{Dutton} A.~A.,  {Macci{\`o}} A.~V.,  {Mendel} J.~T.,   {Simard} L.,  2013,
  \mn@doi [\mnras] {10.1093/mnras/stt608}, \href
  {http://adsabs.harvard.edu/abs/2013MNRAS.432.2496D} {432, 2496}

\bibitem[\protect\citeauthoryear{{Einasto} et~al.,}{{Einasto}
  et~al.}{2012}]{Einasto_2012}
{Einasto} M.,  et~al., 2012, \mn@doi [\aap] {10.1051/0004-6361/201118697},
  \href {http://adsabs.harvard.edu/abs/2012A%26A...540A.123E} {540, A123}

\bibitem[\protect\citeauthoryear{{Evans}}{{Evans}}{1994}]{Evans_1994a}
{Evans} N.~W.,  1994, \mnras, \href
  {http://adsabs.harvard.edu/abs/1994MNRAS.267..333E} {267, 333}

\bibitem[\protect\citeauthoryear{{Evans}, {Wilkinson}, {Perrett}  \&
  {Bridges}}{{Evans} et~al.}{2003}]{Evans_2003}
{Evans} N.~W.,  {Wilkinson} M.~I.,  {Perrett} K.~M.,   {Bridges} T.~J.,  2003,
  \mn@doi [\apj] {10.1086/345400}, \href
  {http://adsabs.harvard.edu/abs/2003ApJ...583..752E} {583, 752}

\bibitem[\protect\citeauthoryear{{Faber} et~al.,}{{Faber}
  et~al.}{2003}]{Faber_2003}
{Faber} S.~M.,  et~al., 2003, in {Iye} M.,  {Moorwood} A.~F.~M.,  eds,  Society
  of Photo-Optical Instrumentation Engineers (SPIE) Conference Series Vol.
  4841, Instrument Design and Performance for Optical/Infrared Ground-based
  Telescopes. pp 1657--1669, \mn@doi{10.1117/12.460346}

\bibitem[\protect\citeauthoryear{{Faifer} et~al.,}{{Faifer}
  et~al.}{2011}]{Faifer_2011}
{Faifer} F.~R.,  et~al., 2011, \mn@doi [\mnras]
  {10.1111/j.1365-2966.2011.19018.x}, \href
  {http://adsabs.harvard.edu/abs/2011MNRAS.416..155F} {416, 155}

\bibitem[\protect\citeauthoryear{{Forbes}, {Lasky}, {Graham}  \&
  {Spitler}}{{Forbes} et~al.}{2008}]{Forbes_2008}
{Forbes} D.~A.,  {Lasky} P.,  {Graham} A.~W.,   {Spitler} L.,  2008, \mn@doi
  [\mnras] {10.1111/j.1365-2966.2008.13739.x}, \href
  {http://adsabs.harvard.edu/abs/2008MNRAS.389.1924F} {389, 1924}

\bibitem[\protect\citeauthoryear{{Forestell} \& {Gebhardt}}{{Forestell} \&
  {Gebhardt}}{2010}]{Forestell_2010}
{Forestell} A.~D.,  {Gebhardt} K.,  2010, \mn@doi [\apj]
  {10.1088/0004-637X/716/1/370}, \href
  {http://adsabs.harvard.edu/abs/2010ApJ...716..370F} {716, 370}

\bibitem[\protect\citeauthoryear{{Gerhard}, {Kronawitter}, {Saglia}  \&
  {Bender}}{{Gerhard} et~al.}{2001}]{Gerhard_2001}
{Gerhard} O.,  {Kronawitter} A.,  {Saglia} R.~P.,   {Bender} R.,  2001, \mn@doi
  [\aj] {10.1086/319940}, \href
  {http://adsabs.harvard.edu/abs/2001AJ....121.1936G} {121, 1936}

\bibitem[\protect\citeauthoryear{{Graham}}{{Graham}}{2013}]{Graham_2013}
{Graham} A.~W.,  2013, {Elliptical and Disk Galaxy Structure and Modern Scaling
  Laws}.
p.~91, \mn@doi{10.1007/978-94-007-5609-0_2}

\bibitem[\protect\citeauthoryear{{Harris}}{{Harris}}{1976}]{Harris_1976}
{Harris} W.~E.,  1976, \mn@doi [\aj] {10.1086/111991}, \href
  {http://adsabs.harvard.edu/abs/1976AJ.....81.1095H} {81, 1095}

\bibitem[\protect\citeauthoryear{{Harris}}{{Harris}}{1986}]{Harris_1986}
{Harris} W.~E.,  1986, \mn@doi [\aj] {10.1086/114062}, \href
  {http://adsabs.harvard.edu/abs/1986AJ.....91..822H} {91, 822}

\bibitem[\protect\citeauthoryear{{Heisler}, {Tremaine}  \& {Bahcall}}{{Heisler}
  et~al.}{1985}]{Heisler_1985}
{Heisler} J.,  {Tremaine} S.,   {Bahcall} J.~N.,  1985, \mn@doi [\apj]
  {10.1086/163584}, \href {http://adsabs.harvard.edu/abs/1985ApJ...298....8H}
  {298, 8}

\bibitem[\protect\citeauthoryear{{Helmi}}{{Helmi}}{2008}]{Helmi_2008}
{Helmi} A.,  2008, \mn@doi [\aapr] {10.1007/s00159-008-0009-6}, \href
  {http://adsabs.harvard.edu/abs/2008A%26ARv..15..145H} {15, 145}

\bibitem[\protect\citeauthoryear{{Humphrey} \& {Buote}}{{Humphrey} \&
  {Buote}}{2010}]{Humphrey_2010}
{Humphrey} P.~J.,  {Buote} D.~A.,  2010, \mn@doi [\mnras]
  {10.1111/j.1365-2966.2010.16257.x}, \href
  {http://adsabs.harvard.edu/abs/2010MNRAS.403.2143H} {403, 2143}

\bibitem[\protect\citeauthoryear{{Hyde} \& {Bernardi}}{{Hyde} \&
  {Bernardi}}{2009}]{Hyde_2009}
{Hyde} J.~B.,  {Bernardi} M.,  2009, \mn@doi [\mnras]
  {10.1111/j.1365-2966.2009.14445.x}, \href
  {http://adsabs.harvard.edu/abs/2009MNRAS.394.1978H} {394, 1978}

\bibitem[\protect\citeauthoryear{{Jarrett}, {Chester}, {Cutri}, {Schneider},
  {Skrutskie}  \& {Huchra}}{{Jarrett} et~al.}{2000}]{Jarrett_2000}
{Jarrett} T.~H.,  {Chester} T.,  {Cutri} R.,  {Schneider} S.,  {Skrutskie} M.,
   {Huchra} J.~P.,  2000, \mn@doi [\aj] {10.1086/301330}, \href
  {http://adsabs.harvard.edu/abs/2000AJ....119.2498J} {119, 2498}

\bibitem[\protect\citeauthoryear{{Kissler-Patig}}{{Kissler-Patig}}{1997}]{Kissler_1997}
{Kissler-Patig} M.,  1997, \aap, \href
  {http://adsabs.harvard.edu/abs/1997A%26A...319...83K} {319, 83}

\bibitem[\protect\citeauthoryear{{Krajnovi{\'c}}, {Emsellem}, {Cappellari},
  {Alatalo}, {Blitz}, {Bois}, {Bournaud}  \& {Bureau}}{{Krajnovi{\'c}}
  et~al.}{2011}]{Krajnovic_2011}
{Krajnovi{\'c}} D.,  {Emsellem} E.,  {Cappellari} M.,  {Alatalo} K.,  {Blitz}
  L.,  {Bois} M.,  {Bournaud} F.,   {Bureau} 2011, \mn@doi [\mnras]
  {10.1111/j.1365-2966.2011.18560.x}, \href
  {http://adsabs.harvard.edu/abs/2011MNRAS.414.2923K} {414, 2923}

\bibitem[\protect\citeauthoryear{{Kroupa}, {Tout}  \& {Gilmore}}{{Kroupa}
  et~al.}{1993}]{Kroupa_1993}
{Kroupa} P.,  {Tout} C.~A.,   {Gilmore} G.,  1993, \mnras, \href
  {http://adsabs.harvard.edu/abs/1993MNRAS.262..545K} {262, 545}

\bibitem[\protect\citeauthoryear{{Lange} et~al.,}{{Lange}
  et~al.}{2015}]{Lange_2015}
{Lange} R.,  et~al., 2015, \mn@doi [\mnras] {10.1093/mnras/stu2467}, \href
  {http://adsabs.harvard.edu/abs/2015MNRAS.447.2603L} {447, 2603}

\bibitem[\protect\citeauthoryear{{Lee} et~al.,}{{Lee} et~al.}{2008}]{Lee_2008}
{Lee} M.~G.,  et~al., 2008, \mn@doi [\apj] {10.1086/522956}, \href
  {http://adsabs.harvard.edu/abs/2008ApJ...674..857L} {674, 857}

\bibitem[\protect\citeauthoryear{{Limber} \& {Mathews}}{{Limber} \&
  {Mathews}}{1960}]{Limber_1960}
{Limber} D.~N.,  {Mathews} W.~G.,  1960, \mn@doi [\apj] {10.1086/146928}, \href
  {http://adsabs.harvard.edu/abs/1960ApJ...132..286L} {132, 286}

\bibitem[\protect\citeauthoryear{{Magorrian} \& {Ballantyne}}{{Magorrian} \&
  {Ballantyne}}{2001}]{Magorrian_2001}
{Magorrian} J.,  {Ballantyne} D.,  2001, \mn@doi [\mnras]
  {10.1046/j.1365-8711.2001.04150.x}, \href
  {http://adsabs.harvard.edu/abs/2001MNRAS.322..702M} {322, 702}

\bibitem[\protect\citeauthoryear{{McDermid} et~al.,}{{McDermid}
  et~al.}{2015}]{Mcdermid_2015}
{McDermid} R.~M.,  et~al., 2015, \mn@doi [\mnras] {10.1093/mnras/stv105}, \href
  {http://adsabs.harvard.edu/abs/2015MNRAS.448.3484M} {448, 3484}

\bibitem[\protect\citeauthoryear{{Mendel}, {Proctor}, {Forbes}  \&
  {Brough}}{{Mendel} et~al.}{2008}]{Mendel_2008}
{Mendel} J.~T.,  {Proctor} R.~N.,  {Forbes} D.~A.,   {Brough} S.,  2008,
  \mn@doi [\mnras] {10.1111/j.1365-2966.2008.13514.x}, \href
  {http://adsabs.harvard.edu/abs/2008MNRAS.389..749M} {389, 749}

\bibitem[\protect\citeauthoryear{{Merrett} et~al.,}{{Merrett}
  et~al.}{2003}]{Merrett_2003}
{Merrett} H.~R.,  et~al., 2003, \mn@doi [\mnras]
  {10.1111/j.1365-2966.2003.07367.x}, \href
  {http://adsabs.harvard.edu/abs/2003MNRAS.346L..62M} {346, L62}

\bibitem[\protect\citeauthoryear{{Morganti}, {Gerhard}, {Coccato},
  {Martinez-Valpuesta}  \& {Arnaboldi}}{{Morganti}
  et~al.}{2013}]{Morganti_2013}
{Morganti} L.,  {Gerhard} O.,  {Coccato} L.,  {Martinez-Valpuesta} I.,
  {Arnaboldi} M.,  2013, \mn@doi [\mnras] {10.1093/mnras/stt442}, \href
  {http://adsabs.harvard.edu/abs/2013MNRAS.431.3570M} {431, 3570}

\bibitem[\protect\citeauthoryear{{Murphy}, {Gebhardt}  \& {Adams}}{{Murphy}
  et~al.}{2011}]{Murphy_2011}
{Murphy} J.~D.,  {Gebhardt} K.,   {Adams} J.~J.,  2011, \mn@doi [\apj]
  {10.1088/0004-637X/729/2/129}, \href
  {http://adsabs.harvard.edu/abs/2011ApJ...729..129M} {729, 129}

\bibitem[\protect\citeauthoryear{{Nagino} \& {Matsushita}}{{Nagino} \&
  {Matsushita}}{2009}]{Nagino_2009}
{Nagino} R.,  {Matsushita} K.,  2009, \mn@doi [\aap]
  {10.1051/0004-6361/200810978}, \href
  {http://adsabs.harvard.edu/abs/2009A%26A...501..157N} {501, 157}

\bibitem[\protect\citeauthoryear{{Napolitano} et~al.,}{{Napolitano}
  et~al.}{2005}]{Napolitano_2005}
{Napolitano} N.~R.,  et~al., 2005, \mn@doi [\mnras]
  {10.1111/j.1365-2966.2005.08683.x}, \href
  {http://adsabs.harvard.edu/abs/2005MNRAS.357..691N} {357, 691}

\bibitem[\protect\citeauthoryear{{Napolitano} et~al.,}{{Napolitano}
  et~al.}{2009}]{Napolitano_2009}
{Napolitano} N.~R.,  et~al., 2009, \mn@doi [\mnras]
  {10.1111/j.1365-2966.2008.14053.x}, \href
  {http://adsabs.harvard.edu/abs/2009MNRAS.393..329N} {393, 329}

\bibitem[\protect\citeauthoryear{{Napolitano} et~al.,}{{Napolitano}
  et~al.}{2011}]{Napolitano_2011}
{Napolitano} N.~R.,  et~al., 2011, \mn@doi [\mnras]
  {10.1111/j.1365-2966.2010.17833.x}, \href
  {http://adsabs.harvard.edu/abs/2011MNRAS.411.2035N} {411, 2035}

\bibitem[\protect\citeauthoryear{{Napolitano}, {Pota}, {Romanowsky}, {Forbes},
  {Brodie}  \& {Foster}}{{Napolitano} et~al.}{2014}]{Napolitano_2014}
{Napolitano} N.~R.,  {Pota} V.,  {Romanowsky} A.~J.,  {Forbes} D.~A.,  {Brodie}
  J.~P.,   {Foster} C.,  2014, \mn@doi [\mnras] {10.1093/mnras/stt2484}, \href
  {http://adsabs.harvard.edu/abs/2014MNRAS.439..659N} {439, 659}

\bibitem[\protect\citeauthoryear{{Navarro}, {Frenk}  \& {White}}{{Navarro}
  et~al.}{1996}]{Navarro_1996}
{Navarro} J.~F.,  {Frenk} C.~S.,   {White} S.~D.~M.,  1996, \mn@doi [\apj]
  {10.1086/177173}, \href {http://adsabs.harvard.edu/abs/1996ApJ...462..563N}
  {462, 563}

\bibitem[\protect\citeauthoryear{{Oko{\'n}} \& {Harris}}{{Oko{\'n}} \&
  {Harris}}{2002}]{Okon_2002}
{Oko{\'n}} W.~M.~M.,  {Harris} W.~E.,  2002, \mn@doi [\apj] {10.1086/338386},
  \href {http://adsabs.harvard.edu/abs/2002ApJ...567..294O} {567, 294}

\bibitem[\protect\citeauthoryear{{Oldham} \& {Auger}}{{Oldham} \&
  {Auger}}{2016}]{Oldham_2016}
{Oldham} L.~J.,  {Auger} M.~W.,  2016, preprint, \href
  {http://adsabs.harvard.edu/abs/2016arXiv160101323O} {} (\mn@eprint {arXiv}
  {1601.01323})

\bibitem[\protect\citeauthoryear{{Oser}, {Ostriker}, {Naab}, {Johansson}  \&
  {Burkert}}{{Oser} et~al.}{2010}]{Oser_2010}
{Oser} L.,  {Ostriker} J.~P.,  {Naab} T.,  {Johansson} P.~H.,   {Burkert} A.,
  2010, \mn@doi [\apj] {10.1088/0004-637X/725/2/2312}, \href
  {http://adsabs.harvard.edu/abs/2010ApJ...725.2312O} {725, 2312}

\bibitem[\protect\citeauthoryear{{Oser}, {Naab}, {Ostriker}  \&
  {Johansson}}{{Oser} et~al.}{2012}]{Oser_2012}
{Oser} L.,  {Naab} T.,  {Ostriker} J.~P.,   {Johansson} P.~H.,  2012, \mn@doi
  [\apj] {10.1088/0004-637X/744/1/63}, \href
  {http://adsabs.harvard.edu/abs/2012ApJ...744...63O} {744, 63}

\bibitem[\protect\citeauthoryear{{Pastorello}, {Forbes}, {Foster}, {Brodie},
  {Usher}, {Romanowsky}, {Strader}  \& {Arnold}}{{Pastorello}
  et~al.}{2014}]{Pastorello_2014}
{Pastorello} N.,  {Forbes} D.~A.,  {Foster} C.,  {Brodie} J.~P.,  {Usher} C.,
  {Romanowsky} A.~J.,  {Strader} J.,   {Arnold} J.~A.,  2014, \mn@doi [\mnras]
  {10.1093/mnras/stu937}, \href
  {http://adsabs.harvard.edu/abs/2014MNRAS.442.1003P} {442, 1003}

\bibitem[\protect\citeauthoryear{{Pinkney}, {Roettiger}, {Burns}  \&
  {Bird}}{{Pinkney} et~al.}{1996}]{Pinkney_1996}
{Pinkney} J.,  {Roettiger} K.,  {Burns} J.~O.,   {Bird} C.~M.,  1996, \mn@doi
  [\apjs] {10.1086/192290}, \href
  {http://adsabs.harvard.edu/abs/1996ApJS..104....1P} {104, 1}

\bibitem[\protect\citeauthoryear{{Pota} et~al.,}{{Pota}
  et~al.}{2013}]{Pota_2013}
{Pota} V.,  et~al., 2013, \mn@doi [\mnras] {10.1093/mnras/sts029}, \href
  {http://adsabs.harvard.edu/abs/2013MNRAS.428..389P} {428, 389}

\bibitem[\protect\citeauthoryear{{Pota} et~al.,}{{Pota}
  et~al.}{2015}]{Pota_2015}
{Pota} V.,  et~al., 2015, \mn@doi [\mnras] {10.1093/mnras/stv677}, \href
  {http://adsabs.harvard.edu/abs/2015MNRAS.450.1962P} {450, 1962}

\bibitem[\protect\citeauthoryear{{Puzia} et~al.,}{{Puzia}
  et~al.}{2004}]{Puzia_2004}
{Puzia} T.~H.,  et~al., 2004, \mn@doi [\aap] {10.1051/0004-6361:20031448},
  \href {http://adsabs.harvard.edu/abs/2004A%26A...415..123P} {415, 123}

\bibitem[\protect\citeauthoryear{{Remus}, {Burkert}, {Dolag}, {Johansson},
  {Naab}, {Oser}  \& {Thomas}}{{Remus} et~al.}{2013}]{Remus_2013}
{Remus} R.-S.,  {Burkert} A.,  {Dolag} K.,  {Johansson} P.~H.,  {Naab} T.,
  {Oser} L.,   {Thomas} J.,  2013, \mn@doi [\apj] {10.1088/0004-637X/766/2/71},
  \href {http://adsabs.harvard.edu/abs/2013ApJ...766...71R} {766, 71}

\bibitem[\protect\citeauthoryear{{Rhode}, {Zepf}, {Kundu}  \& {Larner}}{{Rhode}
  et~al.}{2007}]{Rhode_2007}
{Rhode} K.~L.,  {Zepf} S.~E.,  {Kundu} A.,   {Larner} A.~N.,  2007, \mn@doi
  [\aj] {10.1086/521397}, \href
  {http://adsabs.harvard.edu/abs/2007AJ....134.1403R} {134, 1403}

\bibitem[\protect\citeauthoryear{{Robotham} \& {Obreschkow}}{{Robotham} \&
  {Obreschkow}}{2015}]{Robotham_2015}
{Robotham} A.~S.~G.,  {Obreschkow} D.,  2015, \mn@doi [\pasa]
  {10.1017/pasa.2015.33}, \href
  {http://adsabs.harvard.edu/abs/2015PASA...32...33R} {32, e033}

\bibitem[\protect\citeauthoryear{{Romanowsky}, {Douglas}, {Arnaboldi},
  {Kuijken}, {Merrifield}, {Napolitano}, {Capaccioli}  \&
  {Freeman}}{{Romanowsky} et~al.}{2003}]{Romanowsky_2003}
{Romanowsky} A.~J.,  {Douglas} N.~G.,  {Arnaboldi} M.,  {Kuijken} K.,
  {Merrifield} M.~R.,  {Napolitano} N.~R.,  {Capaccioli} M.,   {Freeman} K.~C.,
   2003, \mn@doi [Science] {10.1126/science.1087441}, \href
  {http://adsabs.harvard.edu/abs/2003Sci...301.1696R} {301, 1696}

\bibitem[\protect\citeauthoryear{{Romanowsky}, {Strader}, {Brodie}, {Mihos},
  {Spitler}, {Forbes}, {Foster}  \& {Arnold}}{{Romanowsky}
  et~al.}{2012}]{Romanowsky_2012}
{Romanowsky} A.~J.,  {Strader} J.,  {Brodie} J.~P.,  {Mihos} J.~C.,  {Spitler}
  L.~R.,  {Forbes} D.~A.,  {Foster} C.,   {Arnold} J.~A.,  2012, \mn@doi [\apj]
  {10.1088/0004-637X/748/1/29}, \href
  {http://adsabs.harvard.edu/abs/2012ApJ...748...29R} {748, 29}

\bibitem[\protect\citeauthoryear{{Salpeter}}{{Salpeter}}{1955}]{Salpeter_1955}
{Salpeter} E.~E.,  1955, \mn@doi [\apj] {10.1086/145971}, \href
  {http://adsabs.harvard.edu/abs/1955ApJ...121..161S} {121, 161}

\bibitem[\protect\citeauthoryear{{Schlegel}, {Finkbeiner}  \&
  {Davis}}{{Schlegel} et~al.}{1998}]{Schlegel_1998}
{Schlegel} D.~J.,  {Finkbeiner} D.~P.,   {Davis} M.,  1998, \mn@doi [\apj]
  {10.1086/305772}, \href {http://adsabs.harvard.edu/abs/1998ApJ...500..525S}
  {500, 525}

\bibitem[\protect\citeauthoryear{{Schwarzschild}}{{Schwarzschild}}{1954}]{Schwarzschild_1954}
{Schwarzschild} M.,  1954, \mn@doi [\aj] {10.1086/107013}, \href
  {http://adsabs.harvard.edu/abs/1954AJ.....59..273S} {59, 273}

\bibitem[\protect\citeauthoryear{{Scott}, {Graham}  \& {Schombert}}{{Scott}
  et~al.}{2013}]{Scott_2013}
{Scott} N.,  {Graham} A.~W.,   {Schombert} J.,  2013, \mn@doi [\apj]
  {10.1088/0004-637X/768/1/76}, \href
  {http://adsabs.harvard.edu/abs/2013ApJ...768...76S} {768, 76}

\bibitem[\protect\citeauthoryear{{S\'ersic}}{{S\'ersic}}{1968}]{Sersic_1968}
{S\'ersic} J.~L.,  1968, {Atlas de galaxias australes}

\bibitem[\protect\citeauthoryear{{Shankar}, {Lapi}, {Salucci}, {De Zotti}  \&
  {Danese}}{{Shankar} et~al.}{2006}]{Shankar_2006}
{Shankar} F.,  {Lapi} A.,  {Salucci} P.,  {De Zotti} G.,   {Danese} L.,  2006,
  \mn@doi [\apj] {10.1086/502794}, \href
  {http://adsabs.harvard.edu/abs/2006ApJ...643...14S} {643, 14}

\bibitem[\protect\citeauthoryear{{Shen} \& {Gebhardt}}{{Shen} \&
  {Gebhardt}}{2010}]{Shen_2010}
{Shen} J.,  {Gebhardt} K.,  2010, \mn@doi [\apj] {10.1088/0004-637X/711/1/484},
  \href {http://adsabs.harvard.edu/abs/2010ApJ...711..484S} {711, 484}

\bibitem[\protect\citeauthoryear{{Sikkema}, {Peletier}, {Carter}, {Valentijn}
  \& {Balcells}}{{Sikkema} et~al.}{2006}]{Sikkema_2006}
{Sikkema} G.,  {Peletier} R.~F.,  {Carter} D.,  {Valentijn} E.~A.,   {Balcells}
  M.,  2006, \mn@doi [\aap] {10.1051/0004-6361:20054606}, \href
  {http://adsabs.harvard.edu/abs/2006A%26A...458...53S} {458, 53}

\bibitem[\protect\citeauthoryear{{Sparre} et~al.,}{{Sparre}
  et~al.}{2015}]{Sparre_2015}
{Sparre} M.,  et~al., 2015, \mn@doi [\mnras] {10.1093/mnras/stu2713}, \href
  {http://adsabs.harvard.edu/abs/2015MNRAS.447.3548S} {447, 3548}

\bibitem[\protect\citeauthoryear{{Spergel} et~al.,}{{Spergel}
  et~al.}{2007}]{Spergel_2007}
{Spergel} D.~N.,  et~al., 2007, \mn@doi [\apjs] {10.1086/513700}, \href
  {http://adsabs.harvard.edu/abs/2007ApJS..170..377S} {170, 377}

\bibitem[\protect\citeauthoryear{{Spiniello}, {Trager}, {Koopmans}  \&
  {Conroy}}{{Spiniello} et~al.}{2014}]{Spiniello_2014}
{Spiniello} C.,  {Trager} S.,  {Koopmans} L.~V.~E.,   {Conroy} C.,  2014,
  \mn@doi [\mnras] {10.1093/mnras/stt2282}, \href
  {http://adsabs.harvard.edu/abs/2014MNRAS.438.1483S} {438, 1483}

\bibitem[\protect\citeauthoryear{{Strader} et~al.,}{{Strader}
  et~al.}{2011}]{Strader_2011}
{Strader} J.,  et~al., 2011, \mn@doi [\apjs] {10.1088/0067-0049/197/2/33},
  \href {http://adsabs.harvard.edu/abs/2011ApJS..197...33S} {197, 33}

\bibitem[\protect\citeauthoryear{{Su}, {Gu}, {White}  \& {Irwin}}{{Su}
  et~al.}{2014}]{Su_2014}
{Su} Y.,  {Gu} L.,  {White} III R.~E.,   {Irwin} J.,  2014, \mn@doi [\apj]
  {10.1088/0004-637X/786/2/152}, \href
  {http://adsabs.harvard.edu/abs/2014ApJ...786..152S} {786, 152}

\bibitem[\protect\citeauthoryear{{Taylor}}{{Taylor}}{2005}]{Taylor_2005}
{Taylor} M.~B.,  2005, in {Shopbell} P.,  {Britton} M.,   {Ebert} R.,  eds,
  Astronomical Society of the Pacific Conference Series Vol. 347, Astronomical
  Data Analysis Software and Systems XIV. p.~29

\bibitem[\protect\citeauthoryear{{Terzi{\'c}} \& {Graham}}{{Terzi{\'c}} \&
  {Graham}}{2005}]{Terzic_2005}
{Terzi{\'c}} B.,  {Graham} A.~W.,  2005, \mn@doi [\mnras]
  {10.1111/j.1365-2966.2005.09269.x}, \href
  {http://adsabs.harvard.edu/abs/2005MNRAS.362..197T} {362, 197}

\bibitem[\protect\citeauthoryear{{Thomas}, {Saglia}, {Bender}, {Thomas},
  {Gebhardt}, {Magorrian}, {Corsini}  \& {Wegner}}{{Thomas}
  et~al.}{2009}]{Thomas_2009}
{Thomas} J.,  {Saglia} R.~P.,  {Bender} R.,  {Thomas} D.,  {Gebhardt} K.,
  {Magorrian} J.,  {Corsini} E.~M.,   {Wegner} G.,  2009, \mn@doi [\apj]
  {10.1088/0004-637X/691/1/770}, \href
  {http://adsabs.harvard.edu/abs/2009ApJ...691..770T} {691, 770}

\bibitem[\protect\citeauthoryear{{Thomas} et~al.,}{{Thomas}
  et~al.}{2011}]{Thomas_2011}
{Thomas} J.,  et~al., 2011, \mn@doi [\mnras]
  {10.1111/j.1365-2966.2011.18725.x}, \href
  {http://adsabs.harvard.edu/abs/2011MNRAS.415..545T} {415, 545}

\bibitem[\protect\citeauthoryear{{Tonry}, {Dressler}, {Blakeslee}, {Ajhar},
  {Fletcher}, {Luppino}, {Metzger}  \& {Moore}}{{Tonry}
  et~al.}{2001}]{Tonry_2001}
{Tonry} J.~L.,  {Dressler} A.,  {Blakeslee} J.~P.,  {Ajhar} E.~A.,  {Fletcher}
  A.~B.,  {Luppino} G.~A.,  {Metzger} M.~R.,   {Moore} C.~B.,  2001, \mn@doi
  [\apj] {10.1086/318301}, \href
  {http://cdsads.u-strasbg.fr/abs/2001ApJ...546..681T} {546, 681}

\bibitem[\protect\citeauthoryear{{Tortora}, {La Barbera}, {Napolitano},
  {Romanowsky}, {Ferreras}  \& {de Carvalho}}{{Tortora}
  et~al.}{2014}]{Tortora_2014}
{Tortora} C.,  {La Barbera} F.,  {Napolitano} N.~R.,  {Romanowsky} A.~J.,
  {Ferreras} I.,   {de Carvalho} R.~R.,  2014, \mn@doi [\mnras]
  {10.1093/mnras/stu1616}, \href
  {http://adsabs.harvard.edu/abs/2014MNRAS.445..115T} {445, 115}

\bibitem[\protect\citeauthoryear{{Trujillo-Gomez}, {Klypin}, {Primack}  \&
  {Romanowsky}}{{Trujillo-Gomez} et~al.}{2011}]{Trujillo_2011}
{Trujillo-Gomez} S.,  {Klypin} A.,  {Primack} J.,   {Romanowsky} A.~J.,  2011,
  \mn@doi [\apj] {10.1088/0004-637X/742/1/16}, \href
  {http://adsabs.harvard.edu/abs/2011ApJ...742...16T} {742, 16}

\bibitem[\protect\citeauthoryear{{Watkins}, {Evans}  \& {An}}{{Watkins}
  et~al.}{2010}]{Watkins_2010}
{Watkins} L.~L.,  {Evans} N.~W.,   {An} J.~H.,  2010, \mn@doi [\mnras]
  {10.1111/j.1365-2966.2010.16708.x}, \href
  {http://adsabs.harvard.edu/abs/2010MNRAS.406..264W} {406, 264}

\bibitem[\protect\citeauthoryear{{Watkins}, {van de Ven}, {den Brok}  \& {van
  den Bosch}}{{Watkins} et~al.}{2013}]{Watkins_2013}
{Watkins} L.~L.,  {van de Ven} G.,  {den Brok} M.,   {van den Bosch} R.~C.~E.,
  2013, \mn@doi [\mnras] {10.1093/mnras/stt1756}, \href
  {http://adsabs.harvard.edu/abs/2013MNRAS.436.2598W} {436, 2598}

\bibitem[\protect\citeauthoryear{{Weijmans} et~al.,}{{Weijmans}
  et~al.}{2009}]{Weijmans_2009}
{Weijmans} A.-M.,  et~al., 2009, \mn@doi [\mnras]
  {10.1111/j.1365-2966.2009.15134.x}, \href
  {http://adsabs.harvard.edu/abs/2009MNRAS.398..561W} {398, 561}

\bibitem[\protect\citeauthoryear{{White} \& {Rees}}{{White} \&
  {Rees}}{1978}]{White_1978}
{White} S.~D.~M.,  {Rees} M.~J.,  1978, \mn@doi [\mnras]
  {10.1093/mnras/183.3.341}, \href
  {http://adsabs.harvard.edu/abs/1978MNRAS.183..341W} {183, 341}

\bibitem[\protect\citeauthoryear{{Wolf}, {Martinez}, {Bullock}, {Kaplinghat},
  {Geha}, {Mu{\~n}oz}, {Simon}  \& {Avedo}}{{Wolf} et~al.}{2010}]{Wolf_2010}
{Wolf} J.,  {Martinez} G.~D.,  {Bullock} J.~S.,  {Kaplinghat} M.,  {Geha} M.,
  {Mu{\~n}oz} R.~R.,  {Simon} J.~D.,   {Avedo} F.~F.,  2010, \mn@doi [\mnras]
  {10.1111/j.1365-2966.2010.16753.x}, \href
  {http://adsabs.harvard.edu/abs/2010MNRAS.406.1220W} {406, 1220}

\bibitem[\protect\citeauthoryear{{Wu}, {Gerhard}, {Naab}, {Oser},
  {Martinez-Valpuesta}, {Hilz}, {Churazov}  \& {Lyskova}}{{Wu}
  et~al.}{2014}]{Wu_2014}
{Wu} X.,  {Gerhard} O.,  {Naab} T.,  {Oser} L.,  {Martinez-Valpuesta} I.,
  {Hilz} M.,  {Churazov} E.,   {Lyskova} N.,  2014, \mn@doi [\mnras]
  {10.1093/mnras/stt2415}, \href
  {http://adsabs.harvard.edu/abs/2014MNRAS.438.2701W} {438, 2701}

\bibitem[\protect\citeauthoryear{{Yencho}, {Johnston}, {Bullock}  \&
  {Rhode}}{{Yencho} et~al.}{2006}]{Yencho_2006}
{Yencho} B.~M.,  {Johnston} K.~V.,  {Bullock} J.~S.,   {Rhode} K.~L.,  2006,
  \mn@doi [\apj] {10.1086/502619}, \href
  {http://adsabs.harvard.edu/abs/2006ApJ...643..154Y} {643, 154}

\bibitem[\protect\citeauthoryear{{Zhu} et~al.,}{{Zhu} et~al.}{2014}]{Zhu_2014}
{Zhu} L.,  et~al., 2014, \mn@doi [\apj] {10.1088/0004-637X/792/1/59}, \href
  {http://adsabs.harvard.edu/abs/2014ApJ...792...59Z} {792, 59}

\bibitem[\protect\citeauthoryear{{Zwicky}}{{Zwicky}}{1937}]{Zwicky_1937}
{Zwicky} F.,  1937, \mn@doi [\apj] {10.1086/143864}, \href
  {http://adsabs.harvard.edu/abs/1937ApJ....86..217Z} {86, 217}

\bibitem[\protect\citeauthoryear{{den Heijer} et~al.,}{{den Heijer}
  et~al.}{2015}]{denHeijer_2015}
{den Heijer} M.,  et~al., 2015, \mn@doi [\aap] {10.1051/0004-6361/201526879},
  \href {http://adsabs.harvard.edu/abs/2015A%26A...581A..98D} {581, A98}

\makeatother
\end{thebibliography}

\appendix
\newpage

\section{Mass estimates at 8~$\re$}
\begin{table}
\centering
{\small \caption{Mass estimates and dark matter fraction ($f_{\rm DM}$) within 8~$\re$ assuming isotropy. Columns 2--4 show the rotationally--supported, pressure--supported (obtained without subtracting $V_{\rm rot}$ from $\vlos$) and total dynamical mass within 8~$\re$, respectively.}\label{tab:appendix_1}}
\begin{tabular}{lcccl}
\hline\hline  
Galaxy & $M_{\rm rot}(<8{\re})$ & $M^{\prime}_{\rm p}(<8{\re})$ & $M_{\rm tot}(<8{\re})$ & $f_{\rm DM} (<8{\re})$ \\
$\rm [NGC]$ & $[10^{10} \Msun]$ & $[10^{11} \Msun]$ & $[10^{11} \Msun]$ &  \\ 
\hline
  720 & 3.2$\pm$2.0 & 5.7$\pm$1.2 & 6.0$\pm$0.6 & 0.65$\pm$0.06\\
  821 & 3.2$\pm$1.8 & 5.5$\pm$1.0 & 5.8$\pm$0.6 & 0.85$\pm$0.01\\
  1023 & 3.8$\pm$1.1 & 2.1$\pm$0.4 & 2.4$\pm$0.2 & 0.63$\pm$0.02\\
  1407 & 0.1$\pm$0.3 & 20.0$\pm$1.6 & 20.0$\pm$0.9 & 0.82$\pm$0.01\\
  2768 & 7.2$\pm$2.4 & 9.8$\pm$1.6 & 10.5$\pm$0.8 & 0.83$\pm$0.01\\
  3115 & 5.2$\pm$0.9 & 1.6$\pm$0.3 & 2.1$\pm$0.2 & 0.56$\pm$0.02\\
  3377 & 0.2$\pm$0.1 & 0.8$\pm$0.1 & 0.9$\pm$0.1 & 0.69$\pm$0.02\\
  3608 & 0.6$\pm$0.6 & 3.7$\pm$1.1 & 3.7$\pm$0.6 & 0.83$\pm$0.02\\
  4278 & 0.3$\pm$0.2 & 4.3$\pm$0.4 & 4.4$\pm$0.3 & 0.83$\pm$0.01\\
  4365 & 1.6$\pm$1.1 & 17.0$\pm$1.6 & 17.2$\pm$0.9 & 0.84$\pm$0.01\\
  4374 & 14.0$\pm$8.8 & 22.0$\pm$5.2 & 23.4$\pm$3.1 & 0.89$\pm$0.01\\
  4473 & 0.4$\pm$0.3 & 2.0$\pm$0.3 & 2.0$\pm$0.2 & 0.64$\pm$0.02\\
  4486 & 2.6$\pm$1.2 & 32.4$\pm$2.1 & 32.7$\pm$1.2 & 0.90$\pm$0.01\\
  4494 & 1.9$\pm$0.5 & 1.3$\pm$0.2 & 1.5$\pm$0.1 & 0.34$\pm$0.04\\
  4526 & 4.7$\pm$1.9 & 3.0$\pm$0.5 & 3.5$\pm$0.3 & 0.54$\pm$0.03\\
  4649 & 6.1$\pm$1.5 & 15.5$\pm$1.1 & 16.1$\pm$0.7 & 0.79$\pm$0.01\\
  3607 & 0.5$\pm$0.6 & 5.4$\pm$1.4 & 5.5$\pm$0.8 & 0.66$\pm$0.03\\
\hline
\end{tabular}
\end{table}

\begin{table*}
\footnotesize
\captionsetup{width=.96\linewidth}
{\small
\caption{Mass estimates ($M_{\rm tot}$) and dark matter fractions ($f_{\rm DM}$) within $5{\re}$ (columns 3 and 4) and $R_{\rm max}$ (columns 5 and 6), respectively, assuming different anisotropy, but with $\alpha\equiv0$. These $M_{\rm tot}$ and $f_{\rm DM}$ are shown in Figure \ref{fig:predict_fdm2} while $R_{\rm max}$ can be found in Table \ref{tab:mass_summary}.} \label{tab:appendix_2}}
\begin{tabular}{lcccccccl}
\hline\hline  
Galaxy & $\beta$ & $M_{\rm tot}(<5{\re})$ & $f_{\rm DM} (<5{\re})$ & $M_{\rm tot}(<R_{\rm max})$ & $f_{\rm DM}(<R_{\rm max})$ \\
$\rm [NGC]$ &  & $[10^{11} \Msun]$ &   &$[10^{11} \Msun]$ & 	\\ 
(1) & (2) &	(3) & (4) & (5) & (6) \\
\hline
   720  &    0 & 3.6 $\pm$0.7 & 0.46$\pm$0.17 & 13.5 $\pm$2.2 & 0.84$\pm$0.05 \\    
       &  0.5 & 3.4 $\pm$0.7 & 0.43$\pm$0.17 & 12.7 $\pm$2.1 & 0.83$\pm$0.04 \\    
       & -0.5 & 3.8 $\pm$0.7 & 0.48$\pm$0.15 & 13.9 $\pm$2.3 & 0.84$\pm$0.04 \\    
  821  &    0 & 4.5 $\pm$0.8 & 0.82$\pm$0.05 & 7.1  $\pm$1.2 & 0.88$\pm$0.03 \\    
       &  0.5 & 4.4 $\pm$0.8 & 0.82$\pm$0.05 & 7.0  $\pm$1.1 & 0.87$\pm$0.03 \\    
       & -0.5 & 4.6 $\pm$0.8 & 0.82$\pm$0.05 & 7.1  $\pm$1.2 & 0.88$\pm$0.03 \\    
  1023 &    0 & 1.8 $\pm$0.3 & 0.52$\pm$0.12 & 5.2  $\pm$0.7 & 0.82$\pm$0.04 \\    
       &  0.5 & 1.8 $\pm$0.2 & 0.51$\pm$0.11 & 5.2  $\pm$0.7 & 0.82$\pm$0.04 \\    
       & -0.5 & 1.8 $\pm$0.3 & 0.52$\pm$0.12 & 5.3  $\pm$0.7 & 0.82$\pm$0.04 \\    
  1400 &    0 & 2.4 $\pm$0.6 & 0.51$\pm$0.28 & 8.7  $\pm$1.5 & 0.85$\pm$0.04 \\    
       &  0.5 & 2.3 $\pm$0.5 & 0.49$\pm$0.18 & 8.4  $\pm$1.5 & 0.84$\pm$0.04 \\    
       & -0.5 & 2.4 $\pm$0.6 & 0.51$\pm$0.18 & 8.9  $\pm$1.5 & 0.85$\pm$0.04 \\    
  1407 &    0 & 11.5$\pm$1.1 & 0.71$\pm$0.07 & 35.4 $\pm$2.7 & 0.89$\pm$0.02 \\    
       &  0.5 & 10.3$\pm$1.0 & 0.67$\pm$0.07 & 31.8 $\pm$2.4 & 0.88$\pm$0.02 \\    
       & -0.5 & 12.0$\pm$1.1 & 0.72$\pm$0.06 & 37.2 $\pm$2.8 & 0.9 $\pm$0.02 \\    
  2768 &    0 & 7.6 $\pm$1.1 & 0.78$\pm$0.07 & 15.9 $\pm$2.1 & 0.89$\pm$0.03 \\    
       &  0.5 & 7.2 $\pm$1.0 & 0.77$\pm$0.07 & 15.1 $\pm$2.0 & 0.88$\pm$0.03 \\    
       & -0.5 & 7.8 $\pm$1.1 & 0.79$\pm$0.06 & 16.3 $\pm$2.2 & 0.89$\pm$0.03 \\    
  3115 &    0 & 2.0 $\pm$0.3 & 0.56$\pm$0.08 & 6.5  $\pm$0.7 & 0.86$\pm$0.02 \\    
       &  0.5 & 2.0 $\pm$0.3 & 0.56$\pm$0.08 & 6.4  $\pm$0.7 & 0.85$\pm$0.02 \\    
       & -0.5 & 2.0 $\pm$0.3 & 0.57$\pm$0.08 & 6.6  $\pm$0.7 & 0.86$\pm$0.02 \\    
  3377 &    0 & 0.8 $\pm$0.1 & 0.67$\pm$0.07 & 2.0  $\pm$0.3 & 0.86$\pm$0.02 \\    
       &  0.5 & 0.8 $\pm$0.1 & 0.68$\pm$0.07 & 2.1  $\pm$0.3 & 0.87$\pm$0.02 \\    
       & -0.5 & 0.8 $\pm$0.1 & 0.67$\pm$0.07 & 2.0  $\pm$0.3 & 0.86$\pm$0.02 \\    
  3608 &    0 & 3.4 $\pm$1.1 & 0.83$\pm$0.45 & 5.3  $\pm$1.5 & 0.88$\pm$0.06 \\    
       &  0.5 & 3.4 $\pm$1.2 & 0.83$\pm$0.56 & 5.3  $\pm$1.4 & 0.88$\pm$0.05 \\    
       & -0.5 & 3.4 $\pm$1.1 & 0.83$\pm$0.64 & 5.3  $\pm$1.5 & 0.88$\pm$0.12 \\    
  4278 &    0 & 2.9 $\pm$0.4 & 0.76$\pm$0.06 & 7.9  $\pm$0.7 & 0.9 $\pm$0.02 \\    
       &  0.5 & 2.9 $\pm$0.4 & 0.76$\pm$0.06 & 7.8  $\pm$0.7 & 0.9 $\pm$0.02 \\    
       & -0.5 & 2.9 $\pm$0.4 & 0.76$\pm$0.06 & 7.9  $\pm$0.7 & 0.9 $\pm$0.02 \\    
  4365 &    0 & 12.1$\pm$1.3 & 0.78$\pm$0.05 & 29.7 $\pm$2.5 & 0.9 $\pm$0.02 \\    
       &  0.5 & 11.1$\pm$1.2 & 0.76$\pm$0.05 & 27.3 $\pm$2.4 & 0.89$\pm$0.02 \\    
       & -0.5 & 12.6$\pm$1.4 & 0.79$\pm$0.05 & 31.0 $\pm$2.6 & 0.91$\pm$0.02 \\    
  4374 &    0 & 14.1$\pm$3.3 & 0.82$\pm$0.06 & 22.7 $\pm$5.1 & 0.88$\pm$0.03 \\    
       &  0.5 & 13.1$\pm$3.2 & 0.81$\pm$0.07 & 21.1 $\pm$4.5 & 0.87$\pm$0.04 \\    
       & -0.5 & 14.6$\pm$3.5 & 0.83$\pm$0.07 & 23.6 $\pm$5.2 & 0.88$\pm$0.04 \\    
  4473 &    0 & 1.5 $\pm$0.3 & 0.55$\pm$0.13 & 4.5  $\pm$0.6 & 0.84$\pm$0.03 \\    
       &  0.5 & 1.5 $\pm$0.3 & 0.55$\pm$0.14 & 4.5  $\pm$0.6 & 0.84$\pm$0.03 \\    
       & -0.5 & 1.5 $\pm$0.3 & 0.55$\pm$0.12 & 4.6  $\pm$0.6 & 0.84$\pm$0.03 \\    
  4486 &    0 & 24.1$\pm$1.7 & 0.88$\pm$0.01 & 141.0$\pm$8.2 & 0.98$\pm$0.0  \\    
       &  0.5 & 22.0$\pm$1.6 & 0.87$\pm$0.01 & 128.0$\pm$6.9 & 0.97$\pm$0.0  \\    
       & -0.5 & 25.2$\pm$1.8 & 0.88$\pm$0.01 & 148.0$\pm$8.2 & 0.98$\pm$0.0  \\    
  4494 &    0 & 1.7 $\pm$0.3 & 0.45$\pm$0.11 & 2.5  $\pm$0.4 & 0.6 $\pm$0.08 \\    
       &  0.5 & 1.7 $\pm$0.2 & 0.44$\pm$0.11 & 2.4  $\pm$0.4 & 0.59$\pm$0.08 \\    
       & -0.5 & 1.7 $\pm$0.2 & 0.46$\pm$0.11 & 2.5  $\pm$0.3 & 0.6 $\pm$0.08 \\    
  4526 &    0 & 3.5 $\pm$0.6 & 0.57$\pm$0.12 & 7.9  $\pm$1.2 & 0.79$\pm$0.05 \\    
       &  0.5 & 3.4 $\pm$0.6 & 0.55$\pm$0.13 & 7.5  $\pm$1.1 & 0.78$\pm$0.06 \\    
       & -0.5 & 3.6 $\pm$0.6 & 0.58$\pm$0.13 & 8.1  $\pm$1.2 & 0.79$\pm$0.05 \\    
  4564 &    0 & 1.3 $\pm$0.3 & 0.73$\pm$0.1  & 1.8  $\pm$0.4 & 0.79$\pm$0.07 \\    
       &  0.5 & 1.3 $\pm$0.3 & 0.74$\pm$0.1  & 1.8  $\pm$0.4 & 0.79$\pm$0.07 \\    
       & -0.5 & 1.3 $\pm$0.3 & 0.73$\pm$0.15 & 1.7  $\pm$0.4 & 0.79$\pm$0.07 \\    
  4649 &    0 & 11.2$\pm$0.9 & 0.72$\pm$0.05 & 52.7 $\pm$3.5 & 0.93$\pm$0.01 \\    
       &  0.5 & 10.2$\pm$0.8 & 0.69$\pm$0.06 & 48.0 $\pm$3.2 & 0.93$\pm$0.01 \\    
       & -0.5 & 11.7$\pm$1.0 & 0.73$\pm$0.05 & 55.1 $\pm$3.5 & 0.94$\pm$0.01 \\    
  4697 &    0 & 9.4 $\pm$2.5 & 0.9 $\pm$0.05 & -- & -- \\    
       &  0.5 & 9.2 $\pm$2.4 & 0.9 $\pm$0.04 & -- & -- \\    
       & -0.5 & 9.5 $\pm$2.4 & 0.9 $\pm$0.04 & -- & -- \\    
  5846 &    0 & 12.4$\pm$1.6 & 0.83$\pm$0.05 & 33.4 $\pm$3.6 & 0.93$\pm$0.02 \\    
       &  0.5 & 11.5$\pm$1.5 & 0.81$\pm$0.05 & 31.0 $\pm$3.4 & 0.92$\pm$0.02 \\    
       & -0.5 & 12.8$\pm$1.7 & 0.83$\pm$0.04 & 34.6 $\pm$3.7 & 0.93$\pm$0.02 \\    
  7457 &    0 & 1.7 $\pm$0.4 & 0.89$\pm$0.04 & 1.8  $\pm$0.4 & 0.9 $\pm$0.04 \\    
       &  0.5 & 1.7 $\pm$0.4 & 0.9 $\pm$0.04 & 1.9  $\pm$0.4 & 0.91$\pm$0.04 \\    
       & -0.5 & 1.6 $\pm$0.4 & 0.89$\pm$0.04 & 1.8  $\pm$0.4 & 0.9 $\pm$0.04 \\    
  3607 &    0 & 2.6 $\pm$0.7 & 0.33$\pm$0.34 & 11.1 $\pm$2.7 & 0.82$\pm$0.06 \\    
       &  0.5 & 2.5 $\pm$0.7 & 0.3 $\pm$0.44 & 10.5 $\pm$2.5 & 0.81$\pm$0.13 \\    
       & -0.5 & 2.7 $\pm$0.7 & 0.35$\pm$0.28 & 11.4 $\pm$2.8 & 0.83$\pm$0.07 \\    
  5866 &    0 & 1.6 $\pm$0.6 & 0.46$\pm$0.45 & -- & -- \\    
       &  0.5 & 1.5 $\pm$0.6 & 0.44$\pm$0.45 & -- & -- \\    
       & -0.5 & 1.6 $\pm$0.6 & 0.46$\pm$0.45 & -- & -- \\    
\hline\end{tabular}
\end{table*}

\begin{table*}
\footnotesize
\captionsetup{width=.96\linewidth}
{\small
\caption{Mass estimates ($M_{\rm tot}$) and dark matter fractions ($f_{\rm DM}$) within $5{\re}$ and $R_{\rm max}$ assuming different anisotropy, obtained with stellar $M/L$ corresponding to a Salpeter IMF from \citet{Cappellari_2013, Cappellari_2013b} (see Section \ref{subs:DMfrac} for details). Columns 3--6 show $M_{\rm tot}$ and $f_{\rm DM}$ obtained by allowing $\alpha$ to vary while in columns 7--10, $\alpha\equiv0$, $R_{\rm max}$ can be found in Table \ref{tab:mass_summary}. \label{tab:appendix_3}}}
\begin{tabular}{lccccccccl}
\hline
 Galaxy & $\beta$  & $M_{\rm tot}(<5{\re})$ & $f_{\rm DM} (<5{\re})$ & $M_{\rm tot}(<R_{\rm max})$ & $f_{\rm DM}(<R_{\rm max})$ & $M_{\rm tot}(<5{\re})$ & $f_{\rm DM} (<5{\re})$ & $M_{\rm tot}(<R_{\rm max})$ & $f_{\rm DM}(<R_{\rm max})$ \\
$\rm [NGC]$ & &$[10^{11} \Msun]$ &   &$[10^{11} \Msun]$ &   &$[10^{11} \Msun]$ &   &$[10^{11} \Msun]$ &   \\
(1) & (2) &	(3) & (4) & (5) & (6) & (7) & (8) & (9) & (10) \\
\hline
  720  &    0  & 3.6 $\pm$ 0.7 & 0.28$\pm$0.41 & 13.5 $\pm$2.3  & 0.78$\pm$0.11 & 3.6  $\pm$0.7 & 0.28$\pm$0.41 & 13.5 $\pm$2.4 & 0.78$\pm$0.11 \\       
       &  0.5  & 3.4 $\pm$ 0.7 & 0.23$\pm$0.43 & 12.7 $\pm$2.2  & 0.77$\pm$0.12 & 3.4  $\pm$0.7 & 0.23$\pm$0.42 & 12.7 $\pm$2.1 & 0.77$\pm$0.11 \\       
       & -0.5  & 3.8 $\pm$ 0.7 & 0.3 $\pm$0.39 & 13.9 $\pm$2.4  & 0.79$\pm$0.11 & 3.8  $\pm$0.8 & 0.3 $\pm$0.4  & 13.9 $\pm$2.4 & 0.79$\pm$0.1  \\       
  821  &    0  & 4.3 $\pm$ 0.8 & 0.77$\pm$0.06 & 6.1  $\pm$1.0  & 0.83$\pm$0.04 & 4.5  $\pm$0.8 & 0.78$\pm$0.06 & 7.1  $\pm$1.2 & 0.85$\pm$0.03 \\       
       &  0.5  & 4.4 $\pm$ 0.8 & 0.78$\pm$0.06 & 6.3  $\pm$1.1  & 0.83$\pm$0.04 & 4.4  $\pm$0.8 & 0.78$\pm$0.06 & 7.0  $\pm$1.2 & 0.85$\pm$0.04 \\       
       & -0.5  & 4.2 $\pm$ 0.8 & 0.77$\pm$0.05 & 6.0  $\pm$1.0  & 0.83$\pm$0.04 & 4.6  $\pm$0.8 & 0.79$\pm$0.05 & 7.1  $\pm$1.2 & 0.85$\pm$0.04 \\       
  1023 &    0  & 1.7 $\pm$ 0.2 & 0.29$\pm$0.15 & 4.2  $\pm$0.5  & 0.7 $\pm$0.05 & 1.8  $\pm$0.3 & 0.34$\pm$0.14 & 5.2  $\pm$0.7 & 0.75$\pm$0.05 \\       
       &  0.5  & 1.7 $\pm$ 0.3 & 0.31$\pm$0.15 & 4.3  $\pm$0.6  & 0.7 $\pm$0.05 & 1.7  $\pm$0.3 & 0.33$\pm$0.14 & 5.1  $\pm$0.7 & 0.75$\pm$0.05 \\       
       & -0.5  & 1.7 $\pm$ 0.2 & 0.29$\pm$0.15 & 4.2  $\pm$0.5  & 0.69$\pm$0.06 & 1.8  $\pm$0.3 & 0.34$\pm$0.13 & 5.3  $\pm$0.7 & 0.75$\pm$0.05 \\       
  1400 &    0  & 2.4 $\pm$ 0.6 & 0.36$\pm$0.41 & 7.7  $\pm$1.4  & 0.78$\pm$0.11 & 2.4  $\pm$0.5 & 0.37$\pm$0.42 & 8.7  $\pm$1.5 & 0.81$\pm$0.1  \\       
       &  0.5  & 2.3 $\pm$ 0.6 & 0.36$\pm$0.42 & 7.7  $\pm$1.3  & 0.78$\pm$0.11 & 2.3  $\pm$0.5 & 0.34$\pm$0.38 & 8.4  $\pm$1.5 & 0.8 $\pm$0.1  \\       
       & -0.5  & 2.4 $\pm$ 0.6 & 0.36$\pm$0.43 & 7.8  $\pm$1.4  & 0.78$\pm$0.11 & 2.4  $\pm$0.6 & 0.38$\pm$0.41 & 8.9  $\pm$1.6 & 0.81$\pm$0.09 \\       
  1407 &    0  & 11.6$\pm$ 1.1 & 0.58$\pm$0.21 & 38.2 $\pm$2.9  & 0.86$\pm$0.07 & 11.5 $\pm$1.1 & 0.58$\pm$0.22 & 35.4 $\pm$2.5 & 0.84$\pm$0.06 \\       
       &  0.5  & 10.0$\pm$ 1.0 & 0.51$\pm$0.25 & 32.8 $\pm$2.3  & 0.83$\pm$0.07 & 10.3 $\pm$1.0 & 0.53$\pm$0.24 & 31.8 $\pm$2.4 & 0.83$\pm$0.08 \\       
       & -0.5  & 12.4$\pm$ 1.3 & 0.61$\pm$0.2  & 41.0 $\pm$3.1  & 0.87$\pm$0.06 & 12.0 $\pm$1.2 & 0.59$\pm$0.22 & 37.2 $\pm$2.7 & 0.85$\pm$0.06 \\       
  2768 &    0  & 7.4 $\pm$ 1.0 & 0.72$\pm$0.06 & 15.0 $\pm$2.0  & 0.85$\pm$0.03 & 7.6  $\pm$1.1 & 0.73$\pm$0.06 & 16.0 $\pm$2.1 & 0.86$\pm$0.03 \\       
       &  0.5  & 7.1 $\pm$ 0.9 & 0.71$\pm$0.06 & 14.5 $\pm$1.8  & 0.84$\pm$0.03 & 7.2  $\pm$0.9 & 0.72$\pm$0.06 & 15.2 $\pm$2.1 & 0.85$\pm$0.03 \\       
       & -0.5  & 7.5 $\pm$ 1.0 & 0.73$\pm$0.05 & 15.3 $\pm$2.0  & 0.85$\pm$0.03 & 7.8  $\pm$1.1 & 0.74$\pm$0.06 & 16.4 $\pm$2.1 & 0.86$\pm$0.03 \\       
  3115 &    0  & 2.0 $\pm$ 0.3 & 0.47$\pm$0.26 & 5.8  $\pm$0.6  & 0.8 $\pm$0.09 & 2.0  $\pm$0.3 & 0.47$\pm$0.26 & 6.5  $\pm$0.7 & 0.82$\pm$0.08 \\       
       &  0.5  & 2.1 $\pm$ 0.3 & 0.48$\pm$0.25 & 5.9  $\pm$0.6  & 0.81$\pm$0.08 & 2.0  $\pm$0.3 & 0.46$\pm$0.25 & 6.4  $\pm$0.7 & 0.82$\pm$0.08 \\       
       & -0.5  & 2.0 $\pm$ 0.3 & 0.46$\pm$0.26 & 5.7  $\pm$0.6  & 0.8 $\pm$0.09 & 2.0  $\pm$0.3 & 0.47$\pm$0.26 & 6.6  $\pm$0.7 & 0.83$\pm$0.08 \\       
  3377 &    0  & 0.6 $\pm$ 0.1 & 0.37$\pm$0.14 & 1.3  $\pm$0.2  & 0.69$\pm$0.06 & 0.8  $\pm$0.1 & 0.49$\pm$0.12 & 2.0  $\pm$0.3 & 0.79$\pm$0.04 \\       
       &  0.5  & 0.7 $\pm$ 0.1 & 0.43$\pm$0.12 & 1.5  $\pm$0.2  & 0.72$\pm$0.05 & 0.8  $\pm$0.1 & 0.5 $\pm$0.11 & 2.0  $\pm$0.3 & 0.79$\pm$0.04 \\       
       & -0.5  & 0.6 $\pm$ 0.1 & 0.34$\pm$0.16 & 1.3  $\pm$0.2  & 0.67$\pm$0.06 & 0.7  $\pm$0.1 & 0.49$\pm$0.11 & 1.9  $\pm$0.3 & 0.79$\pm$0.04 \\       
  3608 &    0  & 3.3 $\pm$ 1.0 & 0.66$\pm$0.24 & 4.6  $\pm$1.2  & 0.74$\pm$0.11 & 3.3  $\pm$1.1 & 0.66$\pm$0.93 & 5.1  $\pm$1.3 & 0.76$\pm$0.11 \\       
       &  0.5  & 3.4 $\pm$ 1.1 & 0.67$\pm$0.42 & 4.7  $\pm$1.3  & 0.74$\pm$0.1  & 3.2  $\pm$1.0 & 0.66$\pm$0.45 & 5.0  $\pm$1.4 & 0.76$\pm$0.12 \\       
       & -0.5  & 3.2 $\pm$ 1.0 & 0.66$\pm$0.26 & 4.5  $\pm$1.3  & 0.74$\pm$0.12 & 3.3  $\pm$1.1 & 0.67$\pm$0.23 & 5.1  $\pm$1.4 & 0.77$\pm$0.18 \\       
  4278 &    0  & 2.8 $\pm$ 0.3 & 0.59$\pm$0.08 & 6.8  $\pm$0.6  & 0.82$\pm$0.03 & 2.8  $\pm$0.3 & 0.6 $\pm$0.07 & 7.7  $\pm$0.7 & 0.84$\pm$0.02 \\       
       &  0.5  & 2.8 $\pm$ 0.3 & 0.6 $\pm$0.07 & 7.0  $\pm$0.6  & 0.82$\pm$0.03 & 2.8  $\pm$0.3 & 0.59$\pm$0.07 & 7.5  $\pm$0.7 & 0.84$\pm$0.02 \\       
       & -0.5  & 2.7 $\pm$ 0.3 & 0.59$\pm$0.08 & 6.7  $\pm$0.6  & 0.82$\pm$0.03 & 2.9  $\pm$0.4 & 0.61$\pm$0.07 & 7.8  $\pm$0.7 & 0.84$\pm$0.02 \\       
  4365 &    0  & 12.3$\pm$ 1.3 & 0.69$\pm$0.05 & 31.8 $\pm$2.9  & 0.87$\pm$0.02 & 12.0 $\pm$1.2 & 0.68$\pm$0.06 & 29.6 $\pm$2.6 & 0.86$\pm$0.02 \\       
       &  0.5  & 11.0$\pm$ 1.2 & 0.65$\pm$0.06 & 28.3 $\pm$2.6  & 0.85$\pm$0.02 & 11.0 $\pm$1.2 & 0.65$\pm$0.06 & 27.1 $\pm$2.4 & 0.84$\pm$0.02 \\       
       & -0.5  & 13.0$\pm$ 1.4 & 0.7 $\pm$0.05 & 33.6 $\pm$3.0  & 0.87$\pm$0.02 & 12.5 $\pm$1.4 & 0.69$\pm$0.05 & 30.9 $\pm$2.7 & 0.86$\pm$0.02 \\       
  4374 &    0  & 14.2$\pm$ 3.5 & 0.75$\pm$0.12 & 23.7 $\pm$5.0  & 0.84$\pm$0.05 & 14.0 $\pm$3.3 & 0.74$\pm$0.09 & 22.7 $\pm$4.9 & 0.83$\pm$0.05 \\       
       &  0.5  & 12.9$\pm$ 3.1 & 0.72$\pm$0.09 & 21.5 $\pm$4.7  & 0.82$\pm$0.05 & 13.0 $\pm$3.2 & 0.72$\pm$0.12 & 20.9 $\pm$4.4 & 0.81$\pm$0.06 \\       
       & -0.5  & 14.9$\pm$ 3.6 & 0.76$\pm$0.08 & 24.8 $\pm$5.5  & 0.84$\pm$0.05 & 14.6 $\pm$3.5 & 0.75$\pm$0.1  & 23.5 $\pm$4.8 & 0.83$\pm$0.05 \\       
  4473 &    0  & 1.4 $\pm$ 0.3 & 0.37$\pm$0.17 & 3.8  $\pm$0.5  & 0.74$\pm$0.05 & 1.5  $\pm$0.3 & 0.39$\pm$0.16 & 4.5  $\pm$0.6 & 0.78$\pm$0.04 \\       
       &  0.5  & 1.5 $\pm$ 0.3 & 0.39$\pm$0.15 & 3.9  $\pm$0.6  & 0.75$\pm$0.05 & 1.5  $\pm$0.3 & 0.39$\pm$0.16 & 4.5  $\pm$0.6 & 0.78$\pm$0.04 \\       
       & -0.5  & 1.4 $\pm$ 0.3 & 0.35$\pm$0.19 & 3.7  $\pm$0.5  & 0.73$\pm$0.05 & 1.5  $\pm$0.3 & 0.4 $\pm$0.16 & 4.5  $\pm$0.6 & 0.78$\pm$0.04 \\       
  4486 &    0  & 23.9$\pm$ 1.7 & 0.8 $\pm$0.03 & 166.0$\pm$9.0  & 0.97$\pm$0.0  & 23.8 $\pm$1.7 & 0.79$\pm$0.03 & 139.0$\pm$7.8 & 0.96$\pm$0.01 \\       
       &  0.5  & 20.6$\pm$ 1.5 & 0.76$\pm$0.04 & 143.0$\pm$8.3  & 0.96$\pm$0.01 & 21.4 $\pm$1.5 & 0.77$\pm$0.03 & 125.0$\pm$7.0 & 0.96$\pm$0.01 \\       
       & -0.5  & 25.6$\pm$ 1.8 & 0.81$\pm$0.03 & 178.0$\pm$10.0 & 0.97$\pm$0.0  & 24.9 $\pm$1.8 & 0.8 $\pm$0.03 & 146.0$\pm$8.3 & 0.96$\pm$0.0  \\       
  4494 &    0  & 1.5 $\pm$ 0.2 & 0.29$\pm$0.14 & 2.2  $\pm$0.3  & 0.46$\pm$0.11 & 1.7  $\pm$0.3 & 0.36$\pm$0.15 & 2.5  $\pm$0.4 & 0.53$\pm$0.09 \\       
       &  0.5  & 1.6 $\pm$ 0.2 & 0.31$\pm$0.14 & 2.2  $\pm$0.3  & 0.47$\pm$0.11 & 1.7  $\pm$0.2 & 0.34$\pm$0.14 & 2.4  $\pm$0.3 & 0.52$\pm$0.1  \\       
       & -0.5  & 1.5 $\pm$ 0.2 & 0.29$\pm$0.15 & 2.2  $\pm$0.3  & 0.45$\pm$0.11 & 1.7  $\pm$0.3 & 0.36$\pm$0.14 & 2.5  $\pm$0.4 & 0.53$\pm$0.09 \\       
  4526 &    0  & 3.5 $\pm$ 0.6 & 0.53$\pm$0.11 & 7.4  $\pm$1.1  & 0.76$\pm$0.05 & 3.6  $\pm$0.7 & 0.54$\pm$0.11 & 8.0  $\pm$1.2 & 0.77$\pm$0.05 \\       
       &  0.5  & 3.5 $\pm$ 0.6 & 0.52$\pm$0.11 & 7.3  $\pm$1.1  & 0.75$\pm$0.05 & 3.5  $\pm$0.6 & 0.52$\pm$0.12 & 7.7  $\pm$1.2 & 0.76$\pm$0.05 \\       
       & -0.5  & 3.5 $\pm$ 0.6 & 0.53$\pm$0.11 & 7.5  $\pm$1.1  & 0.76$\pm$0.05 & 3.7  $\pm$0.6 & 0.55$\pm$0.11 & 8.1  $\pm$1.2 & 0.78$\pm$0.05 \\       
  4564 &    0  & 1.1 $\pm$ 0.2 & 0.56$\pm$0.16 & 1.3  $\pm$0.3  & 0.64$\pm$0.1  & 1.3  $\pm$0.3 & 0.65$\pm$0.13 & 1.7  $\pm$0.4 & 0.72$\pm$0.09 \\       
       &  0.5  & 1.1 $\pm$ 0.3 & 0.59$\pm$0.14 & 1.4  $\pm$0.3  & 0.66$\pm$0.11 & 1.3  $\pm$0.3 & 0.65$\pm$0.12 & 1.8  $\pm$0.4 & 0.73$\pm$0.09 \\       
       & -0.5  & 1.0 $\pm$ 0.2 & 0.55$\pm$0.15 & 1.3  $\pm$0.3  & 0.63$\pm$0.1  & 1.3  $\pm$0.3 & 0.65$\pm$0.13 & 1.7  $\pm$0.4 & 0.72$\pm$0.08 \\       
  4649 &    0  & 11.5$\pm$ 0.9 & 0.57$\pm$0.07 & 63.1 $\pm$4.4  & 0.91$\pm$0.01 & 11.1 $\pm$0.9 & 0.56$\pm$0.07 & 52.1 $\pm$3.5 & 0.89$\pm$0.01 \\       
       &  0.5  & 9.9 $\pm$ 0.8 & 0.5 $\pm$0.08 & 54.4 $\pm$3.7  & 0.9 $\pm$0.01 & 10.0 $\pm$0.8 & 0.51$\pm$0.08 & 47.1 $\pm$3.2 & 0.88$\pm$0.02 \\       
       & -0.5  & 12.3$\pm$ 1.0 & 0.6 $\pm$0.06 & 67.5 $\pm$4.7  & 0.92$\pm$0.01 & 11.6 $\pm$0.9 & 0.57$\pm$0.07 & 54.6 $\pm$3.7 & 0.9 $\pm$0.01 \\       
  4697 &    0  & 9.1 $\pm$ 2.4 & 0.86$\pm$0.07 & --  & -- & 9.3  $\pm$2.4 & 0.86$\pm$0.05 & -- & -- \\       
       &  0.5  & 9.2 $\pm$ 2.4 & 0.86$\pm$0.06 & --  & -- & 9.1  $\pm$2.3 & 0.86$\pm$0.06 & -- & -- \\       
       & -0.5  & 9.1 $\pm$ 2.3 & 0.86$\pm$0.05 & --  & -- & 9.4  $\pm$2.6 & 0.86$\pm$0.06 & -- & -- \\       
  5846 &    0  & 12.4$\pm$ 1.6 & 0.77$\pm$0.04 & 34.0 $\pm$3.6  & 0.91$\pm$0.02 & 12.4 $\pm$1.7 & 0.77$\pm$0.04 & 33.6 $\pm$3.6 & 0.91$\pm$0.02 \\       
       &  0.5  & 11.5$\pm$ 1.5 & 0.75$\pm$0.05 & 31.4 $\pm$3.3  & 0.9 $\pm$0.02 & 11.6 $\pm$1.5 & 0.76$\pm$0.05 & 31.2 $\pm$3.3 & 0.9 $\pm$0.02 \\       
       & -0.5  & 12.9$\pm$ 1.7 & 0.78$\pm$0.04 & 35.3 $\pm$3.8  & 0.91$\pm$0.01 & 12.9 $\pm$1.7 & 0.78$\pm$0.04 & 34.7 $\pm$3.7 & 0.91$\pm$0.02 \\       
  7457 &    0  & 1.1 $\pm$ 0.2 & 0.87$\pm$0.04 & 1.2  $\pm$0.2  & 0.88$\pm$0.04 & 1.7  $\pm$0.4 & 0.92$\pm$0.03 & 1.9  $\pm$0.4 & 0.92$\pm$0.02 \\       
       &  0.5  & 1.3 $\pm$ 0.3 & 0.89$\pm$0.04 & 1.3  $\pm$0.3  & 0.89$\pm$0.03 & 1.8  $\pm$0.4 & 0.92$\pm$0.03 & 1.9  $\pm$0.4 & 0.92$\pm$0.02 \\       
       & -0.5  & 1.0 $\pm$ 0.2 & 0.86$\pm$0.04 & 1.1  $\pm$0.2  & 0.87$\pm$0.04 & 1.6  $\pm$0.4 & 0.91$\pm$0.03 & 1.8  $\pm$0.4 & 0.92$\pm$0.03 \\       
  3607 &    0  & 2.6 $\pm$ 0.7 & 0.01$\pm$0.45 & 10.9 $\pm$2.6  & 0.74$\pm$0.1  & 2.6  $\pm$0.7 & 0.01$\pm$0.45 & 11.0 $\pm$2.7 & 0.74$\pm$0.09 \\       
       &  0.5  & 2.4 $\pm$ 0.6 & -0.0$\pm$0.45 & 10.3 $\pm$2.4  & 0.72$\pm$0.09 & 2.4  $\pm$0.6 & -0.0$\pm$0.45 & 10.3 $\pm$2.5 & 0.72$\pm$0.09 \\       
       & -0.5  & 2.6 $\pm$ 0.7 & 0.04$\pm$0.41 & 11.2 $\pm$2.7  & 0.75$\pm$0.09 & 2.7  $\pm$0.7 & 0.04$\pm$0.45 & 11.4 $\pm$2.7 & 0.75$\pm$0.08 \\       
  5866 &    0  & 1.3 $\pm$ 0.5 & 0.33$\pm$0.45 & --  & -- & 1.6  $\pm$0.6 & 0.46$\pm$0.45 & -- & -- \\       
       &  0.5  & 1.3 $\pm$ 0.5 & 0.36$\pm$0.45 & --  & -- & 1.6  $\pm$0.6 & 0.46$\pm$0.45 & -- & -- \\       
       & -0.5  & 1.2 $\pm$ 0.5 & 0.32$\pm$0.45 & --  & -- & 1.6  $\pm$0.6 & 0.46$\pm$0.45 & -- & -- \\       
\hline\end{tabular}
\end{table*}

\begin{table*}
\footnotesize
\captionsetup{width=.96\linewidth}
{\small
\caption{Mass estimates ($M_{\rm tot}$) and dark matter fractions ($f_{\rm DM}$) within $5{\re}$ and $R_{\rm max}$ assuming different anisotropy, obtained using the best--fit stellar mass--to--light ratios from the  dynamical modelling of \citet{Cappellari_2013, Cappellari_2013b}, i.e., total dynamical mass minus dark matter mass (see Section \ref{subs:DMfrac} for details). Columns 3--6 show $M_{\rm tot}$ and $f_{\rm DM}$ obtained by allowing $\alpha$ to vary while in columns 7--10, $\alpha\equiv0$, $R_{\rm max}$ can be found in Table \ref{tab:mass_summary}. \label{tab:appendix_3}}}
\begin{tabular}{lccccccccl}
\hline
 Galaxy & $\beta$  & $M_{\rm tot}(<5{\re})$ & $f_{\rm DM} (<5{\re})$ & $M_{\rm tot}(<R_{\rm max})$ & $f_{\rm DM}(<R_{\rm max})$ & $M_{\rm tot}(<5{\re})$ & $f_{\rm DM} (<5{\re})$ & $M_{\rm tot}(<R_{\rm max})$ & $f_{\rm DM}(<R_{\rm max})$ \\
$\rm [NGC]$ & &$[10^{11} \Msun]$ &   &$[10^{11} \Msun]$ &   &$[10^{11} \Msun]$ &   &$[10^{11} \Msun]$ &   \\
(1) & (2) &	(3) & (4) & (5) & (6) & (7) & (8) & (9) & (10) \\
\hline
  720  &    0 & 3.5 $\pm$0.7 & 0.46$\pm$0.13 & 11.5 $\pm$2.0 & 0.81$\pm$0.04 & 3.6 $\pm$0.7 & 0.46$\pm$0.14 & 13.2 $\pm$2.3 & 0.84$\pm$0.04 \\      
       &  0.5 & 3.4 $\pm$0.7 & 0.44$\pm$0.15 & 11.1 $\pm$1.8 & 0.8 $\pm$0.04 & 3.3 $\pm$0.7 & 0.42$\pm$0.15 & 12.3 $\pm$2.2 & 0.82$\pm$0.04 \\      
       & -0.5 & 3.6 $\pm$0.7 & 0.47$\pm$0.14 & 11.7 $\pm$2.0 & 0.81$\pm$0.04 & 3.7 $\pm$0.7 & 0.48$\pm$0.14 & 13.7 $\pm$2.3 & 0.84$\pm$0.04 \\      
  821  &    0 & 4.1 $\pm$0.8 & 0.78$\pm$0.06 & 5.7  $\pm$0.9 & 0.83$\pm$0.04 & 4.4 $\pm$0.8 & 0.8 $\pm$0.05 & 7.0  $\pm$1.2 & 0.86$\pm$0.03 \\      
       &  0.5 & 4.3 $\pm$0.8 & 0.79$\pm$0.06 & 5.9  $\pm$1.0 & 0.84$\pm$0.04 & 4.3 $\pm$0.8 & 0.79$\pm$0.05 & 6.8  $\pm$1.1 & 0.86$\pm$0.03 \\      
       & -0.5 & 4.0 $\pm$0.8 & 0.78$\pm$0.06 & 5.6  $\pm$0.9 & 0.83$\pm$0.04 & 4.5 $\pm$0.8 & 0.8 $\pm$0.05 & 7.1  $\pm$1.2 & 0.86$\pm$0.03 \\      
  1023 &    0 & 1.6 $\pm$0.2 & 0.63$\pm$0.08 & 3.5  $\pm$0.4 & 0.81$\pm$0.03 & 1.9 $\pm$0.3 & 0.68$\pm$0.07 & 5.4  $\pm$0.7 & 0.88$\pm$0.02 \\      
       &  0.5 & 1.7 $\pm$0.3 & 0.65$\pm$0.08 & 3.7  $\pm$0.5 & 0.82$\pm$0.03 & 1.8 $\pm$0.3 & 0.68$\pm$0.07 & 5.4  $\pm$0.7 & 0.88$\pm$0.02 \\      
       & -0.5 & 1.5 $\pm$0.2 & 0.61$\pm$0.08 & 3.4  $\pm$0.4 & 0.81$\pm$0.03 & 1.9 $\pm$0.3 & 0.68$\pm$0.07 & 5.4  $\pm$0.7 & 0.88$\pm$0.02 \\      
  1400 &    0 & 2.3 $\pm$0.5 & 0.46$\pm$0.18 & 6.9  $\pm$1.2 & 0.8 $\pm$0.05 & 2.4 $\pm$0.5 & 0.47$\pm$0.16 & 8.6  $\pm$1.5 & 0.84$\pm$0.04 \\      
       &  0.5 & 2.3 $\pm$0.5 & 0.46$\pm$0.37 & 6.9  $\pm$1.2 & 0.8 $\pm$0.05 & 2.2 $\pm$0.5 & 0.45$\pm$0.17 & 8.2  $\pm$1.4 & 0.83$\pm$0.04 \\      
       & -0.5 & 2.3 $\pm$0.5 & 0.46$\pm$0.18 & 6.8  $\pm$1.3 & 0.8 $\pm$0.05 & 2.4 $\pm$0.6 & 0.48$\pm$0.18 & 8.7  $\pm$1.6 & 0.84$\pm$0.04 \\      
  1407 &    0 & 10.8$\pm$1.1 & 0.61$\pm$0.06 & 33.9 $\pm$2.5 & 0.86$\pm$0.02 & 10.8$\pm$1.0 & 0.61$\pm$0.06 & 33.3 $\pm$2.4 & 0.86$\pm$0.02 \\      
       &  0.5 & 9.2 $\pm$0.9 & 0.55$\pm$0.06 & 28.9 $\pm$2.2 & 0.84$\pm$0.02 & 9.3 $\pm$0.9 & 0.55$\pm$0.07 & 28.7 $\pm$2.1 & 0.84$\pm$0.02 \\      
       & -0.5 & 11.6$\pm$1.1 & 0.64$\pm$0.05 & 36.4 $\pm$2.8 & 0.87$\pm$0.01 & 11.5$\pm$1.2 & 0.64$\pm$0.05 & 35.6 $\pm$2.6 & 0.87$\pm$0.02 \\      
  2768 &    0 & 6.8 $\pm$0.9 & 0.57$\pm$0.09 & 13.9 $\pm$1.8 & 0.77$\pm$0.04 & 6.9 $\pm$1.0 & 0.58$\pm$0.09 & 14.6 $\pm$1.8 & 0.78$\pm$0.04 \\      
       &  0.5 & 6.2 $\pm$0.8 & 0.52$\pm$0.1  & 12.6 $\pm$1.7 & 0.74$\pm$0.05 & 6.2 $\pm$0.8 & 0.53$\pm$0.1  & 13.1 $\pm$1.7 & 0.75$\pm$0.05 \\      
       & -0.5 & 7.1 $\pm$1.0 & 0.58$\pm$0.08 & 14.5 $\pm$1.9 & 0.77$\pm$0.04 & 7.3 $\pm$1.0 & 0.6 $\pm$0.08 & 15.3 $\pm$2.0 & 0.79$\pm$0.04 \\      
  3115 &    0 & 2.0 $\pm$0.3 & 0.57$\pm$0.08 & 5.4  $\pm$0.6 & 0.83$\pm$0.02 & 2.0 $\pm$0.3 & 0.57$\pm$0.08 & 6.5  $\pm$0.7 & 0.86$\pm$0.02 \\      
       &  0.5 & 2.1 $\pm$0.3 & 0.59$\pm$0.07 & 5.6  $\pm$0.6 & 0.83$\pm$0.02 & 2.0 $\pm$0.3 & 0.56$\pm$0.08 & 6.4  $\pm$0.7 & 0.85$\pm$0.02 \\      
       & -0.5 & 2.0 $\pm$0.3 & 0.56$\pm$0.08 & 5.3  $\pm$0.6 & 0.82$\pm$0.02 & 2.0 $\pm$0.3 & 0.57$\pm$0.08 & 6.6  $\pm$0.7 & 0.86$\pm$0.02 \\      
  3377 &    0 & 0.6 $\pm$0.1 & 0.57$\pm$0.1  & 1.2  $\pm$0.2 & 0.78$\pm$0.04 & 0.8 $\pm$0.1 & 0.68$\pm$0.07 & 2.0  $\pm$0.3 & 0.87$\pm$0.03 \\      
       &  0.5 & 0.7 $\pm$0.1 & 0.62$\pm$0.09 & 1.4  $\pm$0.2 & 0.81$\pm$0.04 & 0.8 $\pm$0.1 & 0.69$\pm$0.07 & 2.1  $\pm$0.3 & 0.87$\pm$0.02 \\      
       & -0.5 & 0.5 $\pm$0.1 & 0.54$\pm$0.1  & 1.2  $\pm$0.2 & 0.76$\pm$0.05 & 0.8 $\pm$0.1 & 0.67$\pm$0.07 & 2.0  $\pm$0.3 & 0.86$\pm$0.03 \\      
  3608 &    0 & 3.3 $\pm$1.1 & 0.78$\pm$0.25 & 4.3  $\pm$1.2 & 0.81$\pm$0.37 & 3.3 $\pm$1.1 & 0.78$\pm$2.63 & 5.1  $\pm$1.4 & 0.85$\pm$0.18 \\      
       &  0.5 & 3.5 $\pm$1.0 & 0.79$\pm$0.14 & 4.5  $\pm$1.3 & 0.82$\pm$0.22 & 3.3 $\pm$1.0 & 0.78$\pm$0.13 & 5.1  $\pm$1.4 & 0.84$\pm$0.08 \\      
       & -0.5 & 3.2 $\pm$1.1 & 0.77$\pm$0.25 & 4.1  $\pm$1.1 & 0.81$\pm$0.09 & 3.4 $\pm$1.1 & 0.78$\pm$0.31 & 5.2  $\pm$1.4 & 0.85$\pm$0.82 \\      
  4278 &    0 & 2.7 $\pm$0.3 & 0.68$\pm$0.06 & 6.3  $\pm$0.5 & 0.85$\pm$0.02 & 2.8 $\pm$0.3 & 0.69$\pm$0.06 & 7.7  $\pm$0.7 & 0.88$\pm$0.02 \\      
       &  0.5 & 2.9 $\pm$0.3 & 0.69$\pm$0.06 & 6.6  $\pm$0.6 & 0.86$\pm$0.02 & 2.8 $\pm$0.4 & 0.69$\pm$0.06 & 7.5  $\pm$0.7 & 0.87$\pm$0.02 \\      
       & -0.5 & 2.6 $\pm$0.3 & 0.67$\pm$0.06 & 6.1  $\pm$0.6 & 0.85$\pm$0.02 & 2.9 $\pm$0.3 & 0.7 $\pm$0.05 & 7.8  $\pm$0.7 & 0.88$\pm$0.02 \\      
  4365 &    0 & 11.6$\pm$1.2 & 0.78$\pm$0.04 & 27.4 $\pm$2.5 & 0.9 $\pm$0.02 & 11.8$\pm$1.2 & 0.78$\pm$0.04 & 29.1 $\pm$2.6 & 0.9 $\pm$0.02 \\      
       &  0.5 & 10.7$\pm$1.1 & 0.76$\pm$0.04 & 25.3 $\pm$2.3 & 0.89$\pm$0.02 & 10.6$\pm$1.1 & 0.76$\pm$0.04 & 26.2 $\pm$2.3 & 0.89$\pm$0.02 \\      
       & -0.5 & 12.0$\pm$1.2 & 0.79$\pm$0.04 & 28.5 $\pm$2.5 & 0.9 $\pm$0.01 & 12.4$\pm$1.3 & 0.79$\pm$0.04 & 30.5 $\pm$2.8 & 0.91$\pm$0.01 \\      
  4374 &    0 & 13.4$\pm$3.2 & 0.78$\pm$0.07 & 21.1 $\pm$4.5 & 0.85$\pm$0.05 & 13.5$\pm$3.3 & 0.78$\pm$0.09 & 21.7 $\pm$4.8 & 0.85$\pm$0.05 \\      
       &  0.5 & 12.2$\pm$2.9 & 0.76$\pm$0.08 & 19.3 $\pm$4.2 & 0.83$\pm$0.05 & 12.1$\pm$2.9 & 0.75$\pm$0.09 & 19.6 $\pm$4.2 & 0.84$\pm$0.05 \\      
       & -0.5 & 14.0$\pm$3.2 & 0.79$\pm$0.13 & 22.1 $\pm$4.7 & 0.85$\pm$0.04 & 14.2$\pm$3.4 & 0.79$\pm$0.07 & 22.8 $\pm$5.0 & 0.86$\pm$0.05 \\      
  4473 &    0 & 1.4 $\pm$0.3 & 0.6 $\pm$0.11 & 3.4  $\pm$0.5 & 0.82$\pm$0.03 & 1.5 $\pm$0.3 & 0.63$\pm$0.19 & 4.6  $\pm$0.6 & 0.87$\pm$0.02 \\      
       &  0.5 & 1.5 $\pm$0.3 & 0.64$\pm$0.09 & 3.7  $\pm$0.5 & 0.84$\pm$0.03 & 1.5 $\pm$0.3 & 0.63$\pm$0.11 & 4.6  $\pm$0.6 & 0.87$\pm$0.03 \\      
       & -0.5 & 1.4 $\pm$0.3 & 0.59$\pm$0.12 & 3.3  $\pm$0.5 & 0.81$\pm$0.04 & 1.5 $\pm$0.3 & 0.63$\pm$0.11 & 4.6  $\pm$0.6 & 0.87$\pm$0.03 \\      
  4486 &    0 & 22.4$\pm$1.6 & 0.81$\pm$0.03 & 136.0$\pm$7.6 & 0.96$\pm$0.0  & 22.4$\pm$1.7 & 0.81$\pm$0.03 & 131.0$\pm$7.4 & 0.96$\pm$0.0  \\      
       &  0.5 & 19.1$\pm$1.3 & 0.78$\pm$0.04 & 116.0$\pm$6.8 & 0.96$\pm$0.01 & 19.4$\pm$1.4 & 0.78$\pm$0.03 & 113.0$\pm$6.5 & 0.96$\pm$0.01 \\      
       & -0.5 & 24.1$\pm$1.7 & 0.83$\pm$0.03 & 146.0$\pm$8.3 & 0.97$\pm$0.0  & 23.9$\pm$1.7 & 0.83$\pm$0.03 & 140.0$\pm$8.0 & 0.97$\pm$0.0  \\      
  4494 &    0 & 1.5 $\pm$0.2 & 0.51$\pm$0.11 & 2.0  $\pm$0.3 & 0.61$\pm$0.07 & 1.7 $\pm$0.2 & 0.59$\pm$0.08 & 2.5  $\pm$0.4 & 0.7 $\pm$0.06 \\      
       &  0.5 & 1.5 $\pm$0.2 & 0.54$\pm$0.1  & 2.1  $\pm$0.3 & 0.63$\pm$0.07 & 1.7 $\pm$0.2 & 0.58$\pm$0.08 & 2.5  $\pm$0.4 & 0.7 $\pm$0.06 \\      
       & -0.5 & 1.4 $\pm$0.2 & 0.5 $\pm$0.11 & 1.9  $\pm$0.3 & 0.6 $\pm$0.08 & 1.7 $\pm$0.2 & 0.59$\pm$0.08 & 2.6  $\pm$0.4 & 0.7 $\pm$0.06 \\      
  4526 &    0 & 3.3 $\pm$0.6 & 0.53$\pm$0.11 & 6.8  $\pm$1.0 & 0.75$\pm$0.05 & 3.5 $\pm$0.6 & 0.55$\pm$0.11 & 7.7  $\pm$1.1 & 0.78$\pm$0.04 \\      
       &  0.5 & 3.3 $\pm$0.6 & 0.53$\pm$0.12 & 6.7  $\pm$1.0 & 0.75$\pm$0.05 & 3.3 $\pm$0.6 & 0.53$\pm$0.11 & 7.3  $\pm$1.1 & 0.77$\pm$0.05 \\      
       & -0.5 & 3.4 $\pm$0.6 & 0.54$\pm$0.12 & 6.9  $\pm$1.0 & 0.75$\pm$0.05 & 3.6 $\pm$0.6 & 0.57$\pm$0.11 & 7.9  $\pm$1.2 & 0.79$\pm$0.05 \\      
  4564 &    0 & 1.0 $\pm$0.2 & 0.65$\pm$0.11 & 1.3  $\pm$0.3 & 0.71$\pm$0.08 & 1.3 $\pm$0.3 & 0.74$\pm$0.09 & 1.8  $\pm$0.4 & 0.8 $\pm$0.07 \\      
       &  0.5 & 1.1 $\pm$0.3 & 0.68$\pm$0.12 & 1.4  $\pm$0.3 & 0.73$\pm$0.07 & 1.4 $\pm$0.3 & 0.75$\pm$0.09 & 1.8  $\pm$0.4 & 0.8 $\pm$0.06 \\      
       & -0.5 & 0.9 $\pm$0.2 & 0.63$\pm$0.11 & 1.2  $\pm$0.2 & 0.7 $\pm$0.08 & 1.3 $\pm$0.3 & 0.74$\pm$0.11 & 1.8  $\pm$0.4 & 0.79$\pm$0.07 \\      
  4649 &    0 & 10.6$\pm$0.8 & 0.63$\pm$0.05 & 50.9 $\pm$3.5 & 0.91$\pm$0.01 & 10.5$\pm$0.8 & 0.62$\pm$0.06 & 49.6 $\pm$3.2 & 0.91$\pm$0.01 \\      
       &  0.5 & 9.2 $\pm$0.7 & 0.57$\pm$0.07 & 44.2 $\pm$2.9 & 0.9 $\pm$0.01 & 9.2 $\pm$0.7 & 0.57$\pm$0.07 & 43.4 $\pm$2.9 & 0.9 $\pm$0.01 \\      
       & -0.5 & 11.3$\pm$0.9 & 0.65$\pm$0.06 & 54.2 $\pm$3.7 & 0.92$\pm$0.01 & 11.2$\pm$0.9 & 0.65$\pm$0.06 & 52.8 $\pm$3.5 & 0.92$\pm$0.01 \\      
  4697 &    0 & 8.9 $\pm$2.3 & 0.88$\pm$0.26 & -- & -- & 9.3 $\pm$2.3 & 0.89$\pm$0.04 & -- & --  \\      
       &  0.5 & 9.1 $\pm$2.2 & 0.89$\pm$0.04 & -- & -- & 9.0 $\pm$2.2 & 0.89$\pm$0.08 & -- & -- \\      
       & -0.5 & 8.8 $\pm$2.3 & 0.88$\pm$0.05 & -- & -- & 9.4 $\pm$2.4 & 0.89$\pm$0.05 & -- & -- \\      
  5846 &    0 & 11.8$\pm$1.5 & 0.76$\pm$0.05 & 30.7 $\pm$3.3 & 0.9 $\pm$0.02 & 11.7$\pm$1.6 & 0.76$\pm$0.05 & 31.6 $\pm$3.4 & 0.9 $\pm$0.02 \\      
       &  0.5 & 10.7$\pm$1.4 & 0.74$\pm$0.05 & 28.0 $\pm$3.1 & 0.89$\pm$0.02 & 10.5$\pm$1.3 & 0.74$\pm$0.05 & 28.4 $\pm$3.1 & 0.89$\pm$0.02 \\      
       & -0.5 & 12.3$\pm$1.7 & 0.77$\pm$0.04 & 32.1 $\pm$3.4 & 0.9 $\pm$0.02 & 12.3$\pm$1.6 & 0.77$\pm$0.04 & 33.3 $\pm$3.5 & 0.91$\pm$0.02 \\      
  7457 &    0 & 1.0 $\pm$0.2 & 0.92$\pm$0.03 & 1.1  $\pm$0.2 & 0.92$\pm$0.02 & 1.8 $\pm$0.4 & 0.95$\pm$0.02 & 2.0  $\pm$0.4 & 0.95$\pm$0.01 \\      
       &  0.5 & 1.2 $\pm$0.3 & 0.93$\pm$0.02 & 1.3  $\pm$0.3 & 0.93$\pm$0.02 & 1.9 $\pm$0.4 & 0.95$\pm$0.01 & 2.1  $\pm$0.5 & 0.96$\pm$0.01 \\      
       & -0.5 & 1.0 $\pm$0.2 & 0.91$\pm$0.03 & 1.0  $\pm$0.2 & 0.91$\pm$0.02 & 1.7 $\pm$0.4 & 0.95$\pm$0.02 & 1.9  $\pm$0.4 & 0.95$\pm$0.02 \\      
  3607 &    0 & 2.4 $\pm$0.7 & 0.21$\pm$0.45 & 9.3  $\pm$2.2 & 0.78$\pm$0.08 & 2.5 $\pm$0.7 & 0.27$\pm$0.45 & 10.8 $\pm$2.6 & 0.81$\pm$0.08 \\      
       &  0.5 & 2.3 $\pm$0.6 & 0.19$\pm$0.45 & 9.0  $\pm$2.2 & 0.77$\pm$0.1  & 2.4 $\pm$0.6 & 0.21$\pm$0.45 & 10.1 $\pm$2.4 & 0.79$\pm$0.07 \\      
       & -0.5 & 2.4 $\pm$0.6 & 0.23$\pm$0.36 & 9.4  $\pm$2.2 & 0.78$\pm$0.07 & 2.6 $\pm$0.7 & 0.3 $\pm$0.28 & 11.2 $\pm$2.8 & 0.81$\pm$0.07 \\      
  5866 &    0 & 1.2 $\pm$0.5 & 0.24$\pm$0.45 & -- & -- & 1.5 $\pm$0.6 & 0.41$\pm$0.45 & -- & --  \\      
       &  0.5 & 1.2 $\pm$0.5 & 0.27$\pm$0.45 & -- & -- & 1.5 $\pm$0.6 & 0.4 $\pm$0.45 & -- & -- \\      
       & -0.5 & 1.2 $\pm$0.5 & 0.22$\pm$0.45 & -- & -- & 1.6 $\pm$0.6 & 0.42$\pm$0.45 & -- & -- \\      
\hline\end{tabular}
\end{table*}
\bsp

\end{document}